\documentstyle[prd,aps,eqsecnum,epsfig]{revtex}

\def\no{\nonumber \\}
\def\le{\biggl (}
\def\ri{\biggr )}
\def\bml{\begin{mathletters}}
\def\eml{\end{mathletters}}

\begin{document}

\title{Second post-Newtonian gravitational wave  polarizations
 for compact binaries in elliptical orbits }

\author{A. Gopakumar$^{1,2}$ and Bala R. Iyer $^{3}$}

\address{ 
$^1$Department of Physics and McDonnell Center for the Space Sciences,
\\ Washington University, St. Louis, MO 63130, U.S.A\\     
$^2$Physical Research Laboratory, Navrangpura, Ahmedabad 380 009,
India\\
$^3$Raman Research Institute, C.V. Raman Avenue,
Sadashivanagar, Bangalore 560080,
India}

\date{\today}

\maketitle

\begin{abstract}
The second post-Newtonian (2PN) 
contribution to the `plus' and `cross' gravitational
wave polarizations associated with
gravitational radiation from  non-spinning, compact binaries
moving in elliptic orbits is computed. The computation starts from our earlier 
results on 2PN generation,  crucially employs the 2PN accurate
generalized quasi-Keplerian parametrization of elliptic orbits by
Damour, Sch\"afer and Wex and provides 2PN accurate 
expressions modulo the tail terms for 
gravitational wave polarizations incorporating effects of eccentricity
and periastron precession.
\end{abstract}


\section{Introduction}

\label{sec:intro}
Inspiralling compact binaries 
containing
black holes and neutron stars  are one of
the most promising sources of gravitational radiation for both,
almost operational
ground based laser interferometric
gravitational wave detectors
like  LIGO, VIRGO, GEO600 and TAMA300 \cite{ligo}
and the proposed space-based interferometer LISA \cite{lisa}.
To obtain an acceptable signal to noise ratio for detection
in the terrestrial detectors,
one needs to know {\it a priori } the binary's orbital evolution 
in the inspiral waveform \cite {CFPS93} at least upto third
post-Newtonian 
 order beyond the (Newtonian) quadrupole radiation. 
However,
for the 
measurement of distance and position of the binary,
it may be sufficient to know the
two independent gravitational wave polarizations 
$ h_{+}$ and $ h_{\times}$
to only 2PN accuracy \cite{CF94}.
Perturbative computation  via post-Newtonian (PN) 
expansions of the
binary orbit and gravitational wave
(GW) phase   are complete  to  order $v^5$ 
beyond the standard quadrupole formula.
Extension of the PN perturbative
calculations by another two orders, to order $v^7$, is still not complete,
because currently used PN techniques \cite{3PNc}
 leave undetermined a physically crucial parameter
entering at the $v^6$ level in the gravitational wave flux \cite{BIJ01}.
More recently  \cite{DIS98},  it has been shown that 
 by employing  several {\it re-summation techniques}
 -- to improve the convergence of the  PN series --
  one could  make  optimal use of existing 2PN results
to compute 
GW phasing.  Resummed  versions of 
2PN accurate 
search templates may be {\em just sufficient} both 
for the detection and estimation of parameters of gravitational
waves from inspiralling compact binaries of arbitrary mass ratio 
moving in  {\it quasi-circular orbits}.
For inspiralling non-spinning 
compact binaries of arbitrary mass ratio in {\em quasi-circular} 
orbits, both the 2PN accurate  gravitational wave polarizations 
\cite{BIWW96}
and the  associated orbital evolution 
have been explicitly computed 
\cite{BDIWW95,BDI95,WW96}.
A 2.5PN accurate formula for 
the orbital phase as a function of time 
has also  been obtained \cite{LB96}. 
These expressions 
are employed by various data analysis packages
like LAL \cite{LAL}
to search for gravitational waves from inspiralling compact  binaries.

          The purpose of the present work is to 
obtain the `instantaneous'\footnote{
Following \cite{BDI95}, we term contributions to the GW 
waveform which depends only on the state of the binary at the retarded 
instant as its `instantaneous' part.}
 2PN contributions to the 
two gravitational wave polarizations  for 
compact binaries moving in {\em elliptical orbits}.
On the one hand, these expressions for $ h_{+}$ and $ h_{\times} $ 
represent gravitational waves from a binary 
evolving negligibly under gravitational radiation reaction,
incorporating precisely upto 2PN order the effects of
eccentricity and periastron precession, during 
that stage of inspiral when the orbital parameters are essentially 
constant over a few orbital revolutions.
On the other hand, 
it  is the first  
(and the necessary) step in the direction of obtaining `ready to use' 
theoretical templates to search for gravitational waves from 
inspiralling compact binaries  moving in {\em quasi-elliptical} orbits.
The effect of radiation reaction on  orbital evolution  and its  consequence
on these  gravitational waveforms for compact binaries in 
quasi-elliptical orbits is under investigation and will be
discussed  in the future \cite{DGI}.

          Galactic binaries, in general, will be in circular orbits
by the time they reach the final stage of inspiral.
However, there exist astrophysical scenarios where 
compact binaries will have non-negligible
eccentricity during the  final inspiral phase.
We will next  review various such scenarios,
some of them speculative, relevant for
both ground and space based gravitational wave detectors.

              Let us first consider cases that
should be important for ground based interferometers.
Intermediate mass black hole binaries
-- with total masses in the range $ 50\,M_{\odot}
\leq M \leq $ (a~few) $\times 10^2 M_{\odot}$ -- may well be
the first sources to be detected
by LIGO and VIRGO \cite{FH97,lpp97}. 
Many recent astronomical observations, 
involving massive black hole candidates
point to scenarios involving such compact
binaries in eccentric orbits. 
The discovery of numerous  bright compact X-ray sources with 
luminosities $L>10^{39} \,{erg/s}$ in several
starburst galaxies and rapid time variation of their  X-ray fluxes 
implies massive black holes as their central engines.
It is suggested that these observations may be explained by
the merger of globular clusters, containing black holes with 
$ M > 10^3\,M_{\odot}$, with its host galaxy \cite{MH01}. 
However, $10^3\,M_{\odot}$ black hole present in the center 
of the globular cluster will have to be created by 
 many coalescences of    
a $\geq 50\,M_{\odot}$ black hole with lighter ones and    
these binaries, in highly eccentric orbits, 
should be visible to ground based interferometers.
This scenario may be contrasted with the one suggested in \cite{ZM00}
which also involves compact binaries with high eccentricities. 
However, in this case, black hole binaries, weighing a few solar masses and 
residing in star clusters,
get ejected from the cluster by superelastic encounter with other cluster
members. These escaping binaries will have short periods and high 
eccentricities before merging.
It is worth mentioning that numerical simulations 
dealing with
supermassive blackhole formation,
performed 
in the eighties,  from a dense cluster of compact 
stars also indicate creation of 
short period intermediate mass 
black hole binaries in highly eccentric orbits            
\cite{ST85}.

      Recently,  there has been studies suggesting that      
spinning compact binaries may become chaotic \cite{JL00}.
The analysis involves numerical evolution of
two spinning point masses using  2PN
accurate equations of motion.  
The interesting result, observed only for
a very restricted portion of the parameter space,
is that the  outcome of the
evolution is highly sensitive to initial conditions. 
It is   
also observed that binaries 
whose initial orbits are circular may  later become highly eccentric.
These preliminary results  present yet another  
 scenario where eccentricity may become
important.

            Many of the potential sources 
for LISA \cite{lisa}
will be  binaries in `quasi-elliptical' orbits.
We list them below, details and  references to  original papers 
may  be found in \cite{LISA-MC98}.  
 First,
LISA will be sensitive to massive black hole (MBH)
coalescence involving 
$10^3$ to $10^7 M_{\odot}$ 
black holes, 
upto 3 Gpc and beyond. 
It is likely that these binaries
will be in eccentric orbits during inspiral,
as they will be interacting with
dense stellar clusters in the galactic nuclei where they usually reside.
The second candidate involves compact objects orbiting MBH,
where
 compact objects could be   scattered
into very short period eccentric orbits  via gravitational deflections 
by other stars.
Finally, LISA will be sensitive to thousands of binaries
in our galaxy 
and  many of these
short period binaries will also be in 
`quasi-eccentric' orbits. 
Interestingly, LISA will be highly sensitive to black hole binaries 
containing  primordial black holes of mass $\sim 0.5 \,M_{\odot}$.   
These binaries are one of the speculative
candidates for MAssive Compact Halo Objects (MACHOs)
\cite{NSTT97}.  
It is also shown that
\cite{WH98,ITN98} the low frequency
gravitational waves from black hole MACHO binaries
in highly eccentric orbits
would form a strong stochastic background in the
frequency range $ 10^{-5}\,{\rm Hz} < f < 10^{-1}\,{\rm Hz}$,
where LISA will be most sensitive. 

     Finally, we observe that eccentricity will be an important
parameter while searching for continuous gravitational wave sources 
in binary systems.
Recently, it was shown that searching for gravitational waves from
such systems, whose locations are exactly known,
is computationally feasible \cite{DV00}. 
For many such astrophysically interesting systems, 
we note that 
a post-Newtonian orbital description for generic orbits 
will be required.

             The computation of the gravitational wave polarizations  
$  h_{+}$ and $ h_{\times}$  
in terms of the orbital phase and frequency of the binary was 
 discussed by Lincoln and Will \cite {LW90},
using the method of osculating orbital elements from celestial
mechanics and the 2.5PN accurate Damour-Deruelle equations of motion
\cite {DD81,DD85}.
They studied the evolution of general orbits and obtained 
1PN accurate expressions for $ h_{+}$ and $ h_{\times}$ 
for quasi-circular orbits. 
Later Moreno-Garrido, Mediavilla and Buitrago 
obtained polarization waveforms for binaries in 
elliptical orbits at Newtonian order with and without radiation reaction,
studied the effects of  orbital parameters and precession
on gravitational wave amplitude spectrum and implications
for data analysis \cite {MBM94,MBM95}.
Analytic expressions for gravitational wave 
polarizations and far-zone fluxes, for
elliptic binaries were obtained to 1.5PN order by 
Junker and  Sch\" afer, and  Blanchet and  Sch\" afer \cite {JS92,BS93}.
The 2PN accurate gravitational wave polarizations   
for inspiralling compact binaries  moving in {\em quasi-circular}
orbits  was given by  Blanchet, Iyer, Will and Wiseman \cite {BIWW96}.
For the above calculation they employed the 2PN accurate expressions
for $h^{TT}_{ij}$, the transverse traceless part of the 
radiation field representing the deviation of the metric from
the flat spacetime  
and $ \left({d{\cal E}\over dt}\right)$, the far-zone
energy flux obtained independently using two different 
formalisms \cite {B95,BDIWW95,BDI95,WW96}.
In the limiting case of a test particle
orbiting a Schwarzschild black hole, 
perturbative calculations are extended to very high
PN order. For example, in the case of very small mass ratios,
polarization waveforms are
obtained to 4PN order \cite{TS94}.
For the case of spinning compact objects in circular orbits,
precessional, non-precessional and dissipative effects 
on the gravitational waveform due 
to spin-orbit and spin-spin interactions 
 were studied extensively 
\cite {KWW93,ACST94,K95,OTO98}.
We note that using the framework we employ here 
it may be possible to 
extend results of these papers to compact binaries 
of arbitrary mass ratio moving in
elliptical orbits. 
            
             The basic aim of  this paper  is to
 obtain   the instantaneous 2PN corrections to the  `plus' and 
`cross' polarization waveforms for  compact binaries 
of arbitrary mass ratio moving in elliptical orbits
 starting  from the  corresponding 
2PN contributions to  $h^{TT}_{ij}$ \cite {WW96,GI97}.
 As emphasized in \cite{DIS98},
the gravitational wave observations of  inspiralling compact binaries,
  is analogous to the high precision radio-wave
observations of binary pulsars.
 The latter makes use of an accurate relativistic `timing
formula'  based on the  solution  
 - in quasi-Keplerian parametrization -
to the relativistic  equation of
motion  for  a compact  binary moving  in an  elliptical orbit\cite{DD86}.
 In a similar manner,  the former  demands  accurate `phasing',
i.e. an accurate mathematical modeling of the continuous time
evolution of the gravitational waveform.
 This requires for elliptical binaries,
a convenient solution to the 2PN accurate equations  of motion.
 A very elegant 2PN accurate 
generalized quasi-Keplerian parametrization
for elliptical orbits  
has been implemented by Damour, Sch\" afer,
and Wex \cite{DS87,DS88,SW93,NW95}. This   representation
is  thus the most natural and best suited for our purpose to
parametrize the dynamical variables that enter the gravitational
waveforms.
The complete 2PN accurate expressions for 
$  h_{+}$ and $ h_{\times}$ consists   of  the 
`instantaneous' contribution  computed here supplemented by 
tail contributions at 1.5PN and 2PN orders. 
The tail computations  are not considered here; they must  be 
computed and   included in the future.

            The paper is organized as follows: 
In section~\ref{sec:waveforms}, we present the details
of the computation to
obtain `instantaneous' 2PN corrections to 
$ h_{+}$ and $ h_{\times}$ for
inspiralling compact binaries moving in elliptical orbits.
Section~\ref{sec:spc}  deals with the influence 
of the orbital parameters on the
waveform.  
Section~\ref{sec:concl}  comprises our concluding remarks.
 
\section{The 2PN gravitational wave
polarization states}
\label{sec:waveforms}
To compute the two independent
gravitational wave  
polarization states
$h_{+} $ and $ h_{\times} $, 
one needs to choose a convention for the direction and orientation of
the orbit. 
We follow the standard convention of choosing
a triad of unit vectors composed of
${\bf N}$, a unit vector along the radial direction to the observer,
{\bf p}, a unit vector 
along the line of nodes, which coincides with y-axis  and 
{\bf q}, defined  by
${\bf q} = {\bf N} \times {\bf p}$ (see Fig.~\ref{fig:config}).  
The angle between {\bf N} and 
the Newtonian angular momentum vector 
which lies along z-axis
defines the inclination angle $i$ of the orbit.
The orbital phase $\phi$ is measured
from the positive x-axis in a counter clockwise sense,
restricting the values of $i$ from $0$ to ${1 \over 2}\,\pi $. 
The two basic polarization states
$h_{+} $ and $ h_{\times} $ are given by

\begin{mathletters}
\label{hcpf}
\begin{eqnarray}
 h_{+} &=& { 1 \over 2} \biggl ( p_i\,p_j
-q_i\,q_j \biggr )\,h_{ij}^{TT}\,,\\
 h_{\times} & = &{ 1 \over 2} \biggl ( p_i\,q_j
+p_j\,q_i \biggr )\,h_{ij}^{TT} \,,
\end{eqnarray}
\end{mathletters}
where $ h_{ij}^{TT}$ is 
the transverse-traceless (TT) part of the radiation
field representing the deviation of the metric from the flat spacetime.

   From Eqs. (\ref{hcpf}) it is clear that the explicit computation 
of 2PN corrections to 
$  h_{+} $ and $ h_{\times} $ requires the following:
(a) The 2PN corrections to $ h_{ij}^{TT}$,
generally given 
in terms of the dynamical variables of the binary, namely  $\,  
v^2, {G\,m \over r}, \dot r, n_i, v_i, {\bf N. n}, {\bf N.v} $,
where {\bf r} and  {\bf v} are respectively, the relative position and
velocity vectors for the two masses $m_1 $ and $m_2$
in the center of mass frame, 
$ r=|{\bf r}|, v=|{\bf v}|,{\bf n}={ {\bf r} \over r},
 \dot r= {d r\over dt}$ and $
m=m_1 +m_2$. The unit vector {\bf N} lies along the radial direction to
the detector and is given by ${\bf N} = {{\bf R} \over R}$, $R$ being 
the radial distance to the binary;
and (b)  A 2PN accurate orbital representation for elliptical 
orbits to parametrize these dynamical variables.  

 Before explaining in detail the procedure to compute 
2PN contributions to $h_{+} $ and $h_{\times}$, we will first
illustrate that computation by presenting in detail 
the Newtonian computations for $h_{+} $ and $h_{\times}$. 
   
\subsection{The Newtonian GW  Polarizations}

At the  leading Newtonian order, we have 
\begin{equation}
\label{wff0}
(h^{TT}_{km})_{N} = {4\,G\, \mu \,\over c^4\,R}
 {\cal P}_{ijkm}({\bf N})
 \le  v_{ij} -{G\,m \over r}\,n_{ij} \ri \,,
\end{equation}
where $ {\cal P}_{ijkm}({\bf N}) $ is the
usual transverse traceless
projection operator projecting normal to {\bf N},
$v_{ij}=v_iv_j\;,$  $\;n_{ij}=n_in_j$; and
 $\mu$ is the reduced mass of the binary, given 
by $ m_1\,m_2 /m$. 
Note that the  above contribution arises from the mass quadrupole moment
of the binary.

  There is no need to apply the  TT projection in Eq. (\ref{wff0}),
 and Eq. (\ref{hcpf}) at the leading order gives
\begin{mathletters}
\label{hcpd}
\begin{eqnarray}
h_{+} &=& { 2\,G \,\mu\over c^4\,R} 
\biggl \{
 \biggl ( p_i\,p_j -q_i\,q_j \biggr )\,
\biggl ( v_{ij} - {G\,m \over r}\,n_{ij}
 \biggr ) \biggr \}
\,,
\nonumber \\
&= &
{2\,G\,\mu \over c^4\,R} \biggl \{  \biggl ( ({\bf p .v })^2 -
({\bf q.v})^2 \biggr ) + { G\,m \over r}
\biggl ( ({\bf p.n})^2 - ({\bf q. n})^2 \biggr ) \biggr \}\,,
\\ 
h_{\times} & =& { 2\,G\,\mu \over c^4\,R}
\biggl \{
 \biggl ( p_i\,q_j +p_j\,q_i \biggr )\,
\biggl (\,v_{ij} - {G\, m \over r}\,n_{ij}
 \biggr )
\biggr \}
\nonumber \\
&=& 
{4\,G\,\mu \over c^4\,R} \biggl \{
\,({\bf p.v})
\,({\bf q.v}) - {G\,m \over r} \,({\bf p.n})\,({\bf q. n})
\biggr \}\,.
\end{eqnarray}
\end{mathletters}    

      The convention we adopted to define the triad of unit vectors
implies
${\bf p} = ( 0, 1,0),\,$
${\bf q} = ( -\cos i, 0, \sin i),\,$
${\bf N} = ( \sin i, 0, \cos i),\,$
$ {\bf n}= ( \cos \phi, \sin \phi, 0),$ and
${\bf v} =
( \dot r\,\cos \phi - r\,\dot \phi \,\sin \phi,\, \dot r \,\sin \phi
 + r\,\dot \phi \,\cos \phi,\, 0) $, 
where 
$\dot \phi ={d \phi / dt}$. 
With these inputs, Eq. (\ref{hcpd}) becomes,
\begin{mathletters}
\begin{eqnarray}
h_{+} &=& {G\,m\,\eta \,C \over c^4\,R}\,\biggl \{
( 1 +C^2) \biggl [ 
\biggl ( {G\, m \over r} + r^2\,\dot \phi^2 -\dot r^2 
\biggr )
\cos 2\,\phi + 2\,\dot r\, r\,\dot \phi \, \sin 2\,\phi
\biggr ]
\no
&&
- S^2
\,\biggl [ {G\, m \over r} 
-r^2\,\dot \phi^2 -\dot r^2 \biggr ]
\biggr \}\,,
\\
h_{\times} &=& 2 {G\,m\,\eta \,C \over c^4\,R}\,\left \{  
\le {G \,m \over r} + r^2\,{\dot \phi}^2 - \dot r^2 
\ri \sin 2\phi - 2 \dot r \,r\,\dot \phi \,\cos 2\phi 
\right \}\,,
\end{eqnarray}
\label{Eq.2.4}
\end{mathletters}
where $ \eta = \mu /m$  and $ C$ and $ S$ are shorthand notations for $ \cos i$
and $ \sin i$.

   When  dealing with elliptical orbits,
it is convenient and useful to use a representation
to rewrite the dynamical variables $ r, \dot r, \phi $ 
and $ \dot \phi$ in terms of the parameters describing an 
elliptical orbit.
For example, in  Newtonian dynamics,  
the  Keplerian representation in terms of 
eccentricity, semi-major axis,  eccentric, real and mean anomalies
is a convenient  solution to the Newtonian equations of motion
for two masses in  elliptical orbits.
The  Keplerian representation reads:
\begin{mathletters}
\begin{eqnarray}
r &=& a \left ( 1 -e \cos u\right )\,,\\
n( t -t_0 )= l &=& u -e \,\sin u \,,
\\
\phi -\phi_0 &=& v\,,\\
\mbox{ where } v &=& 2\, \tan^{-1}\left \{ \left (
{1 +e \over {1 -e }}
\right)^{ 1\over 2} \, \tan ( { u \over 2}) \right \}\,, 
\end{eqnarray}
\label{admrn}
\end{mathletters}  
where $u, \, l,\, v$ are the eccentric, mean and real 
anomalies 
parametrizing the motion and the constants  
$ a,\, e,\,n,\, t_0,\,\phi_0$ represent semi-major axis, eccentricity, 
mean motion, some initial instant and the orbital 
phase corresponding to that instant respectively.
These constants which characterize a given eccentric orbit 
may be expressed, at the Newtonian order,
in terms of  the  conserved 
energy $E$ and angular momentum per unit reduced mass $|{\bf J}|$ 
as
\begin{mathletters}
\label{Keppar}
\begin{eqnarray}
a &=& { G\,m \over (-2\,E) }\,,\\
e &=& 1 +2\,E\,h^2\,,\\
n &=&{(-2\,E)^{3 \over 2} \over G\,m}\,,
\end{eqnarray}
\end{mathletters}
with $h = { |{\bf J}| \over G\,m}$.
Note that $ n= {2 \, \pi \over T} $, where T is the orbital period.
 
In the case of circular orbits $e=0$  and $v=u=l$, $\phi$ is thus
a linearly increasing  function  of time and $\dot{r}=0$, 
$\dot{\phi}=n={2\,\pi \over T}$. The polarizations are uniquely given
by the  straightforward  substitutions of these simple limiting forms.
The only residual choice is whether one uses the gauge-dependent variable
$\gamma= {G\,m \over {c^2 \,r}}$ or the gauge-independent variable 
$x=(\pi m F_{\rm GW})^{2/3}$. The situation is more involved
in the case of general orbits even at the leading Newtonian order.
Indeed, if $e\neq 0$, then  $v\neq u\neq l$,
$v(u)$ and $u$ are more complicated functions of $l$  and thus 
$\phi$ is not a simple linearly
increasing function of time. This is why a straightforward representation
of the polarizations in terms of $v$ and $u$ or even a more involved one
in terms of $u$ only, which may be
obtained by explicit elimination of $v$ are
inadequate. The clue to the  correct description  follows from the 
analysis of Damour \cite{td83}   for the 
2PN accurate equations of motion of a compact binary.
Here it was shown that  the basic  dynamics   
can  be represented as a function of two variables\footnote{ We denote
by $\lambda$ the variable denoted by $m$ in Ref.\cite{td83} to avoid
confusion with the total mass $m$ in most  current literature including here.}
 $\lambda$ and $l$ and be $2\pi$ -periodic in  both  of them.
The GW polarizations will inherit this double periodicity and we
shall crucially exploit it as follows:
 We will  split $\phi$ into a part $\lambda$ linearly increasing
with time and the remaining part denoted by ${\rm W}(l)$ which is a
periodic function of $l$:
\begin{equation}
\phi = \lambda + {\rm W}(l)\,,
\end{equation}
where $ \lambda $ and $ {\rm W}(l) $ are given by
\begin{mathletters}
\label{lwN}
\begin{eqnarray}
 \lambda &=& \phi_0 + l\,, \\
{\rm W}(l) &=& \left ( v(u(l)) - u(l) + e\, \sin u(l) \right ).
\end{eqnarray}
\end{mathletters}
It is worth emphasizing following \cite{td83} that though one could consider
$\lambda$  to be a function of $l$ given by $\lambda =\phi_0 +l$, 
it is more  advantageous  to   consider $\phi$ and
consequently the GW  polarizations to be independently  periodic in
both $\lambda$ and $l$.
With this observation, as we will see below,
 it is natural to split the $\lambda$ and ${\rm W}(l)$
dependence in the polarization and consider the $\lambda$ dependence as
representing the harmonic  time  dependence and
 the ${\rm W}(l)$ term as representing
the time varying  amplitude modulation (`nutation').  
This clean separation also facilitates 
a simple and precise treatment of the spectral decomposition of the 
GW polarizations as shown in Section~\ref{sec:spc}.
 Finally, we note that this decomposition is not just appropriate to discuss 
effects of eccentricity on the Newtonian waveform  and the periastron
precession at 1PN order but powerful enough to analyze all  PN
effects upto the 2PN order.

Armed with   the above important  conceptual input, 
the computation of $h_+ $ and $h_{\times}$  involves a  routine, 
albeit lengthy  algebra.  
It is straightforward to obtain Newtonian expressions for 
$ r,\, \dot r$ and $\dot \phi$ in terms of $n,~ e$ and
$u$ using Eqs. (\ref{admrn}), (\ref{Keppar}) and relations for 
$E$ and $h^2$, given at Newtonian order by 
$ (-2\,E) = (G\,m\,n)^{2 \over 3}$, 
$ (-2\,E\,h^2 )= 1 -e^2 $.
They are given by
\begin{mathletters}
\begin{eqnarray}
r &=& ({G\,m \over n^2})^{1 \over 3}\,(1 -e\,\cos u) \,,\\
\dot r &=&{( G\,m\,n)^{1 \over 3}\,e\,\sin u \over (1 -e\,\sin u)} \,,\\
\dot \phi &=& { n\, \sqrt {1 -e^2} \over (1 -e\,\sin u)^2} \,.
\end{eqnarray}
\label{Eq.2.9}
\end{mathletters}

Using Eqs. (\ref{Eq.2.9}) and  splitting $\phi$ as 
$\phi = \lambda + {\rm W}(l) $ in 
Eqs. (\ref{Eq.2.4}), we obtain
after some manipulation:

\begin{mathletters}
\label{hpcN}
\begin{eqnarray}
h_{+} &=&
{ G \, m \eta \over c^2\,R}\,  (
{ G\,m\, n \over c^3})^{2/3}\,
{1 \over (1 -e\,\cos u)^2 }
\biggl \{
\left (1+{C}^{2}\right )
\biggl [         
\biggl ( (e\,\cos u)^2 -e\,\cos u
\no
&&
 -2\,e^2 +2 
\biggr )\, \cos 2\,{\rm W}
+ 2\,e\,\sqrt{1 -e^2}\,\sin u\,\sin 2\,{\rm W}
\biggr ]\,\cos 2\,\lambda
\no
&&
+
\left (1+{C}^{2}\right ) \biggl [
 2\,e\,\sqrt{1 -e^2}\, \sin u\, \,\cos 2\,{\rm W}
-
\biggl ( (e\,\cos u)^2 -e\,\cos u 
\no
&&
-2\,e^2 +2 
\biggr )\, \sin 2\,{\rm W}
\biggr ]\,\sin 2\,\lambda
+  S^2\, e\,\cos u\, (1 -e\,\cos u) \biggr \}\,,\\
h_{\times} &=&
{ G \, m \eta \over c^2\,R}\,
({ G\,m\, n \over c^3})^{2/3}\,
{1 \over (1 -e\,\cos u)^2}\,C
\biggl \{ 
\biggl [
-\biggl ( 4\,e\,\sqrt{1 -e^2}\,\sin u\, \biggr )\,
 \cos 2\,{\rm W}
\no
&&
+
\biggl ( 2\,(e\,\cos u)^2 -2\,e\,\cos u -4\,e^2 +4 \biggr )\,
\sin 2\,{\rm W}
\biggr ]\,\cos 2\,\lambda
\no
&&
+\biggl [
\biggl ( 2\,(e\,\cos u)^2 -2\,e\,\cos u -4\,e^2 +4 \biggr )\,
\cos  2\,{\rm W} 
\no
&&
+ \biggl ( 4\,e\,\sqrt{1 -e^2}\,\sin u\, \biggr )
\, \sin 2\,{\rm W} \biggr ]\, \sin 2\,\lambda \biggr \}\,.
\end{eqnarray}
\end{mathletters}

   We observe that terms like  
$ \sin {\rm 2\,W}\,\sin {\rm 2\,\lambda},$
$\cos {\rm 2\,W}\,\cos 2\,\lambda $, and
$\cos {\rm 2\,W}\,\sin {\rm 2\,\lambda}$,
$\sin {\rm 2\,W}\,\cos 2\,\lambda $  are
generated respectively from $\cos 2\,\phi$   and 
$\sin 2\phi $ terms
in Eqs. (\ref{Eq.2.4})
due to the  $ \phi = {\rm W} + \lambda $ split. 
The circular limit of Eqs. (\ref{hpcN}), 
obtained by putting $e= 0 $, agrees with the  Newtonian  terms in
Eqs. (2), (3) and (4) of \cite{BIWW96}. 

\subsection{ The 2PN  GW  Polarizations}

The computation of 2PN corrections to $ h_+ $ and $ h_{\times}$
is similar in principle  to the Newtonian calculation. However, there are 
subtleties and technical details
which will be presented, in some  detail, below.  

   From the  Newtonian calculations, it is easy to note that 
we require a 2PN accurate orbital representation for 
computing 2PN corrections to $ h_+ $ and $ h_{\times}$. 
We employ  the most Keplerian-like solution to the
2PN accurate equations of motion,
obtained by 
Damour, Sch\" afer, and Wex \cite {DS88,SW93,NW95},
given in the usual polar representation 
associated with  the  Arnowit, Deser and Misner (ADM) coordinates. 
It is known  as  the generalized quasi-Keplerian parametrization
and  represents the 2PN motion of a binary containing two compact objects
of arbitrary mass ratio, moving in an elliptical orbit.
The relevant details of the representation is summarized  in what follows.

          Let $r(t),\, \phi(t)$ be the usual polar coordinates 
in the plane of relative motion of the two compact objects
in the ADM gauge.
The radial motion $r(t)$  is conveniently  parametrized  by
\begin{mathletters}
\begin{eqnarray}
r &=& a_r \left ( 1 -e_r \cos u\right )\,,\\
n( t -t_0 ) =l &=& u -e_t \,\sin u + { f_t \over c^4}\sin v
+ { g_t \over c^4}\left ( v -u\right )\,,
\end{eqnarray}
\label{admr}
\end{mathletters}
\noindent where $u$ is the `eccentric anomaly' 
parametrizing the motion and
 the constants $a_r,\, e_r,\, e_t,\, n$ and $t_0$ are
some 2PN semi-major axis,
radial eccentricity,  time eccentricity,  mean motion, and
initial instant respectively. The angular motion $\phi(t)$
is given by
\begin{mathletters}
\begin{eqnarray}
\phi -\phi_0 &=& \left ( 1 +{ k \over c^2} \right )v
+ { f_{\phi} \over c^4} \sin 2v + { g_{\phi} \over c^4}\sin 3v\,,\\
\mbox{ where } v &=& 2\, \tan^{-1}\left \{ \left (
{1 +e_{\phi} \over {1 -e_{\phi}}}
\right)^{ 1\over 2} \, \tan ( { u \over 2}) \right \}\,.
\end{eqnarray}
\label{admphi}
\end{mathletters}
In the above $ v$ is some real anomaly, $\phi_0, k, e_\phi$ are 
some initial phase,  periastron
precession constant, and  angular eccentricity respectively.

  The main difference between the relativistic orbital representation
and the non-relativistic one is the appearance of three 
eccentricities $ e_r,\, e_t,$ and $e_{\phi}$ compared to
one eccentricity  in the Newtonian case. However,  these eccentricities
are related.  
The explicit expressions for the parameters
 $n,\, k,\, a_r,\, e_t,\, e_r,\, e_{\phi},\,
f_t,\, g_t,\, f_{\phi} $ and $g_\phi$ 
in terms  of the 2PN conserved  energy and angular momentum
per unit  reduced mass are given by
Eqs. (38) to (48) of \cite{NW95}.
Though the three eccentricities are related, there is  
the question of selecting a specific one to  present the polarizations.
We have chosen to
present polarization waveforms in terms of $e_t$ since it  explicitly
appears in the equation relating $l$ to $u$, which we will numerically
invert while computing 
its power spectrum. 
The  question  whether, as in pulsar timing, there is a particular
combination of the three eccentricities `a  good eccentricity' 
in terms of which expressions take familiar Newtonian-like forms 
is interesting and open. 

 Exactly as in the Newtonian case, at 2PN order too, 
one can  split $\phi $ into two  parts; a part $\lambda$
linearly increasing with time and a  part ${\rm W}(l)$  periodic in
$l$ but with a more complicated  time variation:
\begin{mathletters}
\label{ioset1}
\begin{eqnarray}
\phi &=& \lambda + {\rm W}(l) \,,\\
\lambda &=& \phi_0 + ( 1 + { k \over c^2})\,l\,, \\
{\rm W}(l)&=& ( 1 + {k \over c^2} )\,\left ( v - u + e_t\, \sin u \right )
\no
&& 
+{ 1 \over c^4}\,\biggl \{ 
f_{\phi}\, \sin 2 v + g_{\phi}\,\sin 3v - f_t\,\sin v -g_t\,(v -u)
\biggr \}.
\end{eqnarray}
\end{mathletters}     
Note that complicated 2PN corrections to $\lambda $ and ${\rm W}(l)$ 
also include $k$ that represents the periastron advance.

          Using Eqs. (\ref{admr}), (\ref{admphi}), and 
Eqs. (38) to (48) of \cite{NW95}, it is  
straightforward to obtain the 2PN 
accurate expressions for 
the dynamical variables
 in terms
of $ \xi ={ G\,m\,n \over c^3} ,e_t $ and $ u$,
using the following relations, easily derivable from
Eqs. (38) to (48) of \cite{NW95}  
\begin{mathletters}
\label{inp2}
\begin{eqnarray}
{-2\,E}&=&{c}^{2}{\xi}^{2/3}\biggl \{ 
1+{ {\xi}^{2/3} \over 12}
\biggl [ 15-\eta
\biggr ]
+{{\xi}^{4/3} \over 24\,}
\biggl[
\biggl ( 15 -15\,\eta -\eta^2
\biggr ) 
\no
&&
+
{1 \over \sqrt{1-e_t^2}}\,
\biggl ( 120 -48\,\eta \biggr )
\biggr ]
\biggr \} \,,
\\           
-2\,E\,h^2 &=& (1 -e_t^2)\,\biggl \{
1+{{\xi}^{2/3} \over 4\,(1 -e_t^2) }\biggl [ -\left (
17-7\,\eta\right ){{ e_t}}^{2}
+9 +\eta  \biggr ]
\no
&&
+ { {\xi}^{4/3} \over 24\,(1 -e_t^2)^2} 
\biggl [
-\left (360-144\,\eta\right ){{e_t}}^{2}
\sqrt {1 -e_t^2}+\left (225-277\,\eta+29\,{\eta}^{2}\right ){{ e_t
}}^{4}
\no
&&
-
\left (210-190\,\eta+30\,{\eta}^{2}\right ){{ e_t}}^{2}+
189-45\,\eta+{\eta}^{2}\biggr ]  \biggr \}\,,
\\
 k  &=& { 3\, \xi^{2/3} \over ( 1 -e_t^2)}+
{\xi^{4/3} \over 4\, (1 -e_t^2)^2 } 
\biggl \{ ( 51 -26\,\eta)\,e_t^2 + (78 -28\,\eta) \biggr \}\,,
\\
e_{\phi} &=& e_{{t}} \biggl \{
1+{\xi}^{2/3}\left (
4-\eta\right )+
{ \xi^{4/3} \over 96\, (1 -e_t^2)^{3/2}}\,
\biggl [ 
\biggl (
-(1152
-656\,\eta
+41\,{\eta}^{2}
 ){e_{{t}}}^{2}
\no
&&
+1968
-1088\,\eta 
-4\,{\eta}^{2}
\biggr )\sqrt {1 -e_t^2}
+ \left ( 720 -288\,\eta \right )\,( 1 -e_t^2)
\biggr ]
\biggr \} \,,
\\
e_r &=& e_{{t}} \biggl \{ 1
+{ \xi^{2/3} \over 2}\,\left( 8 -3\,\eta \right)
+ { \xi^{4/3} \over 24\, (1 -e_t^2)^{3/2}}\, 
\biggl [
\biggl (
 -(288
-242\,\eta
+21\,{\eta}^{2}
){e_{{t}}}^{2}
\no
&&
+390
-308\,\eta
+21\,{\eta}^{2}
\biggr )\,\sqrt { 1 -e_t^2}
+ \left ( 180 -72\,\eta \right )\,( 1 -e_t^2)
\biggr ]
\biggr \}\,.
\end{eqnarray}
\eml
Using Eqs. (\ref{inp2}) in Eqs. (\ref{admr}) and (\ref{admphi}),
we obtain   after some lengthy algebra, expressions for 
$ r, \dot r = {dr \over dt} = {dr \over du}\,
{du \over dt},\phi = \lambda + {\rm W}(l) $ and 
$\dot \phi = {d \phi \over dv}\,{ d v \over du}\, {du \over dt}$,
in terms of $ \xi = {G \, m\,n \over c^3}, e_t, u, ..$
given by: 

\begin{mathletters}
\label{ioset2}
\begin{eqnarray}
r&=&({G\,m \over n^2})^{1/3}\, (1 -e_t\,\cos u)\,\biggl \{ 1
- { {\xi}^{2/3} \over 6\,(1 -e_t\,\cos u)} \biggl [ \left (6
-7\,\eta\right ){e_t}\,\cos u+18-2\,\eta\biggr ]
\no
&&
+
{{\xi}^{4/3} \over 72\,\sqrt { (1 -e_t^2)^3}\,(1 -e_t\,\cos u)}
\biggl [ 
\biggl (
-(72-231
\,\eta+35\,{\eta}^{2} )
(1 -{{ e_t}}^{2})\,e_t\,\cos u
\no
&&
-(
72
+75\,\eta
+8\,{\eta}^{2}
 ){{ e_t}}^{2}
-234
+273\,\eta
+8\,{\eta}^{2}
\biggr )\,\sqrt {1 -e_t^2}
\no
&&
-36\, (1 -e_t^2)\, (5 -2\,\eta)\,( 2 +e_t\,\cos u)
\biggr ]
\biggr \} \,,
\\
\phi &=& l + {\rm W}(l) \,,\\
{l}&=&u-{e_t}\,\sin u-
{{\xi}^{4/3} \over 8\,\sqrt {1-e_t^2}}
\,{1 \over (1 -e_t\,\cos u) }\,
\biggl \{ 
e_t\,\sin u\,\sqrt {1-e_t^2}\,\eta\,\left (4 +\eta\right )
\no
&&
+12\, ( 5 -2\,\eta)\,( u - v) \left ( 1 -e_t\,\cos u \right ) 
\biggr \}\,,
\\
{\rm W} &=& v -u +{e_t}\,\sin u
+ { 3\,{\xi}^{2/3} \over (1 -e_t^2)} \biggl \{  v -u +
e_t\,\sin u \biggr \} 
\no
&&
+ { \xi^{4/3} \over 32\, (1 -e_t^2)^{5/2}}
\,{ 1 \over ( 1 -e_t\,\cos u)^3 }\,
\biggl \{ 
\biggl [  
4\,\sqrt {1 -e_t^2}\,\left (1-e_t\,\cos u \right )^{2}
\biggl (\biggl \{  -(102
\no
&&
-52\,\eta ){{ e_t}}^{2}
-156
+56\,\eta
\biggr \}\,e_t\,\cos u
+\eta\,\left (4+\eta\right ){{ e_t}}^{4}+
\left (
102-60\,\eta -2\,{\eta}^{2}
\right ){{ e_t}}^{2}
\no
&&
+ 156 -52\,\eta+{
\eta}^{2}
\biggr )
+\left (1-e_t^{2}\right )\,
\biggl ( \left ( (3\, e_t^{2}
+12)\,\eta -8
\right ){(e_t\,\cos u)}^{2}
\no
&&
+\left (\left (8-6\,\eta\right )
{ e_t}^{2} +8
-24\,\eta\right )(e_t\,\cos u)
-12\,\eta\,{ e_t}^{4}
-\left (8-
27\,\eta\right ){ e_t}^{2}
\biggr )
\no
&&
\,\eta
\biggr ]\,e_t\, \sin u
+ ( 1 -e_t\,\cos u)^3\, 
\biggl [ 
48\,{(1-e_t^2)}^{2}\left (5-2\,\eta\right )
-8\,
\biggl ((51-26\,\eta ){ e_t}^{2}
\no
&&
+78-28\,\eta\biggr )
\biggr ]\, u  
+ 8\,\left (1-{ e_t}\,\cos u\right )^{3}\,\biggl [ 
\biggl (
\left (51-26\,\eta\right ){{e_t}}^{2}
\no
&&
+78-28\,\eta
\biggr )
\,\sqrt {1-e_t^2}
-(30 - 12\,\eta)\,(1 -2\,e_t^2 +e_t^4)
\biggr ]\,v 
\biggl \}
\,,
\\
{\dot r}&=& { { (G\,m\,n)}^{1 \over 3}\, \over 
{(1 -e_t\,\cos u)}}\,
{{ e_t}\,\sin u }
\biggl \{
1 +{ {\xi}^{2/3} \over 6} \left (6-7\,
\eta \right )
\no
&&
+{\xi^{4/3} \over 72}\,{ 1 \over ( 1 -e_t\,\cos u)^3}
\biggl [ \biggl (
-\left (72-
231\,\eta+35\,{\eta}^{2}\right )\,(e_t\,\cos u)^3
\no
&&
+\left (216
-693\,\eta
+105\,{\eta}^{2}
\right )
\,(e_t\,\cos u)^{2}
+\left (324 +
513\,\eta-96\,{\eta}^{2}
\right )\,{e_t}\,\cos u
\no
&&
- \left ( 36+9\,{\eta}\right )\,\eta\,{{e_t}}^{2} -468
-15\,\eta+35\,{\eta}^{2} \biggr )
\no
&&
+{36 \over \sqrt {1 -e_t^2} }\,
\biggl ( \left( 1-{ e_t}\,\cos u\right )^{2}\left (4-{ e_t}\,\cos u
\right )\left (5-2\,\eta \right )\biggr )
\biggr ]
\biggr \}\,,
\\
{\dot \phi }&=&
{n\,\sqrt {1-e_t^2} \over \left (1-{e_t}\,\cos u\right )^{2}} 
\biggl \{
1+ {\xi^{2/3} \over (1 -e_t^2)\,(1 -e_t\,\cos u) }
\biggl [
 \left (1-\eta
\right ){e_t}\,\cos u
\no
&&
-\left (4-\eta\right ){{e_t}}^{2}+3
\biggr ] 
+ 
{{\xi}^{4/3} \over 12}\,{1 \over (1 -e_t\,\cos u)^3} 
\biggl  [ { 1 \over (1 -e_t^2)^{3/2}}
\biggl \{
 18\,
\left (1-{e_t}\,\cos u\right )^{2}\,
\no
&&
\left ({e_t}\,\cos u-2\,{{e_t}}^{2} +1 \right )
\left (5-2\,\eta
\right )
\biggr \}
+{1 \over  {(1-e_t^2)}^2}
\biggl \{ \biggl (
-(9 -
19\,\eta-14\,{\eta}^{2}
){{e_t}}^{2}
\no
&&
-36
+2\,\eta
-8\,{\eta}^{2}
 \biggr ) \,(e_t\,\cos u)^3
+ \biggl ( 
-(48
-14\, \eta+17\,{\eta}^{2}){{e_t}}^{4}
\no
&&
+\left ( 69-79\,\eta+4\,
{\eta}^{2}\right ){{e_t}}^{2}
+114 +2\,\eta -5\,{\eta}^{2}
 \biggr )\,(e_t\,\cos u)^2
\no
&&
+\biggl (
-(6-32\,\eta-{\eta}^{2} ){{e_t}}^{4}
+\left (93
-19\,\eta
+16\,{\eta}^{2}
\right )e_t^2
\no
&&
-222
+50\,\eta+{\eta}^
{2}  \biggr )\,( e_t\,\cos u)
-6\,\eta\, (1 -2\,\eta )\,e_t^6
\no
&&
+\left (54
-28\,\eta-20\,{\eta}^{2}\right ){{
e_t}}^{4}-\left (
153
-61\,\eta
-2\,{\eta}^{2}
\right )
{{e_t}}^2
+144-48\,\eta 
\biggr \} \biggr ]  \biggr \} \,.
\end{eqnarray}
\eml

     Note that the above   2PN accurate expressions for 
$\lambda$ and ${\rm W}(l)$
in terms of $\xi, e_t, ..{\rm etc}$ are explicitly  needed 
 to explore the spectral  decomposition  of the polarization waveforms. 
They are not meant to be explicated in the 2PN expressions for 
$h_+$ and $h_{\times}$, in terms of $\lambda$ and ${\rm W}(l)$
in Eq. (\ref{hcp2PN}) and (\ref{hcc2PN}).

                 The 2PN corrections to 
$ h_{+} $ and $ h_{\times} $, in a form
similar to Eqs. (\ref{hpcN}), are obtained 
using Eqs. (\ref{hcpf}). 
However, we need  the 
2PN corrections to $h_{ij}^{TT} $ in {\it ADM} coordinates, as
 the parametric expressions for $r,\,\dot r, \,\dot \phi$ and 
the $ \phi = \lambda + {\rm W}(l) $ split, 
given by  Eqs. (\ref{ioset2}), 
are in the {\it ADM} gauge.
However, the 2PN corrections to $ h_{ij}^{TT}$,
given by Eqs (5.3) and (5.4) of \cite{GI97}, 
are available only in {\it harmonic (De-Donder)} coordinates.
Using, in a straightforward manner,
the transformation equations 
of Damour and Sch\" afer \cite {DS85}
to relate the dynamical variables in the 
harmonic and the ADM gauge, we obtain 
the 2PN accurate instantaneous  contributions to 
$ h_{ij}^{TT}$ in the ADM gauge.      
For completeness, we list below the relevant transformation equations
relating the harmonic (De-Donder) variables to the corresponding
ADM ones,
\begin{mathletters}
\label{ctad}
\begin{eqnarray}
{\bf r}_{ \rm D} &=& {\bf r}_{ \rm A}
+ { Gm \over 8\,c^4\,r}\left \{ \left [ \left ( 5 v^2
-\dot r^2 \right )\eta +  2 \,\left ( 1
+12\eta \right ){ Gm\over r} \right ] { \bf r}
\right.\nonumber\\
& & \left.
-  18 \,\eta \, r\dot r\,{\bf v} \right \}\,,\\
t_{\rm D}  &=& t_{ \rm A } - { Gm \over c^4}\,\eta \,\dot r \,,\\
{ \bf v }_{ \rm D} &=& { \bf v }_{ \rm A} - { Gm \dot r\over 8\,c^4\, r^2}
\biggl \{ \biggl [ 7 v^2 + 38 { Gm \over r}
- 3\dot r^2 \biggr ]\eta + 4\,{ G m \over r}\, \biggr \}
{ \bf r}
\nonumber \\
& &
-{ Gm \over 8\,c^4 r} \biggl \{
\biggl [ 5 v^2 - 9 \dot r^2 - 34 { G m\over r} \biggr ]\eta
- 2\,{ Gm \over r} \biggr \}{ \bf v} \,,\\
r _{ \rm D} &=& r _{ \rm A} + { G m \over 8\,c^4}
\left \{  5\, \eta v^2 + 2\,
 \left ( 1 +12\eta \right ){ G m\over r} -  19\, \eta \dot r^2
\right \}\,.
\end{eqnarray}
\end{mathletters}
The subscripts `${\rm D}$' and `${\rm A}$' denote 
quantities in the De-Donder~( harmonic ) and
in the ADM coordinates respectively.
Note that in all the above  equations the differences between the
two gauges are of  2PN order.
As there is no difference between the harmonic and the ADM coordinates
to 1PN  accuracy, no suffix
is  used 
in Eqs. (\ref{ctad})
for the 2PN terms. 
    
      Using Eqs.(\ref{ctad}),
the 2PN corrections to $ h_{ij}^{TT}$ in ADM coordinates
can easily be obtained from 
Eqs. (5.3) and (5.4) of \cite{GI97}.
For economy of presentation, we write 
$ (h_{ij}^{TT})_{\rm A}$ 
 in the following manner,
$ (h_{ij}^{TT})_{\rm A} =  (h_{ij}^{TT})_{\rm O} + $ `$ Corrections $', where
$(h_{ij}^{TT})_{\rm A}$ represent
the metric perturbations in the ADM coordinates.
$ (h_{ij}^{TT})_{\rm O}$ is a short hand notation for
expressions on the r.h.s of Eqs. (5.3) and (5.4) of \cite {GI97},
where $ {\bf N},{\bf n}, {\bf v}, v^2, \dot r, r $ are the ADM variables
${\bf N}_{\rm A},{\rm n}_{\rm A}, {\bf v}_{\rm A},v_{\rm A}^2, 
{\dot r}_{\rm A},
 r_{\rm A} $ respectively.
The `$ Corrections $' represent the differences at the 2PN
order, that arise due to the change of the coordinate system,
given by Eqs. (\ref{ctad}). As the two coordinates are
different only at the 2PN order, the `$ Corrections $' come
only from the leading Newtonian terms in Eqs. (5.3) and (5.4)
of \cite{GI97}.
\begin{eqnarray}
\label{hijA}
(h^{TT}_{ij})^{ }_{\rm A} & =&
(h^{TT}_{ij})^{ }_{\rm O} +
{G \over c^4\,R}\,{ G\,m \over 2\,c^4\,r_{\rm A}} \biggl \{
\biggl [ 
(5 \,v_{\rm A}^2 -55\,{\dot r}_{\rm A}^2)\,\eta
+ 2\left ( 1 +12\,\eta \right )
\,{G\,m \over r_{\rm A}} \biggr ] \,{G\,m\over r_{\rm A}}\,(n_{ij})_{\rm A}^{TT}
\no
& &
- \biggl [ 
(14\,v_{\rm A}^2 -6\,{\dot r}_{\rm A}^2)\,\eta
+8\left ( 1
+5\,\eta \right )\,{G\,m \over r_{\rm A}} \biggr ]
{\dot r}_{\rm A}
\,(n_{(i}v_{j)})_{\rm A}^{TT}
\no
& &
- \biggl [ 
(10 \,v_{\rm A}^2 -18 \,{\dot r}_{\rm A}^2)\,\eta
-\left ( 4 + 68\,\eta \right )\,{ G\,m \over r_{\rm A}}\biggr ]
\,(v_{ij})_{\rm A}^{TT}
\biggr \}\,.
\end{eqnarray}
To check the algebraic correctness of the above transformation, 
we compute the far-zone energy flux  directly in the ADM coordinates
using 
\begin{eqnarray}
\biggl ({d{\cal E}\over dt} \biggr )_{\rm A}
& =& {c^3\,R^2\over 32\pi G} \int \left(
({\dot h^{TT}_{ij}})_{\rm A}\,({\dot h^{TT}_{ij}})_{\rm A}
 \right)d\Omega ({\bf N}).
\end{eqnarray}
 After a careful use of the transformation equations, 
the expression for $(d{\cal E}/dt)_{\rm A}$ calculated above, 
matches with the expression for the far-zone energy flux, 
Eq. (4.7a) of \cite{GI97} obtained earlier. 
This provides a useful  
check  on the transformation from  
$(h^{TT}_{ij})^{ }_{\rm D}$ to 
$(h^{TT}_{ij})^{ }_{\rm A}$.
 
        We now have all the inputs required to compute
the 2PN corrections to $h_{+}$ and $h_{\times}$ 
in terms of a  elegant and convenient parametrization using Eqs. (\ref{hcpf}).
As  mentioned in \cite{WW96,GI97},
 there is no need to apply the {\rm TT } projection
to $ (h^{TT}_{ij})$ given by Eq. (\ref{hijA})  
before contracting with ${\bf p} $ and ${\bf q}$,
as required by Eqs. (\ref{hcpf}).
Thus, we schematically write, 
\begin{equation}
\label{hijsym}
h^{TT}_{ij}= \alpha_{vv} \,v_{ij} +
\alpha_{nn}\,n_{ij}
+ \alpha_{nv} \, n_{(i}v_{j)}\,.
\end{equation}
The polarization states $h_{+} $ and $h_{\times}$, for
Eqs. (\ref{hijsym}) are given by,
\begin{mathletters}
\label{hcpd2}
\begin{eqnarray}
h_{+} &=& { 1 \over 2} \biggl ( p_i\,p_j -q_i\,q_j \biggr )\,
\biggl ( \alpha_{vv} \,v_{ij} + \alpha_{nn}\,n_{ij}
+ \alpha_{nv} \, n_{(i}v_{j)} \biggr )
\,,
\nonumber \\
&= &
{\alpha_{vv} \over 2} \biggl ( ({\bf p .v })^2 -
({\bf q.v})^2 \biggr ) + { \alpha_{nn} \over 2}
\biggl ( ({\bf p.n})^2 - ({\bf q. n})^2 \biggr ) +
{ \alpha_{nv} \over 2}\biggl ( ({\bf p.n}) ({\bf p .v})
- ({\bf q. n}) ({\bf q.v}) \biggr  )\,,
\\
h_{\times} & =& { 1 \over 2} \biggl ( p_i\,q_j +p_j\,q_i \biggr )\,
\biggl ( \alpha_{vv} \,v_{ij} + \alpha_{nn}\,n_{ij}
+ \alpha_{nv} \, n_{(i}v_{j)} \biggr )
\nonumber \\
&=& { \alpha_{vv} }\,({\bf p.v})
\,({\bf q.v}) + \alpha_{nn} \,({\bf p.n})\,({\bf q. n})
+ { \alpha_{nv} \over 2} \biggl ( ({\bf p.n})\,({\bf q.v}) +
({\bf p.v})\,({\bf q. n}) \biggr )\,.
\end{eqnarray}
\end{mathletters}
                             
Before proceeding to a lengthy but straightforward  computation 
of the `instantaneous' 2PN accurate  polarizations 
$ h_{+} $ and $ h_{\times} $,  
we anticipate the structure of the final result by schematically
 examining the functional forms in  the intermediate steps
of the above calculation. 
The polarizations in terms of $\phi$ has the  form

\begin{mathletters}
\begin{eqnarray}
h_{+} &=& \biggl\{ \alpha_{vv}\biggl[(\cdots) 
+ (\cdots) \cos 2\phi + (\cdots)\sin 2\phi\biggr]
+\alpha_{nn}\biggl[(\cdots) + (\cdots)\cos 2\phi\biggr]+\nonumber\\
&&\alpha_{nv}\biggl[(\cdots)+(\cdots)\cos 2\phi 
+(\cdots)\sin 2\phi\biggr]\biggl\}\,,\\
h_{\times} &=& \biggl\{ \alpha_{vv}\bigl[ (\cdots) \cos 2\phi 
+ (\cdots)\sin 2\phi\bigr]
+\alpha_{nn}\biggl[ (\cdots)\sin 2\phi\biggr]+
\alpha_{nv}\biggl[(\cdots)\cos 2\phi 
+(\cdots)\sin 2\phi\biggr]\biggl\}.
\end{eqnarray}
\end{mathletters}
In the above and what follows
 $\bigl(\cdots\bigr)$ denotes a dependence on variables
$e_t, n, m_1, m_2, i$ and $u$.
The structure of the PN expansion of the 
 coefficients $\alpha_{ij}$ above 
is the following:
\begin{eqnarray}
\alpha_{vv}&\sim& 1+
\frac{1}{c} \biggl[\bigl(\cdots\bigr)\cos \phi 
+ \bigl(\cdots\bigr)\sin \phi\biggr]+
\frac{1}{c^2} \biggl[\bigl(\cdots\bigr)
+ \bigl(\cdots\bigr)\cos 2\phi 
+ \bigl(\cdots\bigr)\sin 2\phi\biggr]+\nonumber\\
&&\frac{1}{c^3} \biggl[\bigl(\cdots\bigr)\cos 3\phi 
+ \bigl(\cdots\bigr)\sin 3\phi\biggr]+
\frac{1}{c^4} \biggl[\bigl(\cdots\bigr)
+ \bigl(\cdots\bigr)\cos 4\phi + \bigl(\cdots\bigr)\sin 4\phi\biggr].
\end{eqnarray}
$\alpha_{nn}$ has a similar expansion. $\alpha_{nv}$ also has a similar expansion
but the leading order term is of order $1\over c$. Using this 
information about the functional dependence on $\phi$ and elementary
trigonometry  one can infer in detail  the 
harmonics of $\phi$  
appearing at  each order  and 
consequently the $\lambda$ and ${\rm W}(l)$  dependence on 
display in the equations (\ref{hcp2PN}) and  (\ref{hcc2PN}) below.
The `instantaneous' 2PN accurate  polarizations 
$ h_{+} $ and $ h_{\times} $  
in terms of $ \xi={G\,m\,n \over c^3}, m, \eta, e_t$ and Sines and Cosines 
of $ \lambda, {\rm W}(l), u(l) $ and $i $, using (\ref{hcpd2}), 
(\ref{hijsym}), ( \ref{ioset2}) and (\ref{ioset1}) are finally written as: 
\begin{eqnarray}
\label{hx+f}
(h_{+,{\times}})_{\rm inst} &=& {G\, m\,\eta \over c^2\,R}\, {\xi}^{2/3}
\biggl \{ H^{(0)}_{+,{\times}} + \xi ^{1/2}\,H^{(1/2)}_{+,{\times}}
+\xi \, H^{(1)}_{+,{\times}} + \xi^{3/2}\,H^{(3/2)}_{+,{\times}}
+ \xi^2\,H^{(2)}_{+,{\times}} \biggr \}\,,
\end{eqnarray}
where the curly brackets contain a post-Newtonian expansion.
The expressions for various post-Newtonian terms in  the `plus' and `cross'
polarizations are shown below in a  form emphasizing   the
harmonic  content and the corresponding amplitude modulation.
The various post-Newtonian corrections to the `plus' polarization 
are given by:     
\begin{mathletters}
\label{hcp2PN}
\begin{eqnarray}
H^{(0)}_{+} &=& \biggl \{
\biggl [  \rm P^{0}_{\rm C2C2}\, \cos 2\,W +
\rm P^{0}_{\rm S2C2}\, \sin 2\,W \biggr ] \,\cos 2\,\lambda
+ \biggl [ \rm P^{0}_{\rm C2S2}\,\cos 2\,W 
+ \rm P^{0}_{\rm S2S2}\,
\sin 2\,W \biggr ]
\no
&&
\,\sin 2\,\lambda
+ \rm P^{0}
\biggr \}\,,
\\        
H^{(0.5)}_{+} &=& \biggl \{
\biggl [ \rm  P^{0.5}_{C1C1}\, \cos W +
P^{0.5}_{S1C1}\, \sin W \biggr ]\,\cos \lambda
+ \biggl [ \rm P^{0.5}_{C1S1}\,\cos W
+ P^{0.5}_{S1S1}\,\sin W \biggr ]\,\sin \lambda
\no
&& +
\biggl [ \rm P^{0.5}_{C3C3}\, \cos 3\,W +
P^{0.5}_{S3C3}\, \sin 3\,W \biggr ]\,\cos 3\,\lambda
+ \biggl [ \rm P^{0.5}_{C3S3}\,\cos 3\,W +
P^{0.5}_{S3S3}\,\sin 3\,W \biggr ]
\no
&&
\,\sin 3\,\lambda
\biggr \}\,,
\\      
H^{(1)}_{+} &=& \biggl \{ \rm
\biggl [ \rm  P^{1}_{C2C2}\, \cos 2\,W +
P^{1}_{S2C2}\, \sin 2\,W \biggr ]\,\cos 2\,\lambda
+ \biggl [P^{1}_{C2S2}\,\cos 2\,W
\no
&& 
+\rm  P^{1}_{S2S2}\,\sin 2\,W \biggr ]\,\sin 2\,\lambda
+ \rm \biggl [ P^{1}_{C4C4}\, \cos 4\,W +
P^{1}_{S4C4}\, \sin 4\,W \biggr ]\,\cos 4\,\lambda
\no
&& 
\rm + \biggl [P^{1}_{C4S4}\,\cos 4\,W
\rm + P^{1}_{S4S4}\,\sin 4\,W \biggr ]\,\sin 4\,\lambda
+
{\rm P}^{1}
\biggr \}\,,
\\       
H^{(1.5)}_{+} &=& \biggl \{
 \rm \biggl [  P^{1.5}_{C1C1}\, \cos W +
P^{1.5}_{S1C1}\, \sin W \biggr ]\,\cos \lambda
+ \biggl [P^{1.5}_{C1S1}\,\cos W
+ P^{1.5}_{S1S1}\,\sin W \biggr ]\,\sin \lambda
\no
&& +
\rm \biggl [ P^{1.5}_{C3C3}\, \cos 3\,W +
P^{1.5}_{S3C3}\, \sin 3\,W \biggr ]\,\cos 3\,\lambda
+ \biggl [P^{1.5}_{C3S3}\,\cos 3\,W
\no
&&
\rm + P^{1.5}_{S3S3}\,\sin 3\,W \biggr ]\,\sin 3\,\lambda
+
 \biggl [ {\rm P^{1.5}_{C5C5}\, \cos 5\,W }+
P^{1.5}_{S5C5}\, \sin 5\,W \biggr ]\,\cos 5\,\lambda
\no
&&
+\rm \biggl [P^{1.5}_{C5S5}
\,\cos 5\,W 
+ P^{1.5}_{S5S5}\,\sin 5\,W \biggr ]\,\sin 5\,\lambda
\biggr \}\,,
\\    
H^{(2)}_{+} &=& \biggl \{
\rm \biggl [ P^{2}_{C2C2}\, \cos 2\,W +
P^{2}_{S2C2}\, \sin 2\,W \biggr ]\,\cos 2\,\lambda
+ \rm \biggl [ P^{2}_{C2S2}\,\cos 2\,W
\no
&&
\rm + P^{2}_{S2S2}\,\sin 2\,W \biggr ]\,\sin 2\,\lambda
+ \rm \biggl [ P^{2}_{C4C4}\, \cos 4\,W +
P^{2}_{S4C4}\, \sin 4\,W \biggr ]\,\cos 4\,\lambda
\no
&&
+\rm  \biggl [P^{2}_{C4S4}\,\cos 4\,W
+ P^{2}_{S4S4}\,\sin 4\,W \biggr ]\,\sin 4\,\lambda
+ \rm \biggl [ P^{2}_{C6C6}\, \cos 6\,W 
\no
&&
\rm + P^{2}_{S6C6}\, \sin 6\,W \biggr ]\,\cos 6\,\lambda
+ \rm \biggl [P^{2}_{C6S6}\,\cos 6\,W
+ P^{2}_{S6S6}\,\sin 6\,W \biggr ]\,\sin 6\,\lambda   
+ {\rm P}^{2}
\biggr \}\,,
\end{eqnarray}
\end{mathletters}
and for the `cross' polarization by       
\begin{mathletters}
\label{hcc2PN}
\begin{eqnarray}
H^{(0)}_{\times} &=& \biggl \{
\rm
\biggl [ \rm X^{0}_{C2C2}\, \cos 2\,W +
\rm X^{0}_{S2C2}\, \sin 2\,W \biggr ]\,\cos 2\,\lambda
\rm
+ \biggl [ X^{0}_{C2S2}\,\cos 2\,W 
\no
&&
\rm + X^{0}_{S2S2}\, \sin 2\,W \biggr ]\,\sin 2\,\lambda
\biggr \}\,,
\\      
H^{(0.5)}_{\times} &=& \biggl \{
\rm
\biggl [ X^{0.5}_{C1C1}\, \cos W +
X^{0.5}_{S1C1}\, \sin W \biggr ]\,\cos \lambda
+ \biggl [ X^{0.5}_{C1S1}\,\cos W
+ X^{0.5}_{S1S1}\,\sin W \biggr ]\,\sin \lambda
\no
&& +
\rm \biggl [ X^{0.5}_{C3C3}\, \cos 3\,W +
X^{0.5}_{S3C3}\, \sin 3\,W \biggr ]\,\cos 3\,\lambda
+ \biggl [ X^{0.5}_{C3S3}\,\cos 3\,W
\no
&&
\rm + X^{0.5}_{S3S3}\,\sin 3\,W \biggr ] \,\sin 3\,\lambda
\biggr \}\,,
\\    
H^{(1)}_{\times} &=& \biggl \{
\rm \biggl [ X^{1}_{C2C2}\, \cos 2\,W +
X^{1}_{S2C2}\, \sin 2\,W \biggr ]\,\cos 2\,\lambda
+ \biggl [X^{1}_{C2S2}\,\cos 2\,W
\no
&&
\rm + X^{1}_{S2S2}\,\sin 2\,W \biggr ]\,\sin 2\,\lambda
+\rm \biggl [ X^{1}_{C4C4}\, \cos 4\,W +
X^{1}_{S4C4}\, \sin 4\,W \biggr ]\,\cos 4\,\lambda
\no
&&
\rm + \biggl [X^{1}_{C4S4}\,\cos 4\,W
+ X^{1}_{S4S4}\,\sin 4\,W \biggr ]\,\sin 4\,\lambda
+ {\rm X}^{1}
\biggr \}\,,
\\     
H^{(1.5)}_{\times} &=& \biggl \{
\rm \biggl [ X^{1.5}_{C1C1}\, \cos W +
X^{1.5}_{S1C1}\, \sin W \biggr ]\,\cos \lambda
+ \biggl [X^{1.5}_{C1S1}\,\cos W
\rm + X^{1.5}_{S1S1}\,\sin W \biggr ]
\no
&&
\,\sin \lambda
+\rm \biggl [ X^{1.5}_{C3C3}\, \cos 3\,W +
X^{1.5}_{S3C3}\, \sin 3\,W \biggr ]\,\cos 3\,\lambda
+ \biggl [X^{1.5}_{C3S3}\,\cos 3\,W
\no
&&
\rm + X^{1.5}_{S3S3}\,\sin 3\,W \biggr ]\,\sin 3\,\lambda
+ \rm \biggl [ X^{1.5}_{C5C5}\, \cos 5\,W +
X^{1.5}_{S5C5}\, \sin 5\,W \biggr ]\,\cos 5\,\lambda
\no
&&
\rm + \biggl [X^{1.5}_{C5S5}\,\cos 5\,W
\rm + X^{1.5}_{S5S5}\,\sin 5\,W \biggr ]\,\sin 5\,\lambda
\biggr \}\,,     
\\
H^{(2)}_{\times} &=& \biggl \{
\rm
\biggl [ X^{2}_{C2C2}\, \cos 2\,W +
X^{2}_{S2C2}\, \sin 2\,W \biggr ]\,\cos 2\,\lambda
+ \biggl [ X^{2}_{C2S2}\,\cos 2\,W
\rm + X^{2}_{S2S2}\,\sin 2\,W \biggr ]
\no
&&
\,\sin 2\,\lambda
+ \rm \biggl [ X^{2}_{C4C4}\, \cos 4\,W +
X^{2}_{S4C4}\, \sin 4\,W \biggr ]\,\cos 4\,\lambda
+ \biggl [X^{2}_{C4S4}\,\cos 4\,W
\no
&&     
\rm + X^{2}_{S4S4}\,\sin 4\,W \biggr ]\,\sin 4\,\lambda
+ \rm \biggl [ X^{2}_{C6C6}\, \cos 6\,W +
X^{2}_{S6C6}\, \sin 6\,W \biggr ]\,\cos 6\,\lambda
\no
&&
+ \rm \biggl [X^{2}_{C6S6}\,\cos 6\,W
+ X^{2}_{S6S6}\,\sin 6\,W \biggr ]\,\sin 6\,\lambda
+{\rm X}^{2}
\biggr \}\,,
\end{eqnarray}
\end{mathletters}         
where
$\rm P$'s and $\rm X$'s
are functions of $e_t,\,m_1,\,m_2,\,u(l)$
and $i$. 
The notation ${\rm P^{a}_{CbSc}}$, for instance, denotes the coefficient 
of $\cos {\rm b\,W}\, \sin {\rm c\,\lambda}$ 
at `$a$PN' order
and similar explanation holds for ${\rm X^{a}_{CbSc}}$ too.
The explicit expressions for $\rm P$'s and $\rm X$'s,
{\it i.e }, the coefficients of Sine and Cosine  multiples
of $\lambda $ and ${\rm W}(l)$ appearing in Eqs. (\ref{hcp2PN})
and (\ref{hcc2PN}) are given by
\begin{mathletters}
\begin{eqnarray}
\rm P^{0}_{\rm C2C2} = - \rm P^{0}_{\rm S2S2} &=&     
{1 \over (1 -e_t\,\cos u)^2}\, \biggl \{ (1 +C^2)\,
\biggl [ {(e_t\,\cos u)}^2 -(e_t\,\cos u) -2\, e_t^2 +2 \biggr ] \biggr \} \,, \\
\rm P^{0}_{\rm C2S2} =  \rm P^{0}_{\rm S2C2} &=&     
2\,{1 \over (1 -e_t\,\cos u)^2}\, \biggl \{ 
  (1 + {C}^{2} )\,\sqrt {1 -e_t^2}\,e_{{t}}\,\sin u 
\biggr \}\,, \\
{\rm P^{0} } &=& 
{S^2 \over (1 -e_t\,\cos u)}\,
e_{{t}}{\cos u}\,,\\
\rm P^{0.5}_{\rm C1C1} =  -\rm P^{0.5}_{\rm S1S1} &=&     
{S \over 2}\, { m_1 -m_2 \over m}\,
{1 \over (1 -e_t\,\cos u)}\, \biggl \{  (1-3\,{C}^{2} ) 
\,e_{{t}}\,{\sin u }\, \biggr \}\,,\\
\rm P^{0.5}_{\rm C1S1} =  \rm P^{0.5}_{\rm S1C1} &=&     
{S \over 4}\, { m_1 -m_2 \over m}\,
{1 \over (1 -e_t\,\cos u)^2}\, 
\biggl \{   
\sqrt {1-e_t^2} 
\biggl [ 2\,(e_t\,\cos u) +5
\no
&&
-(6\,(e_t\,\cos u)-1 ){C}^{2}
\biggr ]
\biggr \}\,,\\
\rm P^{0.5}_{\rm C3C3} =  -\rm P^{0.5}_{\rm S3S3} &=&     
-{S \over 2}\,{ m_1 -m_2 \over m}\,
\,{ 1 \over (1 -e_t\,\cos u)^3}
\, 
\biggl \{ (1+{C}^{2} )e_{{t}}\,{\sin u}\, \biggl [
{(e_t\,\cos u)}^{2}-2\,(e_t\,\cos u)
\no
&&
 -4\,{e_{{t}}}^{2} +5 \biggr ]
\biggr \}\,, \\
\rm P^{0.5}_{\rm C3S3} = \rm P^{0.5}_{\rm S3C3} &=&     
{S \over 4}\,{ m_1 -m_2 \over m}\,
\,{1 \over (1 -e_t\,\cos u)^3} 
\biggl \{
 (1+ {C}^{2} )\sqrt {1-e_t^2}
\biggl [
6\,{(e_t\,\cos u)}^{2}-7\,(e_t\,\cos u)
\no
&&
 - 8e_t^2+9 
\biggr ]
\biggr \}\,,\\
\rm P^{1}_{\rm C2C2} = -\rm P^{1}_{\rm S2S2} &=&     
{1 \over 6}\,{ 1 \over (1 -e_t\,\cos u)^3}
\biggl \{ \biggl [
  -(9-5\,\eta ){(e_t\,\cos u)}^{3}+ (18-10\,\eta ){
(e_t\,\cos u)}^{2}
\no
&&
+ \biggl ( (18-10\,\eta ){e_{{t}}}^{2}
+20 -12\,\eta
 \biggr )(e_t\,\cos u) - (33+11\,\eta ){e_{{t}}}^{2}-14+38\,\eta
\biggr ]
\no
&&
+ \biggl [
- (3+13\,\eta ){(e_t\,\cos u)}^{3}
+ (6+26\,\eta ){
(e_t\,\cos u)}^{2}+ \biggl ( (6+26\,\eta ){e_{{t}}}^{2}+33
\no
&&
-51\,\eta
 \biggr )(e_t\,\cos u)
 - (48-34\,\eta ){e_{{t}}}^{2}
+6 -22\,\eta 
\biggr ] {C}^{2}+ (1-3\,\eta ) 
\biggl [
\no
&&
-6\,{(e_t\,\cos u)}^{3}
+12\,{(e_t\,\cos u)}^{2}
+
 (12\,{e_{{t}}}^{2}-13 )(e_t\,\cos u)
-9\,{e_{{t}}}^{2}+4 \biggr ]{C}^{
4} 
\biggr \}\,,\\
\rm P^{1}_{\rm C2S2} = \rm P^{1}_{\rm S2C2} &=&     
{1 \over 6}\,{ e_t\,\sin u \over (1 -e_t\,\cos u)^3}
\biggl \{ { 1 \over \sqrt {1 -e_t^2}}
\biggl [
\biggl ((18-10\,\eta){e_{{t}}}^{2}
-6-2\,
\eta \biggr )(e_t\,\cos u) 
\no
&&
-(39-7\,\eta){e_{{t}}}^{2}
+27+5\,\eta
\biggr ]
+
{ 1 \over \sqrt {1 -e_t^2}}
\biggl [ \biggl ( (6+26\,\eta){e_{{t}}}^{2}+6-38\,\eta
\biggr )(e_t\,\cos u) 
\no
&&
-(30 + 20\,\eta){e_{{t}}}^{2}
+ 18 + 32\,\eta
\biggr ] {C}^{
2}
+3\,\sqrt {1-{e_{{t}}}^{2}}\,(1-3\,\eta)\, \biggl [ -4
\,(e_t\,\cos u)
\no
&&
+5 \biggr ]{C}^{4}
\biggr \}\,,\\
\rm P^{1}_{\rm C4C4} = -\rm P^{1}_{\rm S4S4} &=&     
{S^2 \over 24}\,{ 1 \over (1 -e_t\,\cos u)^4}
\biggl \{
(1-3\,\eta )\,
(1 +C^2)\,
\biggl [
-6\,{(e_t\,\cos u)}^{4}+18\,{(e_t\,\cos u)}
^{3}
\no
&&
+ (48\,{e_{{t}}}^{2}-61 ){(e_t\,\cos u)}^{2}
+ (65-69\,{e_{{t
}}}^{2} )(e_t\,\cos u)
-48\,{e_{{t}}}^{4}+117\,{e_{{t}}}^{2}
\no
&&
-64 
\biggr ]\,\biggr \}\,,\\
\rm P^{1}_{\rm C4S4} = \rm P^{1}_{\rm S4C4} &=&     
{S^2 \over 4}\,{ 1 \over (1 -e_t\,\cos u)^4}
\biggl \{
(1 +C^2)\,
(1-3\,\eta )
\sqrt {1-e_t^2}\,e_t\,\sin u
\biggl [ -4\,{(e_t\,\cos u)}^{2} 
\no
&&
+ 9\,(e_t\,\cos u) 
+8\,{e_{{t}}}^{2}-13 
\biggr ]
\biggr \}\,,\\
\rm P^{1} &=&
{ S^2 \over 24}\,{1 \over  (1-e_{{t}}{\cos u} )^3}
\,\biggl \{  
\biggl [  (30 -2\,\eta ){
(e_t\,\cos u)}^{3}- (60 -4\,\eta ){(e_t\,\cos u)}^{2}
\no
&&
- (57+5\,\eta )
(e_t\,\cos u)
+ (87+3\,\eta ){e_{{t}}}^{2} \biggr ]
+
\biggl [
 (18 -54\,\eta ){(e_t\,\cos u)}^{3} - (36 
\no
&&
-108
\,\eta ){(e_t\,\cos u)}^{2}+ (3-9\,\eta )(e_t\,\cos u)
+ (15-45\,
\eta ){e_{{t}}}^{2} \biggr ] {C}^{2}
\biggr \} \,,\\
\rm P^{1.5}_{\rm C1C1} = -\rm P^{1.5}_{\rm S1S1} &=&     
-{S \over 48}\,{ m_1 -m_2 \over m}\,
{ e_t\,\sin u \over (1 -e_t\,\cos u)^4}
\biggl \{ 
 \biggl [ 48\,{(e_t\,\cos u)}^{3}-144\,{(e_t\,\cos u)}^{2}
\no
&&
+ (33+ 22\,\eta
 )(e_t\,\cos u) 
- (336+12\,\eta ){e_{{t}}}^{2}
+399-10\,\eta
\biggr ]
\no
&&
+
\biggl [  -(108+72\,\eta ){(e_t\,\cos u)}^{3}+ (324 
+ 216\,\eta
 ){(e_t\,\cos u)}^{2}
\no
&&
- (12+240\,\eta )(e_t\,\cos u)
- (144 - 36\,\eta
 ){e_{{t}}}^{2} -60 
\no
&&
+60\,\eta \biggr ]\,{C}^{2}
+5\, (1 -e_t\,\cos u)
 (1-2\,\eta ) \biggl [ 12\,{(e_t\,\cos u)}^{2}
\no
&&
-24\,(e_t\,\cos u)
+5 
\biggr ]{C}^{4} 
 \biggr \} \,,\\
\rm P^{1.5}_{\rm C1S1} = \rm P^{1.5}_{\rm S1C1} &=&     
{S \over 96}\,{ m_1 -m_2 \over m}\,
{1 \over (1 -e_t\,\cos u)^4}\,
\biggl \{ 
{1 \over \sqrt {1 -e_t^2}}
\biggl [   (48 +48\,\eta-96\,{e_{{t}}}^{2}
 ){(e_t\,\cos u)}^{3}
\no
&&
- (24 +48\,\eta
-72\,{e_{{t}}}^{2}\eta ){(e_t\,\cos u)
}^{2}+ \biggl ((396 +104\,\eta ){e_{{t}}}^{2}-204-296\,\eta
 \biggr )(e_t\,\cos u)
\no
&&
+ (123-198\,\eta ){e_{{t}}}^{4}- (546-220
\,\eta ){e_{{t}}}^{2}+303 +98\,\eta
\biggr ]
+ 
{1 \over \sqrt {1 -e_t^2}}
\biggl [  \biggl ( (216
\no
&&
+144
\,\eta ){e_{{t}}}^{2}-72-288\,\eta \biggr ){(e_t\,\cos u)}^{3}- 
\biggl (
 (756+240\,\eta ){e_{{t}}}^{2}
-444 -552\,\eta \biggr )
\no
&&
{(e_t\,\cos u)}^{
2}- \biggl ( (144+96\,\eta ){e_{{t}}}^{2}
-336+96\,\eta 
\biggr )(e_t\,\cos u)
+ (720+144\,\eta ){e_{{t}}}^{4}
\no
&&
- (756 +96\,\eta
 ){e_{{t}}}^{2}+12 -24\,\eta 
\biggr ]
{C}^{2}
- \sqrt {1-e_t^2} \,(1 -2\,\eta)
\biggl [ 120\,{(e_t\,\cos u)}^{3}
\no
&&
-276\,{(e_t\,\cos u)}^{2}+52\,(e_t\,\cos u)
+105\,e_t^2       
-1 
\biggr ]\,
{C}^{4} 
\biggr \}\,,\\
\rm P^{1.5}_{\rm C3C3} = -\rm P^{1.5}_{\rm S3S3} &=&     
{S \over 96}\,{ m_1 -m_2 \over m}\,
{e_t\,\sin u \over (1 -e_t\,\cos u)^4}\,
\biggl \{ 
\biggl [
 (108 - 24\,\eta ){(e_t\,\cos u)}^{3
}- (324 -72\,\eta ){(e_t\,\cos u)}^{2}
\no
&&
- \biggl ( (432 -96\,\eta
 ){e_{{t}}}^{2}
-143-274\,\eta \biggr )(e_t\,\cos u) + (1176 -72\,\eta
 ){e_{{t}}}^{2}-671
\no
&&
-346\,\eta \biggr ] 
+ 
\biggl [  (72+48\,\eta )
{(e_t\,\cos u)}^{3}
- ( 216 + 144\,\eta ){(e_t\,\cos u)}^{2}- \biggl ( (288 
\no
&&
+ 192\,
\eta ){e_{{t}}}^{2}
+88
-736\,\eta \biggr )(e_t\,\cos u)
+ (1152-24
\,\eta ){e_{{t}}}^{2}-632-424\,\eta 
\biggr ] {C}^{2}
\no
&&
+ 5\,( 1 -2\,\eta)\,\biggl [ 
12\,{(e_t\,\cos u)}^{3}-36\,{(e_t\,\cos u)}^{2}
+ (77-48\,{e_{{
t}}}^{2} )(e_t\,\cos u)
\no
&&
+72\,{e_{{t}}}^{2} -77
\biggr ] {C}^{4} 
\biggr \}\,,\\
\rm P^{1.5}_{\rm C3S3} = \rm P^{1.5}_{\rm S3C3} &=&     
{S \over 64}\,{ m_1 -m_2 \over m}\,
{1 \over (1 -e_t\,\cos u)^4}\,
\biggl \{ 
{1 \over \sqrt {1-e_t^2}}
\biggl [
\biggl ( (216-48\,\eta ){e_{
{t}}}^{2}- 120 -48\,\eta \biggr ){(e_t\,\cos u)}^{3}
\no
&&
- \biggl ( (644 -120\,\eta
 ){e_{{t}}}^{2}-436-88\,\eta \biggr ){(e_t\,\cos u)}^{2}- \biggl ( (288 -64
\,\eta ){e_{{t}}}^{4}
- (76
\no
&&
+232\,\eta ){e_{{t}}}^
{2}-340 +424\,\eta \biggr )(e_t\,\cos u)+ (719+2\,\eta ){e_{{t}}}^{4}
- (510+436\,\eta ){e_{{t}}}^{2}
\no
&&
-225+450\,\eta \biggr ]
+{1 \over \sqrt {1-e_t^2}}
 \biggl [  \biggl (
 (144 + 96\,\eta ){e_{{t}}}^{2}-48-192\,\eta \biggr ){(e_t\,\cos u)}^{3
}
\no
&&
- \biggl ( (504 + 160\,\eta ){e_{{t}}}^{2}-296 -368\,\eta
 \biggr ){(e_t\,\cos u)}^{2}- \biggl ( (192 + 128\,\eta ){e_{{t}}}^{4}+
 (96 
\no
&&
-576\,\eta ){e_{{t}}}^{2}
-416 +576\,\eta \biggr )(e_t\,\cos u)+
 (928 - 416\,\eta){e_{{t}}}^{4}
- (1016
\no
&&
-576\,\eta 
){e_{{t}}}^{2} 
+72 -144\,\eta \biggr ] {C}^{2}
- \sqrt { 1 -e_t^2}\,
(1 -2\,\eta ) 
\biggl [ 120\,{(e_t\,\cos u)}^{3}
\no
&&
-276\,{(e_t\,\cos u)}^{2}
- (160\,{e_{{t}}}^{2}-212 )(e_t\,\cos u)
+185\,{e_{{t}}}^{2}-81 
\biggr ]{C}^{4} \biggr \}\,,\\
\rm P^{1.5}_{\rm C5C5} = -\rm P^{1.5}_{\rm S5S5} &=&     
{S^3 \over 96}\,{ m_1 -m_2 \over m}\,
{e_t\,\sin u \over (1 -e_t\,\cos u)^5}\,
(1 +C^2)
\,(1 -2\,\eta) 
 \biggl \{ 12\,{(e_t\,\cos u)}^{4}
\no
&&
-48\,{(e_t\,\cos u)}^{3}
- (144\,{e_{{t}}}^{2}-209
 ){(e_t\,\cos u)}^{2}
+ (360\,{e_{{t}}}^{2} -394)(e_t\,\cos u)
\no
&&
+192\,{e_{{t}}}^{4}
-600\,{e_{{t}}}^{2}
+413 
\biggr \}\,,\\
\rm P^{1.5}_{\rm C5S5} = \rm P^{1.5}_{\rm S5C5} &=&     
-{S^3 \over 192}\,{ m_1 -m_2 \over m}\,
{ \sqrt {1 -e_t^2} \over (1 -e_t\,\cos u)^5}\,
(1 +C^2)
\,(1 -2\,\eta)\,\biggl \{
 120\,{(e_t\,\cos u)}^{4}
\no
&&
-396\,{(e_t\,\cos u)}^{3}
- (480\,{e_{{t}}}^{2} -808
 ){(e_t\,\cos u)}^{2}+ (825\,{e_{{t}}}^{2} 
-773)(e_t\,\cos u)
\no
&&
+384\,{e_{{t}}}^{4}
-1113\,{e_{{t}}}^{2}+625 
\biggl \}\,,\\
\rm P^{2}_{\rm C2C2} = -\rm P^{2}_{\rm S2S2} &=&     
{1 \over 5760}\,{ 1 \over (1 -e_t\,\cos u)^5}\,
\biggl \{ 
{ 2880 \over \sqrt {1 -e_t^2}}\, (1+ {C}^{2})
(1-e_{{t}}{\cos u} )^{2} (5-2\,\eta )
\no
&&
 \biggl [ -2\,{(e_t\,\cos u)}^{3}
+7\,{(e_t\,\cos u)}^{2}
+ (4\,{e_{{t}}}^{2} + 3)(e_t\,\cos u)
-16\,{e_{{t}}}^{2} 
+ 4
\biggr ] 
\no
&&
+ 
{1 \over (1 -e_t^2)}\, \biggl [ 
-(1 -e_t^2)\,(7560
 -15720\,\eta
 -600\,\eta^2   )
\,(e_t\,\cos u)^5
\no
&&
+ (1 -e_t^2)\,
(30240 -62880\,\eta -2400\,{\eta}^{2})\,(e_t\,\cos u)^4
\no
&&
+ \biggl ( 
-(15120  - 31440\,\eta -1200\,{\eta}^{2}
 ){e_{{t}}}^{4}+ (7518 -66070\,\eta
+5670\,{\eta}^{2} 
){e_{{t}}}^{2}
\no
&&
-65838 
+ 82150\,\eta -6870\,{\eta}^{2} \biggr ){(e_t\,\cos u)}^{3}+
 \biggl ( ( 28980  
 + 15420\,\eta 
\no
&&
-35580\,{\eta}^{2}
){e_{{t}}}^{4
}
+ (64686 -152550\,\eta
-6090\,{\eta}^{2} ){e_{{t}}}^{2}
+126654
-
5430\,\eta
\no
&&
+41670\,{\eta}^{2}
 \biggr ){(e_t\,\cos u)}^{2}
+ \biggl ( (
99585+20955\,\eta-19755\,{\eta}^{2} ){e_{{t}}}^{4}
- ( 293136 
\no
&&
  -38160\,\eta - 163200
\,{\eta}^{2}
){e_{{t}}}^{2}
-26769
+83445\,\eta
-143445\,{
\eta}^{2} \biggr )(e_t\,\cos u)
\no
&&
+ (10665 -115245\,\eta+63405\,{\eta}^{2
}
){e_{{t}}}^{6}
- (145440
-277920\,\eta 
+136080\,{\eta}^{2}
){e_{{t}}}^{4}
\no
&&
+ (275607
-212435\,\eta  
+25635\,{\eta}^{2}
){e_{{t}}}^{2}
-67392+2240\,\eta+47040\,{\eta}^{2} \biggr ]
\no
&&
+{1 \over (1 -e_t^2)}\,\biggl [
 \biggl ( (1 -e_t^2)\,( 6120 -21960\,\eta
-11640\,{\eta}^{2} )
 \biggr ){(e_t\,\cos u)}^{5}
- (1 -e_t^2)\,
\no
&&
 \biggl ( 
24480 
-87840\,\eta -46560\,{\eta}^2 
 \biggr ){(e_t\,\cos u)}^{4}
+ \biggl ( ( 12240 -43920\,\eta
\no
&&
-23280\,{\eta}^{2} ){e_{{t}}}^{4} 
- (75006-129750\,
\eta-181050\,{\eta}^{2} ){e_{{t}}}^{2}
-10674
- 38310\,\eta
\no
&&
-157770\,{\eta}^{2}
 \biggr ){(e_t\,\cos u)}^{3}
+ \biggl ( -( 16020 - 113700\,\eta
-56220\,{
\eta}^{2} ){e_{{t}}}^{4}
\no
&&
+ ( 108018 -202650\,\eta
-213270\,{
\eta}^{2} ){e_{{t}}}^{2}
+128322    
-53610\,\eta
\no
&&
+157050\,{\eta}^{2}
 \biggr ){(e_t\,\cos u)}^{2} 
+\biggl ( (  67455 
+ 118965\,\eta+7995\,{\eta}^{2}
 ){e_{{t}}}^{4}
- ( 202608  
\no
&&
+ 171600\,\eta +
89760\,{\eta}^{2}
 ){e_{{t}}}^{2}
-85167+195195\,\eta
+81765\,{\eta}^{2} \biggr )(e_t\,\cos u)
\no
&&
+
(12375 -88515\,\eta  -64125\,{\eta}^{2} ){e_{{t}}}^{6}
-
 (  100800  - 76800\,\eta
-151440\,{\eta}^{2} ){e_{{t}}}^{4}
\no
&&
+ (188361+44835\,\eta -35475\,{\eta}^{2} ){e_{{t}}}^{2}
-
26496-80640\,\eta
-51840\,{\eta}^{2} 
\biggr ]
{C}^{2}
\no
&&
+ 5\, 
\biggl [  -(1800-2856\,\eta -7128\,{\eta}^{2} ){
(e_t\,\cos u)}^{5}+ (7200
-11424\,\eta   
-28512\,{\eta}^{2}
\no
&&
){(e_t\,\cos u)}^{4}+
 \biggl ( (3600 -5712\,\eta-14256\,{\eta}^{2} ){e_{{t}}}^{2}-
4746 
-7342\,\eta
+61614\,{\eta}^{2} 
\no
&&
\biggr ){(e_t\,\cos u)}^{3}+ \biggl (
- (   18684 -43500\,\eta
-
34452
\,{\eta}^{2} ){e_{{t}}}^{2}
- 6+23358\,\eta
\no
&&
-67518\,{\eta}^{2}
 \biggr ){(e_t\,\cos u)}^{2}+ \biggl ( 
(12909   -32529 \,\eta -16815\,{\eta}^{2}
 ){e_{{t}}}^{2}
+8109
\no
&&
 -30609\,\eta  
+19281\,{\eta}^{2}
 \biggr )(e_t\,\cos u)+ (9045
-24345\,\eta
-7335\,{\eta}^{2} ){e_{{t}}}^
{4}
\no
&&
- (15915 -43431\,\eta 
-11289\,{\eta}^{2}
){e_{{t}}}^{2}+
288 -1184\,\eta
+672\,{\eta}^{2} 
\biggr ] {C}^{4}
\no
&&
+15\, (1 -5\,\eta + 5\,{\eta}^{2}
)  
\biggl [ -360\,{(e_t\,\cos u)}^{5}
+1440\,{(e_t\,\cos u)}^{4}
+ \biggl (
720\,{e_{{t}}}^{2}
\no
&&
 -2418 
 \biggr ){(e_t\,\cos u)}^{3}
-
 (2124\,{e_{{t}}}^{2}
 -2178 ){(e_t\,\cos u)}^{2}+ (1553\,{e_{{t}}
}^{2}
\no
&&
-527 )(e_t\,\cos u)
+345\,{e_{{t}}}^{4}
-839\,{e_{{t}}}^{2}
+32 
\biggr ] {
C}^{6} 
\biggr \}\,,\\
\rm P^{2}_{\rm C2S2} = \rm P^{2}_{\rm S2C2} &=&     
{1 \over 192}\,{ e_t\,\sin u \over (1 -e_t\,\cos u)^5}\,
\biggl \{ {1 \over \sqrt{ (1 -e_t^2)^3}}\,
\biggl [
 \biggl ( -(504-1048\,\eta-40\,{\eta}^{2
} ){e_{{t}}}^{4}+ (
144
\no
&&
-592\,\eta +48\,{\eta}^{2} ){e_
{{t}}}^{2}
-1080+216\,\eta
+104\,{\eta}^{2} \biggr ){(e_t\,\cos u)}^{3}+ \biggl (
 ( 1584  
\no
&&
-2800\,\eta
-448\,{\eta}^{2}
){e_{{t}}}^{4}
- ( 1536
-1568\,\eta-416\,{\eta}^{2} ){e_{{t}}}^{2}+
4272-784\,\eta 
\no
&&
-544\,{\eta}^{2}
\biggr ){(e_t\,\cos u)}^{2}
+ \biggl ( (1956 + 908\,\eta-92\,{\eta
}^{2} ){e_{{t}}}^{4}
- (4584 -1736\,\eta 
\no
&&
-760\,{\eta}^{2}
 ){e_{{t}}}^{2}-1692
-628\,\eta   -92\,{\eta}^{2} \biggr )(e_t\,\cos u)+ (
1359
-6699\,\eta 
\no
&&
+ 171\,{\eta}^{2}
 ){e_{{t}}}^{6}- (7113-
20941\,\eta
+13\,{\eta}^{2} ){e_{{t}}}^{4}+ (
10053 -22809\,\eta
\no
&&
-711\,{\eta}^{2
} ){e_{{t}}}^{2}-2859+7895\,\eta+361\,{\eta}^{
2}
\biggr ]
+ 
{1 \over \sqrt{ (1 -e_t^2)^3}}\,
\biggl [  \biggl ( (408
\no
&&
-1464\,\eta-776\,{\eta}^{2} ){e_{{t}}
}^{4} - (1296
- 2896\,\eta-2832\,{\eta}^{2} ){e_{{t}}}^{2}
-552-760\,\eta 
\no
&&
- 1864\,{\eta}^{2}
\biggr ){(e_t\,\cos u)}^{3}+ \biggl ( 
-( 1536-5824\,\eta 
-2000\,\eta^2
){e_{{t}}}^{4}+ (
3648
\no
&&
-11456\,\eta
-7648\,{\eta}^{2} ){e_{{t}}}^{2}
+2208+3616
\,\eta
+5072\,{\eta}^{2} 
 \biggr ){(e_t\,\cos u)}^{2} 
\no
&&
+\biggl ( (
1692 +2740\,\eta 
-3012\,{\eta}^{2}
 ){e_{{t}}}^{4}
- (3096+5768\,\eta  -9480\,{\eta}^{2} )
{e_{{t}}}^{2}
\no
&&
-2916
+5044\,\eta 
-5892\,{\eta}^{2}
\biggr )(e_t\,\cos u)
- (1167 -987\,\eta 
-1605 \,{\eta}^{2}
){e_{{t}}}^{6}
\no
&&
+ (2937-10061\,
\eta-3027\,{\eta}^{2} ){e_{{t}}}^{4}
- ( 2757
-
17289\,\eta
-151\,{\eta}^{2} 
 ){e_{{t}}}^{2}
\no
&&
+2427
-8887\,\eta + +1079\,{\eta}^{2}
 \biggr ]\,{C}^{2}
+{1 \over \sqrt{1 -e_t^2}}\, 
\biggl [  \biggl ( (600 
-952\,\eta-2376\,{\eta}^{2} ){e_{{t}}}^{2}
\no
&&
-216 -584\,\eta
+3528\,{\eta}^{2}
\biggr ){(e_t\,\cos u)}^{3}+ 
\biggl ( -(2688
-5440\,\eta
-7296\,{\eta}^{2}  
 ){e_{{t}}}^{2}
\no
&&
+1440-448\,\eta 
-11040\,{\eta}^{2}
\biggr ){(e_t\,\cos u)
}^{2}+ \biggl ( (1684
-2596\,\eta
-6940\,{\eta}^{2}
 ){e_{{t}}}^
{2}
\no
&&
-340 -2780\,\eta+10972\,{\eta}^{2} \biggr )(e_t\,\cos u)+ (
975
-3195\,\eta
+795\,{\eta}^{2}
){e_{{t}}}^{4}
\no
&&
- (1546
-4498 \,\eta 
-430\,{\eta}^{2}
){e_{{t}}}^{2}+91
+617\,\eta 
-2665\,{\eta}^{2}
\biggr ]
{C}^{
4}
+5\, (1 -5\,\eta 
\no
&&
+5\,{\eta}^{2} ) \,
\sqrt {1 -e_t^2}\, 
 \biggl [
-72\,{(e_t\,\cos u)}^{3} +240\,{(e_t\,\cos u)}^{2}-220\,(e_t\,\cos u)
\no
&&
+ 3\,{e_{{t}}}^{2}+49
\biggr ] {C}^{6} \biggr \}
+ (1 +C^2)\,(5 -2\,\eta)\,
{1 \over (1 -e_t^2)}\,
{e_t\,\sin u \over (1 -e_t\,\cos u)^4}\,
\biggl \{  
\no
&&
(2\,{e_{{t}}}^{2} +1)(e_t\,\cos u)
-8\,{e_{{t}}}^{2} 
+5
\biggr \}\,,\\
\rm P^{2}_{\rm C4C4} = -\rm P^{2}_{\rm S4S4} &=&     
{S^2 \over 2880}\,
\,{1 \over (1 -e_t\,\cos u)^5}\,
\biggl \{  
\biggl [ (2160-6960\,\eta+720\,{\eta}^{2}
 ){(e_t\,\cos u)}^{5} 
\no
&&
-\biggl (8640
-27840\,\eta
+2880\,{\eta}^{2}
\biggr ){
(e_t\,\cos u)}^{4}+ \biggl ( 
-(17280 -55680\,\eta 
\no
&&
+5760\,{\eta}^{2} ){e_{
{t}}}^{2}
+14238
-37370\,\eta
-25350\,{\eta}^{2} \biggr ){(e_t\,\cos u)}^{3}+ 
 \biggl (
( 63000
\no
&&
-199800\,\eta +20520\,{\eta}^{2})\,e_t^2
-31314
+78150\,\eta
+59850\,{\eta}^{2} \biggr ){(e_t\,\cos u)}^{2}
\no
&&
+ \biggl ( (17280
-55680\,\eta 
+ 5760\,{\eta}^{2}
){e_{{t}}}^{4}
- (24105
 -60435\,\eta
-49125\,{\eta}
^{2}
 ){e_{{t}}}^{2}
\no
&&
-23661
+107055\,\eta
-114375
\,{\eta}^{2} \biggr )(e_t\,\cos u)- 
(49905
-148875\,\eta
\no
&&
+7155\,{\eta}^{2}
 ){e_{{t}}}^{4}
+ ( 43635 -102705\,\eta
-61095\,{\eta}^{2}
 ){e_{{t}}}^{2}
+ 14592 
-75520\,\eta
\no
&&
+80640\,{\eta}^{2}
\biggr ]
+ \biggl [ 
 (
1800  -5160\,\eta -1080\,{\eta}^{2}){(e_t\,\cos u)}^{5}
- (7200 
-20640\,\eta    
\no
&&
-4320\,{\eta}^{2}
){(e_t\,\cos u)}^{4}
+ \biggl ( -(14400
-41280\,\eta
- 8640\,{ \eta}^{2}
 ){e_{{t}}}^{2}
+9660 
\no
&&
-14480\,\eta 
-48240\,{\eta}^{2}
\biggr ){(e_t\,\cos u)}^{3}
+ \biggl ( 
(58140 -175500\,\eta -3780\,{\eta}^{2}
){e_{{t}}}^{2}
\no
&&
-26400
+53580\,\eta +84420\,{\eta}^{2}
 \biggr ){(e_t\,\cos u)}^{2}
+ \biggl ( 
(14400 -41280\,\eta 
\no
&&
-8640\,{\eta}^{
2} ){e_{{t}}}^{4}
- (21780 - 48810\,\eta 
-60750\,{\eta}^{2}
 ){e_{{t}}}^{2}
-22080+99150\,\eta
\no
&&
-106470\,{\eta}^{2} \biggr )(e_t\,\cos u)
- (62460 -211650\,\eta
+69930\,{\eta}^{2} ){e_{{t}}}^{4}
+
 (74160
\no
&&
-255330\,\eta+91530\,{\eta}^{2} ){e_{{t}}}^{2}
-3840
+16640\,\eta -11520\,{\eta}^{2}
 \biggr ] {C}^{2}
\no
&&
+ 3\,(1-5\,\eta+5
\,{\eta}^{2} ) \biggl [ 360\,{(e_t\,\cos u)}^{5}
-1440\,{(e_t\,\cos u)}^{4}
- (2880
\,{e_{{t}}}^{2}
\no
&&
-4578 ){(e_t\,\cos u)}^{3}
+ (
7740\,{e_{{t}}}^{2}
-7794
 ){(e_t\,\cos u)}^{2}
+ (2880\,{e_{{t}}}^{4}
\no
&&
-8885\,{e_{{t}}}^{2}
+ 4979 )(e_t\,\cos u)-4245\,{e_{{t}}}^{4}+6755\,{e_{{t}}}^{2} 
-2048
\biggr ] {C}^{4
}
\biggr \}\,,\\ 
\rm P^{2}_{\rm C4S4} = \rm P^{2}_{\rm S4C4} &=&     
{S^2 \over 48}\,
{e_t\,\sin u \over (1 -e_t\,\cos u)^5}\,
\biggl \{ 
{1 \over \sqrt {1-e_t^2}}\,
\biggl [
 \biggl ( -(144 -464\,\eta +
48\,{\eta}^{2}
 ){e_{{t}}}^{2}
\no
&&
+96
-272\,\eta 
-96\,{\eta}^{2}
\biggr )
{(e_t\,\cos u)}^{3}+ \biggl ( (
558 -1790\,\eta 
+198\,{\eta}^{2} ){e_{{t}
}}^{2}
\no
&&
-402+1166\,\eta+270\,{\eta}^{2} \biggr ){(e_t\,\cos u)}^{2}
+ \biggl ( (288
-928\,\eta+96\,{\eta}^{2} ){e_{{t}}}^{4}
\no
&&
- ( 680 -1936\,\eta
-632\,{\eta}^{2} ){e_{{t}}}^{2}
 + 224 - 336\,\eta
-1232\,{\eta}^{2
} \biggr )(e_t\,\cos u)
\no
&&
 - (
879
-2665\,\eta
+93\,{\eta}^{2}
 ){e_{{t}}}^{4
}+ (1448
-4084\,\eta
-788\,{\eta}^{2}
 ){e_{{t}}}^{2}
-509+1179
\,\eta
\no
&&
+1061\,{\eta}^{2} \biggr ] 
+
{1 \over \sqrt {1 -e_t^2}}
\biggl [  \biggl ( -(120
-344\,\eta-72\,{\eta}^{2}
){e_{{t}}}^{2}
+72
-152\,\eta
\no
&&
-216\,{\eta}^{2}
 \biggr ){(e_t\,\cos u)}^{3}
+ \biggl ( (486 -1430\,\eta-162\,{\eta}^{2} ){e_{{t}}}^{2}-
330+806\,\eta
\no
&&
+630\,{\eta}^{2} \biggr ){(e_t\,\cos u)}^{2}
+ \biggl ( (240 -688\,
\eta-144\,{\eta}^{2} ){e_{{t}}}^{4}
- (540
 -1236\,\eta
\no
&&
-1332\,
{\eta}^{2} ){e_{{t}}}^{2}
 + 132 + 124\,\eta
 -1692\,{\eta}^{2}
 \biggr )(e_t\,\cos u)
- (936 -2950\,\eta
\no
&&
+378\,{\eta}^{2} ){e_{{t}}}^{4}
+ (1566-4674\,\eta-198\,{\eta}^{2} ){e_{{t}}}^{2}
-570 +1484\,\eta
+756\,{ \eta}^{2}
 \biggr ] {C}^{2}
\no
&&
+ \sqrt {1-{e_{{t}}}^{2} }\,
 (1 -5\,\eta +
5\,{\eta}^{2} ) \biggl [ 72\,{(e_t\,\cos u)}^{3}
-240\,{(e_t\,\cos u)}^
{2}
\no
&&
- (144\,{e_{{t}}}^{2} -364)(e_t\,\cos u)
+249\,{e_{{t}}}^{2}-301
 \biggr ] {C}^{4} 
\biggr \}\,,\\
\rm P^{2}_{\rm C6C6} = -\rm P^{2}_{\rm S6S6} &=&     
{S^4 \over 1920}\,(1 +C^2)\,
(1 -5\,\eta +5\,\eta^2)\,
{1 \over (1 -e_t\,\cos u)^6}\,
\biggl \{ 120\,{(e_t\,\cos u)}^{6}
\no
&&
-600\,{(e_t\,\cos u)}^{5}
- (2160\,{e_{{t}}}^{2}-3206 ){(e_t\,\cos u)}^{4}+
 (7860\,{e_{{t}}}^{2}
\no
&&
 -8444 ){(e_t\,\cos u)}^{3}
+ (
5760\,{e_{{t}}}^{4}
-19135\,e_t^2
+13051 ){(e_t\,\cos u)}^{2}
- (
11475\,{e_{{t}}}^{4}
\no
&&
-23240\,{e_{{t}}}^{2}
+11269
 )(e_t\,\cos u)
-3840\,{e_{{t}}}^{6}
+17235\,{e_{{t}}}^{4}
-21325\,e_t^2
+7776 
\biggr \}\,,\\
\rm P^{2}_{\rm C6S6} = \rm P^{2}_{\rm S6C6} &=&     
{\sqrt {1-e_t^2} \over 192}\, S^4\,(1 +C^2)\,
(1 -5\,\eta +5\,\eta^2)\,
{e_t\,\sin u \over (1 -e_t\,\cos u)^6}\,
\biggl \{ 72\,{(e_t\,\cos u)}^{4}
\no
&&
-312\,{(e_t\,\cos u)}^{3}
- (384\,{e_{{
t}}}^{2}-844 ){(e_t\,\cos u)}^{2}+ (1053\,{e_{{t}}}^{2} 
-1325)
(e_t\,\cos u)
\no
&&
+384\,{e_{{t}}}^{4}
-1437\,{e_{{t}}}^{2}  
+1105 
\biggr \}\,,\\
\rm P^{2} &=&     
{S^2 \over 192}\,
{1 \over (1 -e_t\,\cos u)^5}
\biggl \{
\biggl [
 (120 -120\,\eta-8\,{\eta}^{2} ){
(e_t\,\cos u)}^{5}+ (3360
\no
&&
-2608\,\eta 
+416\,{\eta}^{2}
){(e_t\,\cos u)}^{4}
- (
11058-7718\,\eta + 1086\,{\eta}^{2} ){(e_t\,\cos u)}^{3}
\no
&&
+ \biggl (
 (168 + 1240\,\eta+152\,{\eta}^{2} ){e_{{t}}}^{2}
+ 13710 -8746\,\eta
+738\,{\eta}^{2} \biggr ){(e_t\,\cos u)}^{2}
\no
&&
+ \biggl ( -(3441
-139\,\eta 
-277\,\eta^2
){e_{{t}}}^{2} -5181 
+2751\,\eta
-423\,{\eta}^{2}
 \biggr )(e_t\,\cos u)
\no
&&
+ (135
-477\,\eta-363\,{\eta}^{2} ){e_{{t}}}^{4}
+ (3003 -425\,\eta
+297\,{\eta}^{2}
){e_{{t}}}^{2}-816+528\,
\eta
\biggr ]
\no
&&
+2\, \biggl [  (180 -516\,\eta-108\,{\eta}^{2}){(e_t\,\cos u)}^{5}
-
 (720
-2064\,\eta
-432\,{\eta}^{2} ){(e_t\,\cos u)}^{4}
\no
&&
+ (
678-1928\,\eta -504\,\eta^2){(e_t\,\cos u)}^{3}+ \biggl ( (150 -430\,\eta
-90\,{\eta}^{2} ){e_{{t}}}^{2}
+336
\no
&&
-1010\,\eta+90\,{\eta}^{2}
 \biggr ){(e_t\,\cos u)}^{2}+ \biggl ( -(930
-2705\,\eta -315\,{\eta}^{2}  
){e_{{t}}}^{2}
\no
&&
-96
+283\,\eta+9\,{\eta}^{2} \biggr )(e_t\,\cos u)+ ( 378
-1107\,\eta 
-81\,\eta^2
){e_{{t}}}^{4}+ (24
-61\,\eta 
\no
&&
-63\,{\eta}^{2}
){e_{{t}}}^{2} \biggr ]{C}^{2}+ 
(1 -5\,\eta +5\,\eta^2)\,
\biggl [
120\,{(e_t\,\cos u)}^{5}-480\,{(e_t\,\cos u)}^{4}
\no
&&
+566\,{(e_t\,\cos u)}^{3}
+\biggl ( 84\,e_t^2
-102
\biggr ){(e_t\,\cos u)}^{2}
-\biggl ( 
343\,e_t^2-1
\biggr )(e_t\,\cos u)
\no
&&
+105\,e_t^4
+49\,{ e_t}^{2}
\biggr ]
{C}^{4} 
\biggr \}
+{ S^2\over 12}
{1 \over (1-e_{{t}}{\cos u} )^{2} }
 \biggl \{ 
{1 \over  {(1 -e_t^2)}} \,\biggl [ 
 \biggl (
\no
&&
(1 -e_t^2)\,
(240
-193\,\eta
+24\,{\eta}^{2}) \biggr )(e_t\,\cos u)
+51
-33\,\eta \biggr ] 
+{ e_t\, \cos u\over \sqrt { 1 -e_t^2}}\, 
\no
&&
\biggl [ 
-(60 -24\,\eta)\,e_t\,\cos u
+ 
150 -60\,\eta\,
 \biggr ] \biggr \}\,,\
\end{eqnarray}
\end{mathletters}

\begin{mathletters}
\begin{eqnarray}
{\rm X^{0}_{\rm C2C2} }= -\rm X^{0}_{\rm S2S2} &=& 
-4\,C\,{ \sqrt {1 -e_t^2} \over (1 -e_t\,\cos u)^2}
\,
e_{{t}}\,{\sin u} \,, \\
{\rm X^{0}_{\rm C2S2} }= \rm X^{0}_{\rm S2C2} &=& 
2\,C\,{ 1 \over (1 -e_t\,\cos u)^2}\biggl \{
  (e_t\,\cos u)^2 -(e_t\,\cos u) - 2\,({e_{{t}}}^{2} -1)
 \biggr \}\,, \\
{\rm X^{0.5}_{\rm C1C1} }= -\rm X^{0.5}_{\rm S1S1} &=& 
{C\,S \over 2}\, {m_1 -m_2 \over m}\,
{\sqrt {1-e_t^2}  \over (1 -e_t\,\cos u)^2}\biggl \{
2\,(e_t\,\cos u)-3  \biggr \}\,,\\
{\rm X^{0.5}_{\rm C1S1} }= \rm X^{0.5}_{\rm S1C1} &=& 
-C\,S\, { m_1 -m_2 \over m}\,
{ 1 \over (1 -e_t\,\cos u)}\,e_t\,\sin u \,,\\
{\rm X^{0.5}_{\rm C3C3} }= -\rm X^{0.5}_{\rm S3S3} &=& 
-{C\,S \over 2}\,{ m_1 -m_2 \over m}\,
{ 1 \over (1 -e_t\,\cos u)^3}\biggl \{
\sqrt {1-e_t^2} \biggl [6\,{(e_t\,\cos u)}^
{2}-7\,(e_t\,\cos u) 
\no
&&
 -8\,{e_{{t}}}^{2} +9 \biggr ]
\biggr \}\,,\\
{\rm X^{0.5}_{\rm C3S3} }= \rm X^{0.5}_{\rm S3C3} &=& 
-{C\,S\,}\, {m_1 -m_2 \over m}\,
{ 1 \over (1 -e_t\,\cos u)^3}\biggl \{
e_{{t}}\,{\sin u}\, 
\biggl [ {(e_t\,\cos u)}^{2}-2
\,(e_t\,\cos u) 
\no
&&
-4\,{e_{{t}}}^{2} +5 \biggr ] \biggr \}\,,\\
{\rm X^{1}_{\rm C2C2} }= -\rm X^{1}_{\rm S2S2} &=&   
{C \over 3}\,{e_t\,\sin u \over (1 -e_t\,\cos u)^3}\,
\biggl \{ 
{ 1 \over \sqrt {1 -e_t^2}}
\,\biggl [
\biggl ( -(12 +8\,\eta ){e_{{t}}}^{2}+20
\,\eta \biggr )(e_t\,\cos u)
\no
&&
+(33 +11\,\eta ){e_{{t}}}^{2}-21-23\,\eta
 \biggr ]
+3\,\sqrt {1 -e_t^2}\, 
\,
(1-3\,\eta ) \,( 2 \,(e_t\,\cos u) -3)\,{C}^{2} 
\biggr \}\,,\\
{\rm X^{1}_{\rm C2S2} }= \rm X^{1}_{\rm S2C2} &=&   
{C \over 6}\,{1 \over (1 -e_t\,\cos u)^3}\,
\biggl \{ 
\biggl [
-(12 +8\,
\eta ){(e_t\,\cos u)}^{3}
+(24+16\,\eta ){(e_t\,\cos u)}^{2}
\no
&&
+ \biggl (
(24+16\,\eta ){e_{{t}}}^{2}+53-63\,\eta \biggr )(e_t\,\cos u)
-(69 +
13\,\eta ){e_{{t}}}^{2}
\no
&&
 -20 +52\,\eta
\biggr ]
+(1-3\,\eta )
\biggl [ 
-6\,(e_t\,\cos u)^3 +12\,(e_t\,\cos u)^2
\no
&&
 -(13 -12\,e_t^2)\,(e_t\,\cos u) 
-21\,e_t^2 +16
\biggr ]
{C}^{2} 
\biggr \}\,,\\
{\rm X^{1}_{\rm C4C4} }= -\rm X^{1}_{\rm S4S4} &=&   
{ C\,S^2 \over 2}\,{ e_t\,\sin u \over (1 -e_t\,\cos u)^4}\,
\sqrt {1-e_t^2}\,
(1-3\,\eta )\biggl \{ 4\,{
(e_t\,\cos u)}^{2} -9\,(e_t\,\cos u)
\no
&&
 -8\,{e_{{t}}}^{2} +13 \biggr \}
\,,\\
{\rm X^{1}_{\rm C4S4} }= \rm X^{1}_{\rm S4C4} &=&   
{C\,S^2 \over 12}\,{ 1 \over (1 -e_t\,\cos u)^4}\,
(1-3\,\eta )
\biggl \{
-6\,{(e_t\,\cos u)}^{4}
\no
&&
+18\,{(e_t\,\cos u)}^{3}+(48\,{e_{{t}}}^{2}-61 )
{(e_t\,\cos u)}^{2}
-(69\,{e_{{t}}}^{2}-65 )(e_t\,\cos u)
\no
&&
-48\,{e_{{t}}}^{4} 
+117\,{e_{{t}}}^{2}-64
\biggr \}\,,\\
{\rm X^{1} }&=&
{C\,S^2 \over 2}\,{ e_{{t}}\,{\sin u} \over (1 -e_t\,\cos u)^3}\,
\sqrt {1-e_t^2}\,
(1-3\,\eta)\,,\\
{\rm X^{1.5}_{\rm C1C1} }=- \rm X^{1.5}_{\rm S1S1} &=& 
{C\,S \over 48 }\,{m_1 -m_2 \over m}\,
{ 1 \over (1 -e_t\,\cos u)^4}\,
\biggl \{
{1 \over \sqrt {1-e_t^2}}
\biggl [
-( 96\,{e_{{t}}}^{2} 
-48 -48\,\eta
){(e_t\,\cos u)}^{3}
\no
&&
+ \biggl ( (
432-24\,\eta ){e_{{t}}}^{2}-264-144\,\eta \biggr ){(e_t\,\cos u)}^{2}-
 \biggl ( (84 +88\,\eta ){e_{{t}}}^{2}
\no
&&
+108 -280\,\eta \biggr )
(e_t\,\cos u)- (477-138\,\eta ){e_{{t}}}^{4}
+ (702 -164\,\eta
 ){e_{{t}}}^{2}
\no
&&
-153-46\,\eta
\biggr ]
+ \sqrt {1-{e_{{t}}}^{2} }
 (1-2\,\eta ) 
\biggl [ 
24\,{(e_t\,\cos u)}^{3}-84\,{(e_t\,\cos u)}^{2}
\no
&&
 +68\,(e_t\,\cos u) 
-3\,{e_{{t}}}^{2}
-5  \biggr ] {C}^{2} 
\biggr \}\,,\\
{\rm X^{1.5}_{\rm C1S1} }= \rm X^{1.5}_{\rm S1C1} &=& 
{C\,S \over 24 }\,{m_1 -m_2 \over m}\,
{ e_t\,\sin u \over (1 -e_t\,\cos u)^4}\,
\biggl \{
\biggl [ 48\,{(e_t\,\cos u)}^{3}-144\,{(e_t\,\cos u)}^{2}
+ (33
\no
&&
+22\,\eta )(e_t\,\cos u)
+ (216+36\,\eta ){e_{{t}}}^{2}-153-58\,\eta
\biggr ]
+ (1-2\, \eta ) \biggl [ 
\no
&&
12\,{(e_t\,\cos u)}^{3}
-36\,{(e_t\,\cos u)}^{2}
+29\,(e_t\,\cos u)
+24\,{e_{{t}}}^{2}-29
 \biggr ]{C}^{2} 
\biggr \}\,,\\
{\rm X^{1.5}_{\rm C3C3} }= -\rm X^{1.5}_{\rm S3S3} &=& 
-{C\,S \over 32 }\,{m_1 -m_2 \over m}\,
{ 1 \over (1 -e_t\,\cos u)^4}\,
\biggl \{ {1 \over \sqrt {1 -e_t^2}}
\biggl [
 \biggl ( (168+48\,\eta ){e_{{t}}}^{2}-72
\no
&&
 -144\,\eta \biggr ){
(e_t\,\cos u)}^{3}
+ \biggl ( -(540+88\,\eta ){e_{{t}}}^{2}+332
+296\,\eta
 \biggr ){(e_t\,\cos u)}^{2}
\no
&&
+ \biggl ( -(224+64\,\eta ){e_{{t}}}^{4}
-
 (60-504\,\eta ){e_{{t}}}^{2}+ 412-568\,\eta \biggr )(e_t\,\cos u)
\no
&&
+
 (725 -10\,\eta ){e_{{t}}}^{4}
- (570+316\,\eta 
){e_{{t}}}^{2}-171+342\,\eta 
\biggr ]
\no
&&
+ \sqrt {1 -e_t^2}\,  (1-2
\,\eta ) 
\biggl [
-72\,{(e_t\,\cos u)}^{3}
+172\,{(e_t\,\cos u)}^{2}
+ \biggl (96\,e_t^2 
\no
&&
-140
\biggr )(e_t\,\cos u)
-191\,{e_{{t}}}^{2}+135 
\biggr ]
{C}^{2} 
\biggr \}\,,\\
{\rm X^{1.5}_{\rm C3S3} }= \rm X^{1.5}_{\rm S3C3} &=& 
{C\,S \over 48 }\,{m_1 -m_2 \over m}\,
{ e_t\,\sin u \over (1 -e_t\,\cos u)^4}\,
\biggl \{
\biggl [
 (84+24\,\eta ){(e_t\,\cos u)}^{3}- \biggl ( 252
\no
&&
+72\,
\eta \biggr ){(e_t\,\cos u)}^{2}
- \biggl ( (336+96\,\eta ){e_{{t}}}^{2}
+11 -582\,\eta \biggr )(e_t\,\cos u) 
\no
&&
+ (1080+120\,\eta ){e_{{t}}}^{2}
- 565 - 558\,\eta
\biggr ]
+ 3\,(1-2\,\eta ) 
\biggl [ 12\,{(e_t\,\cos u)}^{3}
\no
&&
-36\,{(e_t\,\cos u)}^{2}
- (48\,{e_{{t}}}^{2}-77 )(e_t\,\cos u)
+88\,{e_{{t}}}^{2}-93 
\biggr ] {C}^{2} 
\biggr \}\,,\\
{\rm X^{1.5}_{\rm C5C5} }= -\rm X^{1.5}_{\rm S5S5} &=& 
{C\,S^3 \over 96 }\,{m_1 -m_2 \over m}\,
{ 1 \over (1 -e_t\,\cos u)^5}\,
(1-2\,\eta ) 
\biggl \{ \sqrt {1 -e_t^2} \biggl [
\no
&&
120\,{(e_t\,\cos u)}^{4}
-396\,{(e_t\,\cos u)}^{3}-
 (480\,{e_{{t}}}^{2} 
-808
){(e_t\,\cos u)}^{2}+ \biggl (825\,{e_{{t}}}
^{2}
\no
&&
-773 \biggr )(e_t\,\cos u)
+384\,{e_{{t}}}^{4}
-1113\,{e_{{t}}}^{2}
+625 
\biggr ] \biggr \}\,,\\
{\rm X^{1.5}_{\rm C5S5} }= \rm X^{1.5}_{\rm S5C5} &=& 
{C\,S^3 \over 48 }\,{m_1 -m_2 \over m}\,
{ e_t\,\sin u \over (1 -e_t\,\cos u)^5}\,
 (1-2\,\eta )
 \biggl \{ 12\,{(e_t\,\cos u)}^{4}
\no
&&
-48\,{(e_t\,\cos u)}^{3}
- (144\,
{e_{{t}}}^{2} -209){(e_t\,\cos u)}^{2} + (360\,{e_{{t}}}^{2}-394 )
(e_t\,\cos u)
\no
&&
+192\,{e_{{t}}}^{4} 
-600\,{e_{{t}}}^{2}
+413
\biggr \}\,,\\
{\rm X^{2}_{\rm C2C2} }= -\rm X^{2}_{\rm S2S2} &=& 
2\,C\,{ e_t\,\sin u \over (1 -e_t\,\cos u)^3}\,
\biggl \{  { 5 -2\,\eta  \over  {1 -e_t^2}}
 \biggl [  
 -(2\,{e_{{t}}}^{2} +1 )(e_t\,\cos u)+
8\,{e_{{t}}}^{2}-5 
\biggr ] \biggr \}
\no
&&
+
{C \over 96}\,{e_t\,\sin u \over (1 -e_t\,\cos u)^5}\,
 \biggl \{ {1 \over (1 -e_t^2)^{3\over 2} }\,\biggl [
\biggl ( -(24-568\,\eta
-8\,{\eta}^{2} ){e_{{t}}}^{4}
\no
&&
+
 (
720-1872\, \eta 
-720\,{\eta}^{2}){e_{{t}}}^{2}
+744 
+632\,\eta+520\,{\eta}^{2}
\biggr ){(e_t\,\cos u)}^{
3}
\no
&&
+ \biggl ( (336-3056\,\eta+
352\,{\eta}^{2} ){e_{{t}}}^{4}-
 (1728-7840\,\eta
\no
&&
-1504\,{\eta}^{2} ){e_{{t}}}^{2}
-2928
-2768\, \eta 
-1280\,{\eta}^{2} \biggr ){(e_t\,\cos u)}^{2}+ \biggl (
-( 2652
\no
&&
-988\,\eta 
-852\,{\eta}^{2}
){e_{{t}}}^{4}+ (5400 - 3224\,\eta
-4008\,{\eta}^{2}
 ){e_{{t}}}^{2}
\no
&&
+1572 +220\,\eta +2580\,{\eta}^{2}
\biggr )(e_t\,\cos u)
- (885-5145\,\eta+297\,{\eta}^{2} ){e_{{t}}}^{6}
\no
&&
+
 (4995-13935\,\eta-321\,{\eta}^{2} ){e_{{t}}}^{4}
- (
7047-12691\,\eta
\no
&&
-2333\,{\eta}^{2} ){e_{{t}}}^{2}+1497
-3229\,\eta
-1523\,{ \eta}^{2}
 \biggr ]
+{2 \over \sqrt {1 -e_t^2}}\,  
\biggl [  \biggl (
- (216
\no
&&
-568\,\eta-264\,{\eta}^{2} ){e_{{t}}}^{2}
+120
-184\,\eta -552\,{\eta}^2
 \biggr ){(e_t\,\cos u)}^{3}+ \biggl ( (
936
\no
&&
 -2552\,\eta
-840\,{\eta}^{2}
){e_{{t}}}^{2} -600
+1208\,\eta+1848\,{\eta}^{2}
 \biggr ){(e_t\,\cos u)}^{2}+ \biggl ( -(868
\no
&&
-2220\,\eta
-1220\,{\eta}^{2}  
 ){e_{{t}}}^{2}+484-684\,\eta-2372\,{\eta}^{2} \biggr )(e_t\,\cos u) 
-(
597
\no
&&
-1737\,\eta 
-303\,{\eta}^{2}
){e_{{t}}}^{4}+ (1342
-3710\,\eta 
-1250 \,{\eta}^{2}
){e_{{t}}}^{2}-601
\no
&&
+1397\,\eta+1379\,{
\eta}^{2} 
\biggr ]
{C}^{2}
+ \sqrt {1 -e_t^2}\, (1
-5\,\eta + 5\,\eta^2)\, 
\biggl [
120\,{(e_t\,\cos u)}^{3}
\no
&&
-432\,{(e_t\,\cos u)}^{2}
+484\,(e_t\,\cos u)
+75\,{
e_{{t}}}^{2}
-247 
\biggr ]
{C}^{4} 
\biggr \}\,,\\
{\rm X^{2}_{\rm C2S2} }= \rm X^{2}_{\rm S2C2} &=& 
{C \over 2880 }\,{ 1 \over (1 -e_t\,\cos u)^5}\,
\biggl \{  
{1 \over ( 1-e_t^2)}\,\biggl [
  \biggl ( 
(1 -e_t^2)\, (360-8520\,\eta
\no
&&
-120\,{\eta}^{2}) \biggr 
){(e_t\,\cos u)}^{5}
-
 \biggl ( 
(1 -e_t^2)\,( 1440-34080\,\eta
- 480\,{\eta}^{2})
\no
&&
 \biggr ){(e_t\,\cos u)}^{4}+ \biggl ( (
720-17040\,\eta 
-240\, {\eta}^{2}
){e_{{t}}}^{4}
\no
&&
- (43158-78910\,
\eta-46290\,{\eta}^{2} ){e_{{t}}}^{2}
-31002 - 14350\,\eta
\no
&&
-46050\,{\eta}^{2}
 \biggr ){(e_t\,\cos u)}^{3}+ \biggl ( -(
16020
-146340\,\eta+21540\,{
\eta}^{2} ){e_{{t}}}^{4}
\no
&&
+ (
119994
-295890\,\eta
-79710\,{
\eta}^{2} ){e_{{t}}}^{2}
+116346+6990\,\eta
\no
&&
+101250\,{\eta}^{2}
 \biggr ){(e_t\,\cos u)}^{2}+ \biggl ( (
110835
-10935\,\eta-18105\,{\eta}^{2}
 ){e_{{t}}}^{4}
\no
&&
-( 276384+12480\,\eta 
-145680\,{\eta}^{2} 
 ){e_{{t}}}^{2}-54771
+165975\,\eta
\no
&&
-127575\,{\eta}^{2}
 \biggr )(e_t\,\cos u)
+ (26955-161775\,\eta+45135\,{\eta}^{2} ){e_{{t}}}^{6}
\no
&&
- (176400-366960\,\eta+95520\,{\eta}^{2} ){e_{{t}}}^{4}+
 (279333
\no
&&
-230305\,\eta+23505\,{\eta}^{2} ){e_{{t}}}^{2}
-56448
-22400\,\eta
+26880\,{\eta}^{2} 
\biggr ]
\no
&&
+{1 \over \sqrt { (1 -e_t^2)^3}}\, \biggl [ \biggl (
-(1 -e_t^2)\,(28800 -11520\,\eta )
 \biggr ){(e_t\,\cos u)}^{5}
\no
&&
+ \biggl (
(1 -e_t^2)\,(158400 -63360\,\eta)
\biggr ){(e_t\,\cos u)}^{4}
- \biggl (
(57600
-23040\,\eta ){e_{{t}}}^{4}
\no
&&
- (
244800
-97920\,\eta 
 ){e_{{t}}}^{2}+
187200
-74880\,\eta 
\biggr ){(e_t\,\cos u)}^{3}
\no
&&
+ \biggl ( (
345600 -138240\,\eta
 ){e_{{t}}}^{4}
- (417600-167040\,\eta ){e_{{t}}}^{2}
\no
&&
+72000-28800\,\eta \biggr ){(e_t\,\cos u)}^{2}+ 
\biggl ( -(518400-207360\,\eta
 ){e_{{t}}}^{4}
+ (590400 
\no
&&
-236160\,\eta ){e_{{t}}}^{2}
-72000+28800\,\eta \biggr )(e_t\,\cos u)
+ (230400-92160\,\eta ){e_{{t}}}^{4}
\no
&&
- (288000
-115200\,\eta
){e_{{t}}}^{2}
+
57600
-23040\,\eta
\biggr ]
+2\,  
\biggl [
- \biggl (
3240  -8520\,\eta
\no
&&
-3960\,{ \eta}^{2}
\biggr ){(e_t\,\cos u)}^{5}+ (12960 -34080\,\eta
-15840\,{\eta}^ {2} ){(e_t\,\cos u)}^{4}
\no
&&
+ \biggl ( ( 6480 
-17040\,\eta-7920\,{\eta}^{2} ){e_{{t}}}^{2}
-12582 
+24070\,\eta
+43770\,{\eta}^{2}
\biggr )
\no
&&
{(e_t\,\cos u)}^{3}+ \biggl ( -(35820
-101340\, \eta 
-21060\,{\eta}^{2}
){e_{{t}}}^{2}
+6846
\no
&&
+4530\,\eta
-78930\,{\eta}^{2} \biggr ){
(e_t\,\cos u)}^{2}+ \biggl ( (28455 
-74415\,\eta-33825\,{\eta}^{2} ){e_{{t}}}^{2}
\no
&&
+12159
-64695\,\eta
+85575\,{\eta}^{2} 
 \biggr )(e_t\,\cos u)+ (20655
-68535\,\eta 
\no
&&
+21735\,{\eta}^{2}
){e_{{t}}}^{4}
- (40425-
127185\,\eta+22785\,{\eta}^{2} ){e_{{t}}}^{2}
+4512
-6880\,\eta
\no
&&
-16800\,
{\eta}^{2} 
\biggr ]
{C}^{2}+15\,(1 -5\,\eta +5\,{\eta}^
{2}) 
\biggl [
-120\,{(e_t\,\cos u)}^{5}
+480\,{(e_t\,\cos u)}^{4}
\no
&&
+ (240\,{
e_{{t}}}^{2}-806 ){(e_t\,\cos u)}^{3}
- (900\,{e_{{t}}}^{2}
-918
 ){(e_t\,\cos u)}^{2}
\no
&&
+ (955\,{e_{{t}}}^{2}
-613 )(e_t\,\cos u)-45\,{e_{{t}
}}^{4}-205\,{e_{{t}}}^{2}+96 
\biggr ]{C}^{4} 
\biggr \}\,,\\
{\rm X^{2}_{\rm C4C4} }=-\rm X^{2}_{\rm S4S4} &=& 
{C\,S^2 \over 24 }\,
{ e_t\,\sin u \over (1 -e_t\,\cos u)^5}\,
\biggl \{ {1 \over \sqrt {1 -e_t^2}}\,\biggl [
 \biggl ( (120 -344\,\eta 
\no
&&
-72\,{\eta}^{2}
){e_{{t}}}^{2} -72 +152\,\eta
+216\,{\eta}^{2} 
\biggr ){(e_t\,\cos u)}^{3}+ \biggl ( -(480-1400\,\eta
\no
&&
-192\,{\eta}^{2} ){e_{{t }}}^{2}
+324
-776\,\eta
-660\,{\eta}^{2}
 \biggr ){(e_t\,\cos u)}^{2}+ \biggl ( -(
240-688\,\eta
\no
&&
-144\,{\eta}^{2} ){e_{{t}}}^{4}+
(
518-1126\,\eta 
-1442\,{\eta }^{2} 
){e_{{t}}}^{2}-110
-234\,\eta 
\no
&&
+1802\,{\eta}^{2}
\biggr )(e_t\,\cos u)+ (831
-2425\,\eta 
-147\,{\eta}^{2}
){e_{{t}}}
^{4}
- (1340
\no
&&
-3544\,\eta
-1328\,{\eta}^{2}
 ){e_{{t}}}^{2}
+449-879\,\eta
-1361\,{\eta}^{2}
\biggr ]
+ \sqrt {1 -e_t^2}\,
(1-5\,\eta
\no
&&
+5\,{\eta}^{2} )
\biggl [ -48\,{(e_t\,\cos u)}^{3}+162\,{(e_t\,\cos u)}^{2}+
 (96\,{e_{{t}}}^{2} -250 )(e_t\,\cos u)
\no
&&
-201\,{e_{{t}}}^{2} +241
\biggr ] {C}^{2} 
\biggr \}\,,\\
{\rm X^{2}_{\rm C4S4} }=\rm X^{2}_{\rm S4C4} &=& 
{C\,S^2 \over 720 }\,
{ 1 \over (1 -e_t\,\cos u)^5}\,
\biggl \{ \biggl [
 (
900
-2580\,\eta
-540\,{\eta}^{2} 
){(e_t\,\cos u)}^{5}
\no
&&
- (
3600-10320\,\eta 
-2160\,{\eta}^{2} 
){(e_t\,\cos u)}^{4}+
 \biggl ( -(7200
-20640\,\eta 
\no
&&
-4320\,{\eta}^{2}
){e_{{t}}}^{2}
+4830
- 7240\,\eta
-24120\,{\eta}^{2} \biggr ){(e_t\,\cos u)}^{3}+ \biggl ( (27990
\no
&&
-82350\,\eta 
-7290\,{\eta}^{2}
){e_{{t}}}^{2}
-12120
+21390\,\eta+47610\,
{\eta}^{2}
 \biggr ){(e_t\,\cos u)}^{2}
\no
&&
+ \biggl ( (7200
-20640\,\eta
-4320\,{\eta}^{2}
 ){e_{{t}}}^{4}
- (
8430-12105\, \eta 
-42675\,{\eta}^{2}
){e_{{t}}}^{2}
\no
&&
-13500
+61875\,\eta
-65535\,{\eta}^{2}
 \biggr )
(e_t\,\cos u)
- (24930
-74325\,\eta
\no
&&
+3465\,{\eta}^{2}
 ){e_{{t}}}^{4}+
 (23100
-57765\,\eta
-24135\,{\eta}^{2}
 ){e_{{t}}}^{2}+5760
-30080\,\eta 
\no
&&
+32640\,{\eta}^{2}
\biggr ]
+ 3\,( 1 -5\,\eta +5\,\eta^2)\,  
\biggl [ 120\,{(e_t\,\cos u)}^{5}-480\,{(e_t\,\cos u)}^{4}
\no
&&
+ (1526-960\,{e_{{t}}}^{2}
 ){(e_t\,\cos u)}^{3}
+ (2700\,e_t^2
-2718
 ){(e_t\,\cos u)}^{2}+
 (960\,{e_{{t}}}^{4}
\no
&&
-3235\,{e_{{t}}}^{2}+1933 )(e_t\,\cos u)
-2115\,{e_{{t}}}^{4}+3805\,{e_{{t}}}^{2}-1536 \biggr ] \,{C}^{2} 
\biggr \}\,,\\
{\rm X^{2}_{\rm C6C6} }=-\rm X^{2}_{\rm S6S6} &=& 
{C\,S^4 \over 96 }\,
\,(1+5\,{\eta}^{2}-5\,\eta )\,\sqrt {1 -e_t^2}\,
{ e_t\,\sin u \over (1 -e_t\,\cos u)^6}\,
\biggl \{ 
-72
\,{(e_t\,\cos u)}^{4}
\no
&&
+312\,{(e_t\,\cos u)}^{3}+ (384\,{e_{{t}}}^{2}-844 ){(e_t\,\cos u)}^
{2}
- (1053\,{e_{{t}}}^{2}
\no
&&
 -1325
 )(e_t\,\cos u)-384\,{e_{{t}}}^{4}
+1437
\,{e_{{t}}}^{2}-1105 
\biggr \}\,,\\
{\rm X^{2}_{\rm C6S6} }=\rm X^{2}_{\rm S6C6} &=& 
{C\,S^4 \over 960 }\,
\,(1-5\,\eta + 5\,\eta^2)
\,{1 \over (1 -e_t\,\cos u)^6}\,
 \biggl \{ 120\,{(e_t\,\cos u)}^{6}
\no
&&
-600\,{(e_t\,\cos u)}^{5}-
 (2160\,{e_{{t}}}^{2}-3206 ){(e_t\,\cos u)}^{4}+ (
7860\,{ e_{{t}}}^{2} 
\no
&&
-8444
){(e_t\,\cos u)}^{3}
+ (5760\,{e_{{t}}}^{4}
-19135\,
{e_{{t}}}^{2}
+13051
 ){(e_t\,\cos u)}^{2}
\no
&&
- (
11475\,{e_{{t}}}^{4}
-23240\,{e_{{t}}}^{2}
+11269 )(e_t\,\cos u)
-3840\,{e_{{t}}}^{6} 
+17235\,{e_{{t}}}^ {4}
\no
&&
-21325\,{e_{{t}}}^{2}
+7776
\biggr \}\,,\\
{\rm X^{2} }&=& 
{ C\,S^2 \over 24}
\,{e_t\,\sin u \over (1-e_t\,\cos u)^5}\,
\biggl \{ {1 \over \sqrt {1-e_t^2}}
\biggl [ 
 \biggl ( -(36 -116\,\eta+12\,{\eta }^{2} ){e_{{t}}}^{2}
\no
&&
+24-68\,\eta
-24\,{\eta}^{2} \biggr ){(e_t\,\cos u)}^{2}
+ \biggl ( (174 
-538\,\eta
+10\,{\eta}^{2}
 ){e_{{t}}}^{2}
-150
\no
&&
+442\,\eta
+62 \,{\eta}^{2}
 \biggr )(e_t\,\cos u) 
- (153
-459\, \eta 
-21\,{\eta}^{2}
){e_{{t}}}^{4}+ (168
-496\,\eta 
\no
&&
-40\,{\eta}^{2}
){e_{{t}}}^{2}-27+85\,\eta-17\,{\eta}^{2}
\biggr ]
+ \sqrt {1 -e_t^2}\,(1+5\,{\eta}^{2}-5\,\eta
 ) 
\no
&&
\biggl [
6\,{(e_t\,\cos u)}^{2}
-22\,(e_t\,\cos u)  
+15\,{e_{{t}}}^{2}
+1
\biggr ]\,C^{2}
\biggr \}\,,
\end{eqnarray}
\end{mathletters}
where $C= \cos i $ and $S= \sin i$.
The relations between the coefficients like  $P^i_{CnCn}=P^i_{SnSn}$
or $X^j_{CnSn}=-X^j_{SnCn}$ 
are a trivial  trigonometric  consequence of
the  $ \phi = \lambda + {\rm W}$ split 
in the expression for GW polarizations in terms of $\phi$.

     To compare with the earlier {\it  {\rm  2PN} accurate gauge independent 
expressions for $h_{\times}$ and $h_{+}$ for binaries in 
circular orbits}, we proceed  as follows.
First, we set $ e_t =0 $ in Eqs. (\ref{hx+f}) and
rewrite the resulting
expressions for $ h_{+} $ and $ h_{\times} $ 
in terms of the `gauge independent' orbital angular frequency $\omega $
for circular orbits.
The  2PN accurate relation connecting the mean motion
$\,n$ to $\omega $ may be derived from 
 Eqs. (39), (44) and (46) of \cite {NW95} and it reads:
\begin{eqnarray}
\label{ntoomg}
\xi &=& \tau \biggl \{ 1 - {3\, \tau^{2\over 3} \over ( 1 -e_t^2) }
+ { \tau^{4 \over 3} \over 4\,(1 -e_t^2)}
\biggl [ ( 51 - 26\,\eta ) - {1 \over (1 -e_t^2)}
( 69 - 54\,\eta ) \biggr ]
\biggr \}\,,
\end{eqnarray}
where $\tau = {G\,m\,\omega \over c^3} $.
Next, we use the  following angular transformation relation    
$ \phi \equiv \lambda  + {\rm W}(l)
 = \left ( \phi_{\rm BIWW} -{\pi \over 2} \right ) $, where 
$\phi_{\rm BIWW}$ the orbital phase variable appearing in
\cite{BIWW96}.
The expressions for $h_{\times}$ and $h_{+}$ thus obtained
agree \cite{fn3} with Eqs. (2), (3) and (4)
of \cite{BIWW96} modulo the tail terms.

   All the computations to obtain Eqs. (\ref{hx+f}) 
are performed  using MAPLE\cite{MAPLE}.
This completes the  calculation of the 2PN  accurate GW polarizations  
for compact binaries moving on elliptic orbits, 
modulo the tail terms 
\footnote{ A {\em C} or {\em Fortran} version
of the above $h_{\times}$ and $h_{+}$  expressions  is  available  on request
from {\em gopu@wugrav.wustl.edu} }. 
Though in principle the required equations for the tails 
are available in \cite{BS93}, the explicit expressions
for the tail contribution to 
$h_{\times}$ and $h_{+}$ for eccentric binaries have not  been obtained.
As mentioned earlier, this  should be  computed and included
to  write down  the complete 2PN polarizations.

\section { Influence of the orbital parameters on the
waveform}
\label{sec:spc} 

In this section, we 
investigate the dominant effects of eccentricity, 
orbital inclination and other orbital elements 
on  $h_{\times}(t)$ and  $ h_{+}(t) $. 
For this purpose, 
the one sided power spectral density of the   Newtonian
contributions to the polarization waveforms 
are computed,
by taking the squared-modulus of 
their respective
discrete Fourier transforms,  
sampled over an orbital period.
The results thus describe  the influence of orbital 
elements on the power spectrum of  
Newtonian waveforms when gravitational radiation-reaction
is negligible  and referred to here as a `non-evolving' waveform. 

          To relate  earlier  studies done at  Newtonian order
to the present one, we   proceed in two stages. 
In the  first instance, to compare with the  results of \cite{MBM94,MBM95},
the orbital motion is  restricted to the leading  Newtonian order, 
and the   periastron advance is 
mimicked by the introduction
of an arbitrary constant  shift   parameter $k$ in the $\phi$ variable.
In the second case,
the orbital motion is taken to be   2PN accurate. In this case, the 
periastron advance is fully included in the formalism
and explicitly  defined in terms of the 
binary's
 parameters like
the  masses and eccentricity.
In both the cases mentioned above, 
 only the leading  Newtonian part of the   GW polarizations
is  considered.

     Let us  begin with  the 
`${\times}$' polarization.
 For the ease of  presentation, $h^{\rm N}_{\times}$,
 the Newtonian part of $h_{\times} (t)$, is  written  compactly  below   as:
\begin{mathletters}
\label{sp1}
\begin{eqnarray}
h^{\rm N}_{\times} &=& {G\,m\,\eta \over c^2\,R}\,( {G\,m\,n \over c^3} )^{2/3}\, H^{(0)}_{\times}\,,\\
H^{(0)}_{\times} &=& \biggl \{ {\rm A_{2S}}(l)\,\sin 2\lambda
+ {\rm A_{2C}}(l)\,\cos 2\lambda  \biggr \}\,,
\end{eqnarray}
\end{mathletters}
where $ {\rm A_{2S}}(l) \equiv {\rm A_{2S}}(u(l))
 = {\rm X^{0}_{C2S2}}\, \cos 2\,{\rm W}(l) 
+ {\rm X^{0}_{S2S2}}\,\sin 2\,{\rm W}(l) $,
and $ {\rm A_{2C}}(l) \equiv {\rm A_{2C}}(u(l))
= {\rm X^{0}_{C2C2}}\, \cos 2\,{\rm W}(l) 
+ {\rm X^{0}_{S2C2}}\,\sin 2\,{\rm W}(l)$.
Note that ${\rm A_{2S}}(l) $ and $ {\rm A_{2C}}(l)$ are real and  periodic 
functions  of $l$ with period given by $ {2 \,\pi \over n}$.
The spectral analysis of
$h_{\times} $ will be performed using
$H^{(0)}_{\times}$,
{\it the scaled  $\times$ polarization waveform},
because 
since we are dealing with non-evolving binaries,
${G\,m\,\eta \over c^2\,R}\,( {G\,m\,n \over c^3} )^{2/3}$ 
essentially remains a constant over a few orbital periods. 
Similar arguments hold for $h_+$ too. 

\subsection{Newtonian orbital motion}

In this section, we restrict the dynamics of the binary to 
Newtonian order.
This implies we are using Eqs. 
(\ref{lwN}) for $\lambda $ and ${\rm W}(l)$ 
in the 
$\phi= \lambda + {\rm W}(l)$ split.
However, following \cite{MBM94,MBM95}, we  introduce an arbitrary 
periastron advance parameter $k$ into the definition of
$\lambda$ so that  $ \lambda = \phi_0 + (1 +k)\,l$ and
${\rm W}(l) = ( v -u + e\,\sin u)\,(1 +k) $.
Note that with these forms for $ \lambda $ and ${\rm W}(l)$,
 the scaled GW polarization waveforms are entirely 
specified by $e$, $k$ and $\phi_0$. 

    As mentioned earlier,
 ${\rm A_{2S}}(l)$ and ${ \rm A_{2C}}(l)$ are periodic 
functions of  $l = n\,(t -t_0)$, where $ n = 2\,\pi\,f_r$, 
$f_r$ being the
frequency associated with the `radial  period' \cite{DS88}  
{\it i.e.} the time of return to the periastron.  
Consequently, they can be expanded in 
{\it Fourier series} as follows 
\bml
\label{Nsp2}
\begin{eqnarray}
{\rm A_{2C}}(l) &=& 
\sum_{j=-\infty}^{\infty} C_j \,e^{i\,j\,l}\,, \\
{\rm A_{2S}}(l) &=& 
\sum_{j=-\infty}^{\infty} S_j \,e^{i\,j\,l}\,.
\end{eqnarray}
\eml  

Employing,  Eqs. (\ref{Nsp2}) and 
 $ \lambda= \phi_0 + (1 +k) \,l $, in (\ref{sp1}), we get
\begin{equation}
\label{sp5}
H^{(0)}_{\times} = 
\sum_{j=-\infty}^\infty \biggl ( \bar S_j\,e^{i\,\omega^{+}_j\,l}
+ \bar C_j \,e^{i\,\omega^{-}_j\,l} \biggr )\,,
\end{equation}
where
\bml
\label{sp6}
\begin{eqnarray}
\bar S_j &\equiv& { e^{i\,2\,\phi_0} \over 2} \left( C_j -i\,S_j \right )\,,\\
\bar C_j &\equiv& { e^{-i\,2\,\phi_0} \over 2} \left( C_j +i\,S_j \right )\,,\\
\omega^{+}_j &\equiv& (j + 2\,p )\,,\\
\omega^{-}_j &\equiv& (j - 2\,p )\,,\\
p&\equiv& ( 1 +k)\,.
\end{eqnarray}
\eml
Eq. (\ref{sp5})  may be re-written as 
\begin{eqnarray}
\label{sp7}
H^{(0)}_{\times} &=&
 \sum_{j=0}^{\infty}  \biggl [ \bar S_j\,e^{i\,\omega^{+}_j\,l}
+ \bar C_{-j} \,e^{-i\,\omega^{+}_j\,l}  
+ \bar S_{-j} \,e^{-i\,\omega^{-}_j\,l}  
+ \bar C_{j} \,e^{i\,\omega^{-}_j\,l} \biggr ]
\no
&&
-\biggl [ \bar S_{0} \,e^{i\,2\,p\,l}
+  \bar C_{0} \,e^{-i\,2\,p\,l} \biggr ] 
 \,.
\end{eqnarray}
Recalling $l=n(t-t_0)$, with  $n={ 2 \pi \,f_r}$,  
the frequency  content and the associated
intensities may be read off from the above. 
     From Eqs.(\ref{sp7}), it  follows  that the Fourier spectrum
of  $ H^{(0)}_{\times}(l) $, 
consists of lines at frequencies  
 $\omega^{+}_j\,f_r$ and $\omega^{-}_j\,f_r$  with 
 powers $( | \bar S_j|^2 + |\bar C_{-j}|^2 )$  and   
$ (| \bar S_{-j}|^2 + |\bar C_{j}|^2) $ respectively. 
Using the reality of $\phi_0$, ${\rm A_{2S}}$ and ${\rm A_{2C}}$, though 
$S_j$ and $C_j$ are complex numbers, 
it is easy to show that
 $ | \bar S_{j}|^2 = |\bar C_{-j}|^2$ implying that
power in the line with frequency $\omega^{+}_j\,f_r$ will be
$ 2\,| \bar S_j|^2 $.            
Similarly,  $ | \bar C_{j}|^2 = |\bar S_{-j}|^2$,
and  power in the  $\omega^{-}_j\,f_r$ line   is $ 2\,| \bar C_j|^2 $.

       Thus, the Fourier series for Newtonian 
part of $h_{\times}$ effectively reduces to 
\begin{equation}
\label{sp8}
H^{(0)}_{\times} = 
{  \sqrt 2\,} 
\,\biggl \{
 \sum_{j=1}^{\infty}  \biggl [ \bar S_j\,e^{i\,\omega^{+}_j\,l}
+ \bar C_j \,e^{i\,\omega^{-}_j\,l} \biggr ] 
+ {\bar S_0}\,e^{i\,2\,p\,l}
\biggr \}\,.
\end{equation}
        The  `one sided  power spectrum' 
for the Newtonian $h_{\times}$  may be written as 
\begin{equation}
\label{sp9}
H^{(0)}_{\times} = 
{ \sqrt 2\,}\,
\biggl \{
 \bar S_0 \,e^{i\,2\,p\,l} +
 \sum_{j=1}^{\infty}  \biggl [ \bar S_j\,e^{i\,(j +2\,p)\,l}
+ \bar C_j \,e^{i\,| (j -2\,p)|\,l} \biggr ] 
\biggr \}\,,
\end{equation} 
where explicitly  the  sum is over   positive  frequencies. 
In the generic case,
the $\omega^{+}_j=(j + 2\,p )$ part
gives lines at frequencies
 $(1+2\,p)\,f_r, (2+2\,p)\,f_r, (3+2\,p)\,f_r,......$ 
with strengths $\sim$  
$|\bar S_1|^2,\,|\bar S_2|^2,\,|\bar S_3|^2, ....$ respectively.
Similarly, the  $\omega^{-}_j=(j - 2\,p ) $ part of Eq. (\ref{sp8})
creates lines at frequencies  
$  (1-2\,p)\, f_r, (2-2\,p)\,f_r, (3-2\,p)\,f_r,....{\rm etc}$
 with strengths proportional to
$|\bar C_1|^2,\,|\bar C_2|^2,\,|\bar C_3|^2, ....{\rm etc}$ 
respectively. 
There will be also a line at frequency 
$2\,p\,f_r$ with strength $\sim  |\bar S_0|^2 = |\bar C_0|^2$ \cite{fn1}.

These observations are easy to understand.
At  Newtonian order, in the absence of the  periastron
precession, {\em i.e.} $k=0$,
there is only one time scale
in the problem,
given by the orbital period
and the spectrum consists of lines at multiples of
the orbital frequency.
When periastron precession is introduced, $k\neq 0$, a second slower
time-scale enters the problem,
which splits and shifts original
spectral lines from their earlier positions, thereby 
lifting the degeneracy associated with the
non-precessing orbit.

        A caveat is worth noting: 
The discussion after Eq. (\ref{sp9}) is valid 
only if all  the terms corresponding to frequencies 
 $(j_s+2p)\,f_r$ and $(|j_c-2p|)\,f_r$ are linearly independent,
where $ j_c $ and $j_s$ are summation index $j$ for 
$\bar C_j$ and $\bar S_j$.
This is in general true except when $j_s+2p=|j_c-2p|$, which corresponds to
values of $k$ equal to $0$, $0.25$ and $0.5$ \cite{fn2}.
For these values of $k$, power in  a given spectral line will  
have contributions both  from $\bar C_j$ and $\bar S_j$ for different
values of $j$ given by $ j_c-j_s-4=4\,k$. 
These special values of $k$ are interesting
in that they can provide useful 
checks on the numerical accuracy of the analytical procedure
outlined above. 
This is because for values of $k =0, 0.5, 0.25$, 
the full time domain waveform [ and not just the parts 
${\rm A_{2S}}(l), {\rm A_{2C}}(l)$ ]
$H^{0}_{\times}(l)$ is {\em exactly} periodic over 
$2\pi, 2\pi$ and $4\pi$ intervals respectively.
Consequently, one may alternatively  compute the desired
power spectrum by a direct  Fourier transform
of the full  $ H^{(0)}_{\times}(l) $, 
without going via  
Eq. (\ref{sp9}), which exploits double 
periodicity of $\phi$ in $l$ and $\lambda$.
Similar arguments hold true for the $+$ polarization.

       It is clear from Eq. (\ref{sp9}) that
the strengths of the 
different Fourier components are determined by the coefficients
$\bar{C}_j$ and $\bar{S}_j $ which are 
given in terms of $C_j$
and $S_j$,  
the discrete Fourier transforms of  ${\rm A_{2C}}(l)$ and
 ${\rm A_{2S}}(l)$ 
 \footnote{ Using the  Fourier integral theorem, it is easy to
   show that ${\rm A_{2C}}(l)$ may be written in terms of the Fourier
transform of ${\rm
 A_{2C}}(l)$, the discretized version of which allows us to express ${\rm
 A_{2C}}(l)$ in terms its discrete Fourier transform. Similar arguments
 apply to
 ${\rm A_{2S}}(l)$.}. 
This has become possible   since we have  exploited 
the double-periodicity of the motion in angles $l$ and $\lambda$.
Thus the calculation reduces to the numerical 
implementation 
of $S_j$ and $C_j$ which we turn to next.

       Though the power spectrum for the Newtonian part of
$h_{\times}$ can be obtained using Eq. (\ref{sp9}), its
implementation is not straightforward
due to following reasons. 
First, the Discrete Fourier Transforms $C_j$ and $S_j$
 can be evaluated using
 standard Fast Fourier Transform routines  as in 
{\em Numerical Recipes}
\cite{NR}  only  after 
$\rm A_{2S} $ and $\rm A_{2C}$
are written as  explicit functions of $l$. 
However,   in our analysis they are  explicit  functions   of 
$u$  and thus implicit functions of $l$  via 
$ l= u -e\, \sin u $.
Consequently, we must first compute $u(l)$ and substitute
it in Eq. (\ref{sp1}) to proceed.  
Secondly, ${\rm  W}(u) = ( v -u + e\,\sin u)$ 
will not be a smooth function of $u$ if we numerically implement
$ v =2\, \tan^{-1}\left \{ \left (
{1 +e \over {1 -e}}
\right)^{ 1\over 2} \, \tan ( { u \over 2}) \right \} $.
 We will also need to use a smooth functional
relation connecting $v$ and $u$ to obtain a well behaved
$H^{0}_{\times} (u(l) $ and $H^{0}_{+} (u(l) $. 

      Let us first consider the implementation  of $u(l)$.   
There are two independent ways to obtain $u(l)$ 
from $ l= u -e\, \sin u $. 
The first method is widely used, for analytical treatments, 
in standard textbooks of celestial mechanics \cite{TDT}.
The idea here is to  
expand the eccentric anomaly $u$ in terms of
the mean anomaly $l$.  At the Newtonian order, it  is given by     
\begin{equation}
\label{utoM}
u = l + \sum_{s=1}^{{\infty}} \biggl ( {2 \over s} \biggr )\,
J_s(s\,e)\, \sin sl\,,
\end{equation}
where $ J_s(s\,e)$ is the Bessel function of
the first kind of order $s$ with  $s
\geq 1$.

     Alternatively, we can numerically invert
Eq.(\ref{admrn}) connecting the  mean and  eccentric anomalies,
using the  Newton-Raphson method implemented by {\em rtsafe}
routine of {\em Numerical Recipes}\cite{NR},
and  obtain $u(l)$.
We compute $u(l)$ using both methods
to make sure that they give consistent results,
for the  parameter values  we are dealing here.   

  We now turn to the numerical implementation of ${\rm W}(u)$.
In text books of celestial mechanics, the   
transcendental relation connecting true and eccentric anomalies  
is expressed as a series given by
\begin{equation}
v - u = 2\,\sum_{j=1}^{\infty} \biggl ( 
{{\beta}^{j} \over j}\, \sin j\,u \biggr )
\,,\\
\label{eq:vu}
\end{equation}
where $ \beta = {1 \over e}\, \left ( 1 -\sqrt{1-e^2} \right )$.
We use above  expansion  of $ v -u$ in ${\rm W}(u) = v -u +e\,\sin u$,
to circumvent artificial discontinuities in  ${\rm W}(u)$  
as a  function of the  eccentric anomaly $u$.

      Using the  above inputs, we compute 
$u(l)$ and W(u(l)) at a  finite number of  points  by sampling $l$. 
Next, we use 
the {\em realft} routine of \cite{NR} 
to compute 
the discrete Fourier transforms  $S_j $ and $C_j$ of the 
discretely sampled periodic
functions ${\rm A_{2S}}(l)$ and ${\rm A_{2C}}(l)$.
We then compute the  `one sided power spectrum'
for the Newtonian $h_{\times}$ using Eq. (\ref{sp9})
for various values of
$e$, $ k $ and $i$.  We now have all the inputs to
 investigate  the 
influence of orbital elements on the Newtonian part of
the ${\times}$ polarization waveform.  
The results and discussions are postponed to the end of this
section. 

       The spectral analysis for $h_+$  is similar to that
for $h_{\times}$ and we only quote the main results without
any further  details.
\begin{equation}
\label{spP1}
H^{(0)}_{+} =
P_0(l) +{\rm P_1}(l)\,\sin 2\lambda
+ {\rm P_2}(l)\,\cos 2\lambda \,, 
\end{equation}
where $ P_1 = {\rm P^{0}_{C2S2}}\, \cos 2\,{\rm W} 
+ {\rm P_{S2S2}}\,\sin 2\,{\rm W}$,
$ P_2 = {\rm P_{C2C2}}\, \cos 2\,{\rm W} 
+ {\rm P_{S2C2}}\,\sin 2\,{\rm W}$
and $ P_0 = {\rm P^{0}}= ( 1 -C^2)\, e\,\cos u\, (1 -e\,\cos u) $.
   The Fourier series for Eq. (\ref{spP1}) is given by
\begin{equation}
\label{spP2}
H^{(0)}_{+} = 
\sum_{j=-\infty}^{\infty} 
\biggl [ \bar S^{+}_j\,e^{i\,\omega^{+}_j\,l}
+ \bar C^{+}_j \,e^{i\,\omega^{-}_j\,l} 
+  \bar P^{0}_j\, e^{i\,j\,l}  \biggr  ] 
\,,
\end{equation} 
where $\bar S^{+}_j $ and $\bar C^{+}_j$ are defined similar to
$\bar S_j $ and $\bar C_j$ but with $A_{\rm 2S} $ and $A_{\rm 2C}$ replaced by
$P_1 $ and $P_2$. Similarly,
$\bar P^{0}_j $ is the discrete Fourier transform of $P_{0}(l)$.
Using arguments similar to the ones used for the  $H^{0}_{\times}$ analysis,
we relate  $ \bar S^{+}_j $,  $ \bar C^{+}_{-j} $ and 
 $ \bar S^{+}_{-j} $, $ \bar C^{+}_j $ and obtain the 
`one sided power spectrum'
for Newtonian $h_+$ as:   
\begin{equation}
\label{spP4}
H^{(0)}_{+} = 
\sqrt 2\,
\biggl \{
 \sum_{j=1}^{\infty}  \biggl [  \bar S^{+}_j\,e^{i\,(j +2\,p)\,l}
+ \bar C^{+}_j \,e^{i\,| (j -2\,p)|\,l}
+ \bar P^{0}_j\, e^{i\,j\,l} \biggr ] 
+ \bar C^{+}_0\, e^{i\,2\,p\,l}\,
\biggr \}
+ \bar P^{0}_0 \,.
\end{equation}  
         From Eq. (\ref{spP4}), it follows 
that for the `$+$' polarization  
 there will be lines at frequencies
 $0,\, 2p\,f_r,\, | 1 -2 p|\,f_r,\, f_r, (1+2p)\,f_r,\,$
$ |2 -2p|\,f_r,\, 2\,f_r,\, (2+2p)\,f_r,\,...$
with relative strengths 
$\sim\, \, {1 \over 2}\,|\bar P^{0}_0|^2,\, 
|\bar S^{+}_0|^2,\,|\bar C^{+}_1|^2,\,|\bar P^{0}_1|^2,\,
$ $ |\bar S^{+}_1|^2,\, 
|\bar C^{+}_2|^2,\,|\bar P^{0}_2|^2,\, |\bar S^{+}_2|^2, .....$
respectively. 
Note that there are lines unaffected by introduction of $k$.
These arise from the non-$\lambda$ term
in $H^0_{+}(l)$.
The values of  $k =0, 0.25, 0.5$ are special and require a treatment analogous
to the corresponding one in the cross  polarization case.

   Using the  above inputs, we plot the time-domain waveforms
$H^{0}_{\times, +}(l)$ 
and the associated  normalized relative power spectrum
$ \frac {( H^{(0)}_{{\times},{+}})^{j} }{\sum_j\, (H^{(0)}_{{\times},+})^{j}} $
in Figs. (\ref{fig:c_N_k_0.1})-(\ref{fig:c_check}).
The combined influence of the orbital parameters
like  eccentricity $e$,  
 periastron advance parameter $k$ and orbital 
inclination $i$, on the time-domain waveform
 and the associated  power spectrum of the Newtonian
$h_{\times}$ and $h_{+}$ 
using  Newtonian accurate orbital motion is
summarized below.

\begin{itemize}
\item {\bf Eccentricity, $e$ }.

The effects of $e$ on $H^{0}_{\times}$ are 
explored in Fig. (\ref{fig:c_N_k_0.1}).       
In the limit of low values of eccentricity  $e$, as expected,
the dominant  contribution to the  power  spectrum,
comes from the second harmonic.
However, as the  eccentricity $e$ increases,
higher harmonics appear in the spectrum with comparable strengths.
For a given value of the periastron advance  $k$ and inclination
angle  $i$, the position of the dominant
harmonic changes as the value of 
 $e$ increases.
The shape of the waveform also changes significantly as we increase $e$.
For moderate and high values of  $e$  there is a stronger
burst of radiation  near values  $l =0 $ and $2\,\pi$
corresponding to the  periastron passage, since  
near the   periastron, the two  masses are
closest to each other and their relative  velocity is a maximum.
In the frequency domain this results in the broad peak
containing many frequencies. 
The line feature in the frequency domain 
on the other hand corresponds to the average orbital  motion of the binary. 

\item {\bf The `arbitrary' periastron advance parameter, $k$.\,}
The observation made here are based on  Fig. (\ref{fig:c_N_e_0.5}).  
A careful inspection of $H^{0}_{\times,+}(l)$
with $ k \neq 0$ indicates that 
in general
they are {\rm not} $2\,\pi$ periodic. This is  expected
as $k$ is a measure of the angle of return to the periastron.
As mentioned earlier, in the power spectrum, the main effect of
including an arbitrary $k$
is a `splitting' and subsequent `shifting' of
the position of each spectral line from its
integer multiple value in units of $f_r$, the radial frequency.
The shift  is appreciable for medium and high eccentricities  and leads to
a shift of the dominant  harmonic in the spectrum.           

\item {\bf Orbital inclination, $i$.\, }
 A change in the orbital inclination  changes
only the magnitude of $H^{0}_{\times}$ and its
power spectrum, keeping the
relative distribution of 
spectral lines
 the same.
This is easy to see as the  dependence of
orbital inclination angle $i$ is easily factored out
in the expression for $h_{\times}$.
However, the shape of $H^{0}_+$ and its power spectrum is
influenced by $i$
as seen in Fig. (\ref{fig:p_N_0.5_0.1}).

     If  the polarizations    of the gravitational wave
$h_{\times}$ and $h_{+}$ are available, 
the orbital inclination can be inferred 
by computing the ratio of the total power measured in each 
polarization.
For circular orbits, the result analytically follows since 
$ { | h^{N}_{\times} |^2  \over | h^{N}_+ |^2 } $ is only a function of
$i$ and is given by
$ {4 \, \cos^2 i \over (1 + \cos^2 i)^2}$.
In the general eccentric case,   
to explore  this, 
we plot $ \frac { | h^{N}_{\times} |^2 }{| h^{N}_+ |^2 } $ as a function
of $i$ for various values of the $e$ and $k$ in 
Fig. (\ref{fig:i_N}). 
The  plots are identical for different  values  of  $e$ and $k$ but
vary  with the inclination angle  providing support for the claim
made in the beginning of this paragraph.
\end{itemize}

    Even though in general,  an arbitrary periastron 
advance parameter $k$ at Newtonian order
destroys the  $2\pi $-periodicity of $H^{0}_{\times,+}(l)$, 
we  may  choose $k$ values {\em exactly} equal to $0$ or $0.5$
so that $H^{0}_{\times}(l)$ is still   {\em  $2\,\pi$} periodic.
 These particular  values  of $k$ allow us to perform
useful  numerical  checks on our analytical procedure. 
In this case, we can compute the  power spectrum directly from
$H^{0}_{\times}(l)$ by numerically implementing the  
 Discrete Fourier Transform of Eqs. (\ref{sp1}).
We can also implement  Eq. (\ref{sp9})
to obtain the  power spectrum, after
adding contributions from various
$\bar C_j$ and $\bar S_j$
to {\em a given } harmonic, which is now some integer    
multiple of the radial frequency, $f_r$.
The results are displayed in 
Fig. (\ref{fig:c_check}).  
For  better  comparison, in these  figures,
we normalize  relative to  the power in  the dominant harmonic
rather than relative to the total power as in other   figures.
We next, choose $k =0.25$,  so that now 
$H^{0}_{\times}(l)$, is  now   $4\,\pi$ periodic.
Again, comparison  with the power spectrum computed directly from 
Eq. (\ref{sp1}) and via Eq. (\ref{sp9})
is possible. 
The results for $k=0, 0.25$ and $0.5$ via these two methods 
are compared and found to be
identical up to numerical errors as seen in Fig. (\ref{fig:c_check}),
providing important checks on our  analysis and routines that compute
the  one sided power spectrum via Eq.(\ref{sp9}).
We observe a  similar behavior  for $H^{0}_{+}$.

\subsection{  The  2PN accurate orbital description}

The spectral analysis discussed in the previous section may be
extended to 2PN accurate orbital motion with minor technical
modifications.
The expressions for $ {\rm W}(l) $ and $\lambda$
in the  $\phi = \lambda +  {\rm W}(l)$ split 
are now given by Eqs. (\ref{ioset1}).  Moreover,  
the orbital elements appearing in $H^{(0)}_{\times}$ and
$H^{(0)}_{+}$ are now 2PN accurate.
These changes will modify expressions for $S_j$, $C_j$ in 
Eq. (\ref{sp9}) for the `${\times}$' waveform  and
the corresponding expressions for the `$+$' waveform.

        To  implement the  2PN accurate spectral analysis,
we note the following:
First,  at the  2PN level  the  simpler   approach to obtain the 
$u(l)$ relation, connecting the  mean and  
eccentric anomalies, is to numerically solve for
$u(l)$ from Eq. (\ref{admr}) because 2PN accurate 
analytic expression for $u(l)$ similar to Eq. (\ref{utoM}) 
is not available in the literature.
However, we may employ Eq.(\ref{eq:vu}) with $e_{\phi}$ in the place
$e$ to get $v -u$ at 2PN order.
 This is because in the  generalized quasi-Keplerian representation,
the relation connecting true anomaly $v$ to eccentric anomaly $u$
has the same structural form as for the Keplerian case.
 Secondly,  there are post-Newtonian corrections to the relations
connecting  $e_r $ and $e_{\phi}$ to $e_t, m_1, m_2 $ and $n$. 
In our analysis, only those values of $e_t$ are
 considered which lead to  $e_{\phi}, e_r $  less  than one. 
Finally, in this  2PN accurate orbital description,
the periastron precession constant $k$ is no longer arbitrary  but 
uniquely  determined  by $m_1,m_2, n $ and $e_t$   as  given by  in 
 Eq. (\ref{inp2}).

           We explore the 
effects of 2PN accurate orbital motion on 
the power spectrum for `${\times}$'
polarization in 
Figs. (\ref{fig:c_2_1.4_1.4_100}), 
and (\ref{fig:c_2_e_0.4}).
In Fig. (\ref{fig:c_2_1.4_1.4_100}), we
explore the  influence of $e_t$
on the relative power spectrum and the behavior  is  qualitatively 
similar to the Newtonian case. 
We explore, in Fig. 
(\ref{fig:c_2_e_0.4}),  
the effect of changing values for $k$ by 
varying $m_1, m_2$ and $ n$ after fixing the  value  of  $e_t$.  
We see that  the  behavior is similar 
to the Newtonian case when we vary values of $k$  for a given $e$.
This is required as at 2PN order, for a given $e_t$,
$k$ is uniquely determined by
$m_1,m_2$ and $n$.
 However,  there are quantitative differences 
in that positions and strengths of various
harmonics are different in the  Newtonian and the  2PN cases.     

         A quantitative comparison between    
the spectral analysis with Newtonian and 
2PN motion is presented in 
Figs. (\ref{fig:p_2_N_comp}) and  (\ref{fig:I_p_2_N_comp}).
Note that we can perform this comparison as we are using
scaled polarization waveforms, $H^{0}_{\times} $ and
$H^{0}_{+}$.
In Fig. (\ref{fig:p_2_N_comp}), we plot both $H^{0}_{+}(l) $ and its power
spectrum using Newtonian and 2PN accurate orbital motion. 
We choose the  arbitrary parameter   $k$, introduced in the  Newtonian    case 
to match the  2PN accurate $k$ associated with the  generalized quasi-Keplerian
representation. In this manner, we force orbital elements for
Newtonian and 2PN dynamics to be the same.
Though, qualitatively similar, the  plots for  the 
Newtonian and the  2PN orbital motion are 
quantitatively different in that strengths of spectral lines
are different by a few parts in thousand
in most cases.
 
Finally, in Fig. (\ref{fig:I_p_2_N_comp}), we plot
 $H^{0}_{+}(l) $ and its power spectrum
as a function of orbital inclination angle 
for Newtonian and 2PN accurate orbital motion.
The value of $k$ for the Newtonian runs are again chosen so that
it is comparable to the  actual 2PN accurate $k$ value.
It is clear from these figures that inclusion of PN corrections to
orbital motion changes distribution of spectral lines, though 
the position of the dominant (maximum amplitude) harmonic 
is roughly the same.
This figure also shows how the  orbital inclination angle $i$ slowly
modulates the spectral lines for Newtonian and 2PN orbital motion.

\section {Conclusions}
\label {sec:concl}

\subsection{ Summary of results}

In this paper we have computed all the 
`instantaneous' 2PN  
contributions to 
$h_{+} $ and $ h_{\times}$ for two compact objects 
of arbitrary mass ratio
moving in elliptical orbits, using 
2PN corrections to $h^{TT}_{ij}$ 
and the generalized quasi-Keplerian representation
for the 2PN motion.
The expressions for $ h_{+} $ and $ h_{\times} $
obtained here represent gravitational radiation
from an elliptical binary during that stage of inspiral
when orbital parameters are essentially 
the same over a few orbital periods, in other words when the
gravitational radiation reaction is negligible.
We investigate the effect of eccentricity,
advance of  periastron
and orbital inclination on the power 
spectrum of the 
Newtonian part of  $ h_{+} $ and $ h_{\times} $.
 The 2PN accurate generalized    quasi-Keplerian representation
is used in conjunction with  two angular variables
$l$ and $\lambda$  chosen to facilitate the subsequent  analysis
of the waveform evolving under gravitational radiation reaction.
 These expressions thus form the  first step
in the direction of obtaining `ready-to-use' theoretical templates 
for inspiralling  compact bodies moving 
 in {\em quasi-elliptical} orbits. 

\subsection{Future directions}

 There are several issues that remain open for further investigation.
We list them below.

\begin{enumerate}

\item

The next natural step is to obtain evolving $h_{\times}$ and $h_+$, when
lowest order radiation reaction effects are included in the evolution of
orbital elements. This is currently under investigation \cite{DGI}.

\item 

 There are tail contributions to $h_+ $ and $h_{\times}$ appearing
at 1.5PN and 2PN orders. Though the formal expressions for tail terms
are available in \cite{BS93,BDI95}, they need to be written down 
in a form similar to `instantaneous' contributions to 
$h_+ $ and $h_{\times}$ presented in this paper.

\item 
After computing `ready-to-use' search templates for 
inspiralling binaries in `quasi-elliptical' orbits, one will be able to
address a variety of data analysis issues
related to the observations of gravitational radiation from
eccentric binaries in great detail.      
These could  include defining a `restricted post-Newtonian' waveform,
extending to 2PN accuracy the   effect of eccentricity on detection
discussed  in \cite{MBM94,KP99} where currently  the  orbital dynamics
is  restricted to leading   Newtonian   order only.

\item
 
Finally, the  analysis of the present paper may be useful for 
detecting  continuous gravitational waves from known  sources in binaries. 
Recent analysis \cite{DV00} employing 
a Keplerian representation for the  binary's orbital motion
indicates that the  computational cost
required to search  such sources is  affordable.
The  present analysis and \cite{DGI}   may  be crucial  for 
including the  relevant relativistic effects.

\end{enumerate}

{\em  Note added in  proof:\,\,}
We observe that the results of our spectral analysis are in 
agreement with \cite{MBM95} for low values of $k$.
Since the effect of  the
periastron precession  on the amplitude of spectral lines 
is not fully taken into consideration 
in \cite{MBM95}, the  strength of the  spectral lines in our analysis
differs from theirs, for high values of $k$.
It should be noted that for a given $k$, both methods give 
the same frequency shift for spectral lines, for all values of $k$.

\section*{Acknowledgments}
We  thank   T.  Damour for   discussions and 
insights that led us towards the final form for the 
gravitational wave  polarizations  presented here 
and D. Bhattacharya  
for critical comments  clarifying the  implementation 
and results of the  spectral analysis of waveforms.
We are grateful to L. Blanchet, S. Iyer, B. S. Sathyaprakash, 
G. Sch\"afer and B. Schutz 
for their  comments at different  stages  of this project.
One of us (AG) is supported in part by NSF grant No. PHY 96-00049 
and is grateful to C. M. Will for 
encouragement.

\newpage
\begin{figure}
\epsfxsize=6in \epsfbox{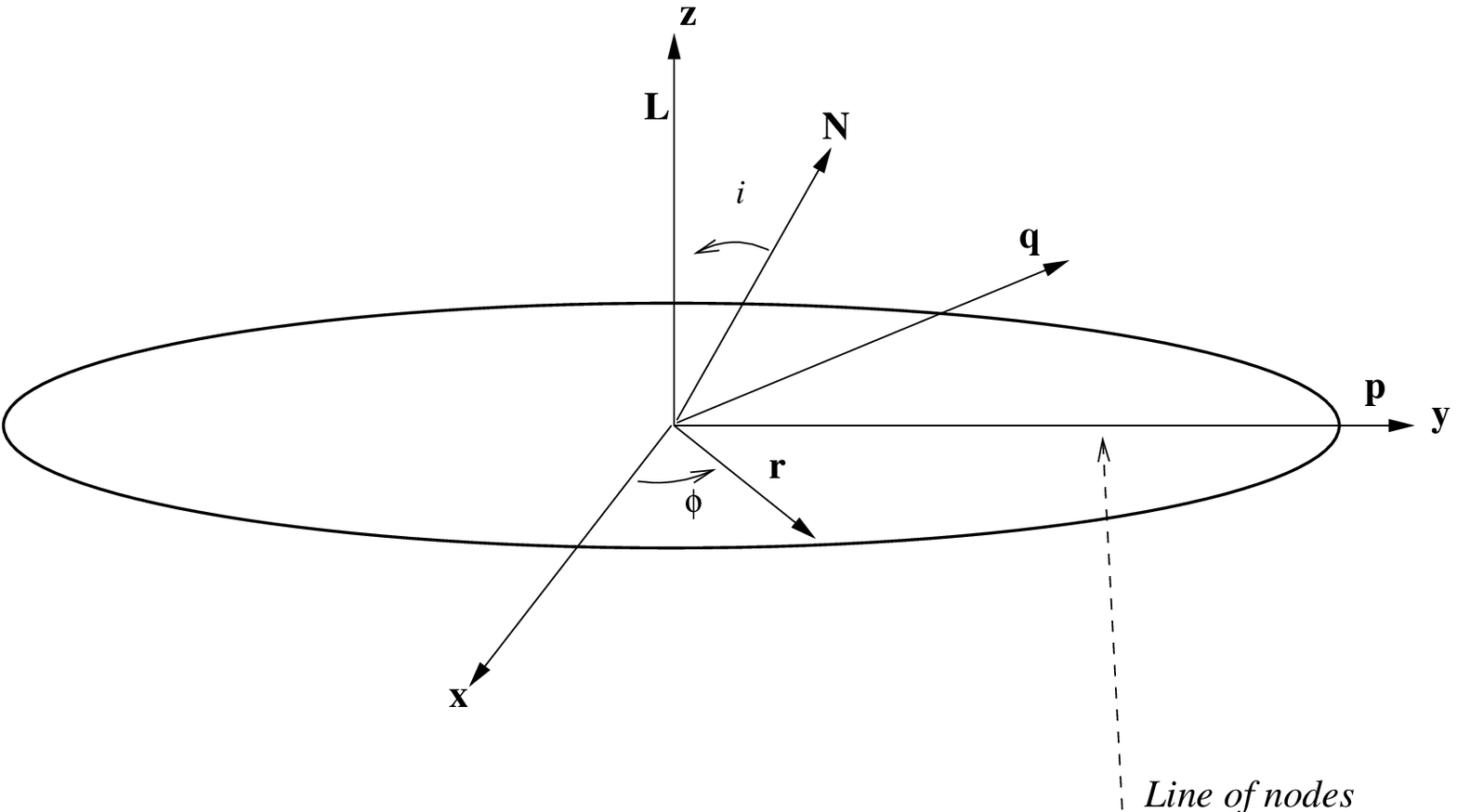}
\bigskip
\caption{ The orientation of unit vectors, which defines
$\times$ and $+$ polarization waveforms.
The unit vectors {\bf p} and {\bf q} are  
the gravitational wave's {\em principal axes} 
with ${\bf q} = $ {\bf N} $\times$ {\bf p}. 
Note that {\bf N} is 
a unit vector lying along the radial direction to the detector
and {\bf p} lies
along the line of nodes.
The  Newtonian angular momentum vector ${\bf L }
= \mu\, {\bf r} \,\times\,{\bf v} $ is normal to the orbital plane
and helps to define orbital inclination angle $i$.
In this paper, the origin for 
$\phi = \lambda +{\rm  W}$ is $+$ve x-axis, hence it is related to 
$\phi_{\rm BIWW}$ by $\phi = \phi_{\rm BIWW} -{\pi \over 2}$.
} 
\label{fig:config}
\end{figure}
\newpage       
\newpage
\begin{figure}
\epsfxsize=6.5in \epsfbox{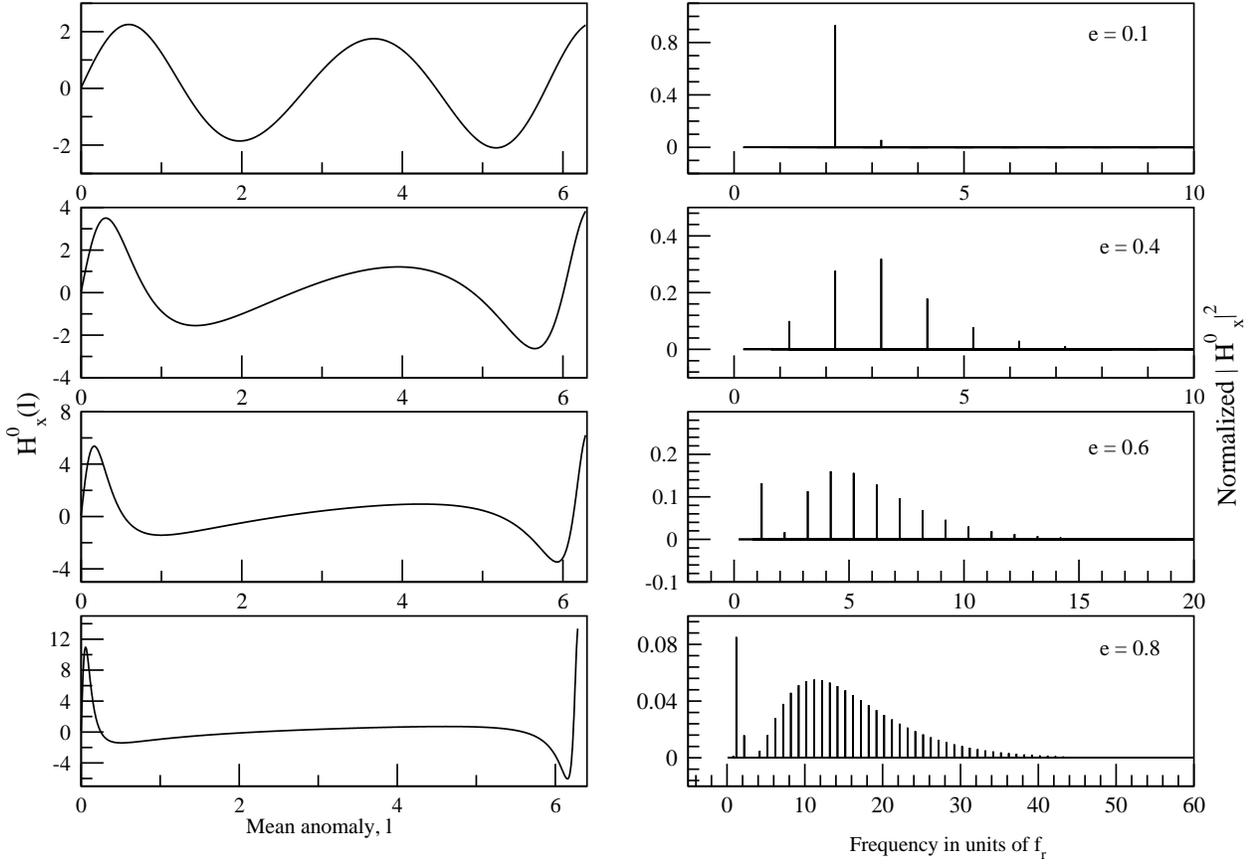}
\bigskip
\caption{
Plots for scaled GW polarization waveform, $H^{0}_{\times}$
as a function of the  mean anomaly, $l$  and
the corresponding 
normalized relative power spectrum using 
Newtonian orbital motion, for various values of  eccentricity $e$.
Note 
in $H^{0}_{\times}(l)$  
a `burst' of GW  emission  near  periastron passage
and a shift in the position of the dominant harmonic in the power
spectrum as $e$ increases.  
In the  Fourier domain, the former results in a broad frequency rich peak.
In all panels, the (arbitrary) periastron precession constant  
and the orbital inclination angle take values 
$0.1$ and $ {\pi \over 3} $ respectively.
} 
\label{fig:c_N_k_0.1}
\end{figure}
\newpage       

\newpage
\begin{figure}
\epsfxsize=6.5in \epsfbox{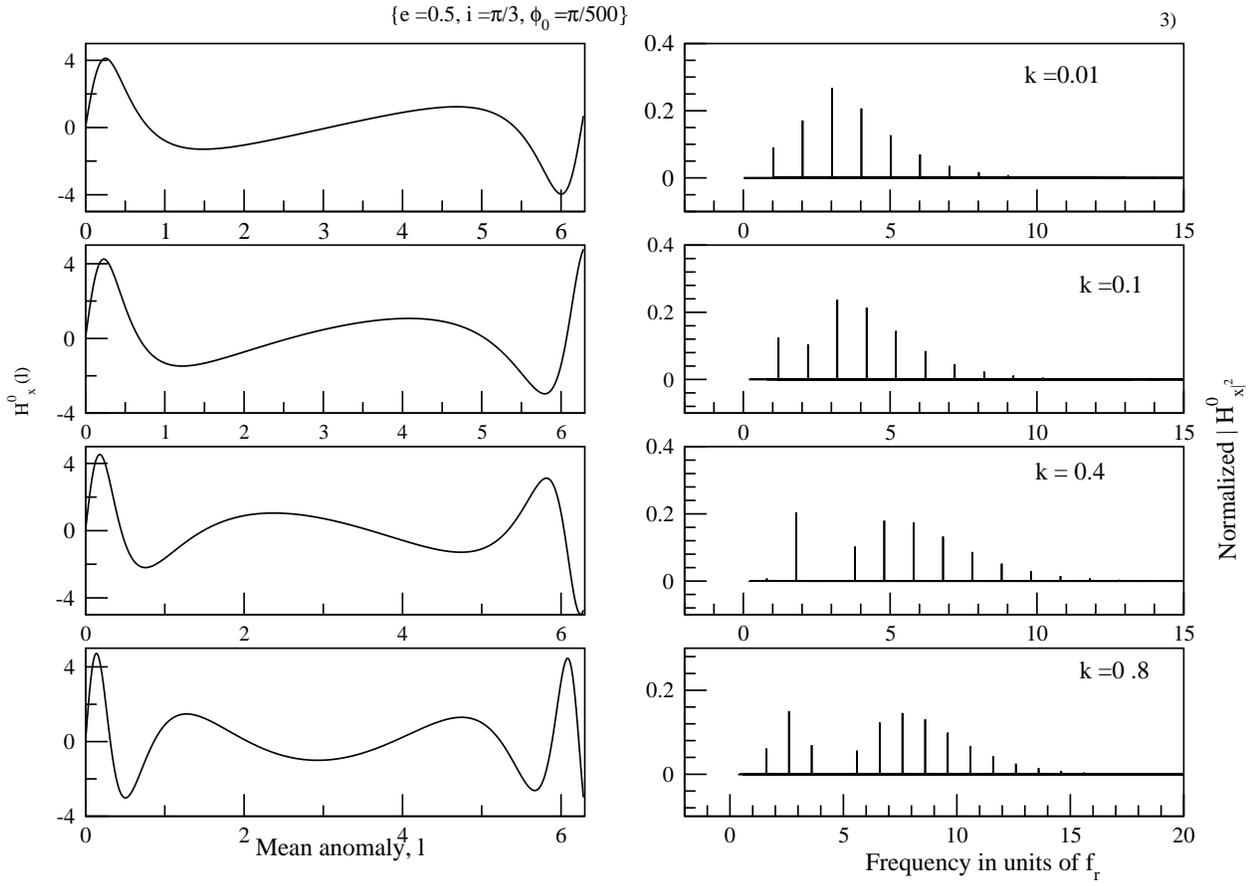}
\bigskip
\caption{
The configuration is similar to Fig.\ref{fig:c_N_k_0.1}, but 
in the  panels,  $k$, the (arbitrary) periastron precession constant
is varied for fixed  eccentricity $e=0.5$ and orbital 
inclination angle  $i = {\pi \over 3} $.
Note the splitting and shifting of  
spectral lines from 
integer multiple values of $f_r$ as $k$ is increased.}
\label{fig:c_N_e_0.5}
\end{figure}

\newpage
\begin{figure}
\epsfxsize=6.5in \epsfbox{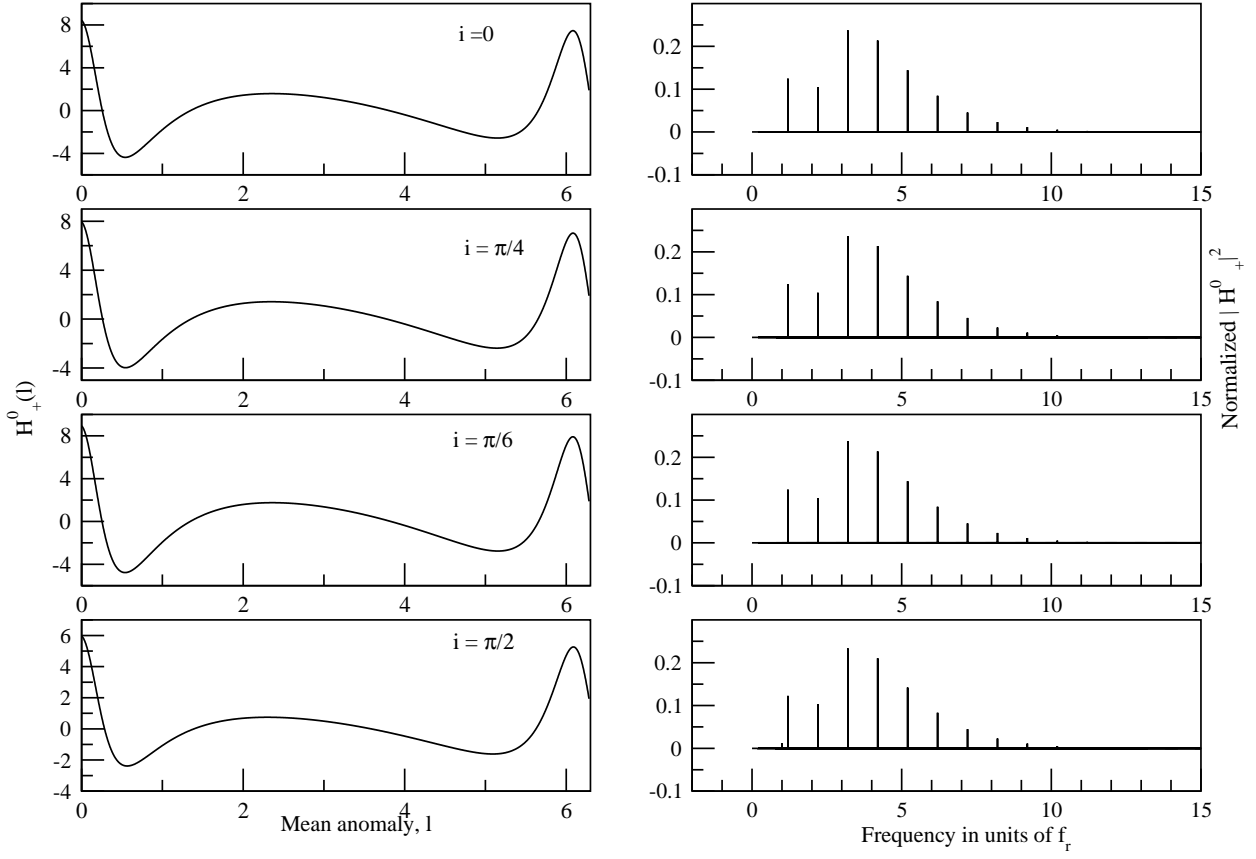}
\bigskip
\caption{
Plots of scaled GW polarization waveform, $H^{0}_{+}$
as a function of mean anomaly, $l$  and
its  corresponding
normalized relative power spectrum for
Newtonian orbital motion, when the orbital inclination angle $i$ is varied.
In all frames, eccentricity  $e=0.5$ and 
periastron precession constant, $k=0.1$.}
\label{fig:p_N_0.5_0.1}
\end{figure}

\newpage
\begin{figure}
\epsfxsize=6.5in \epsfbox{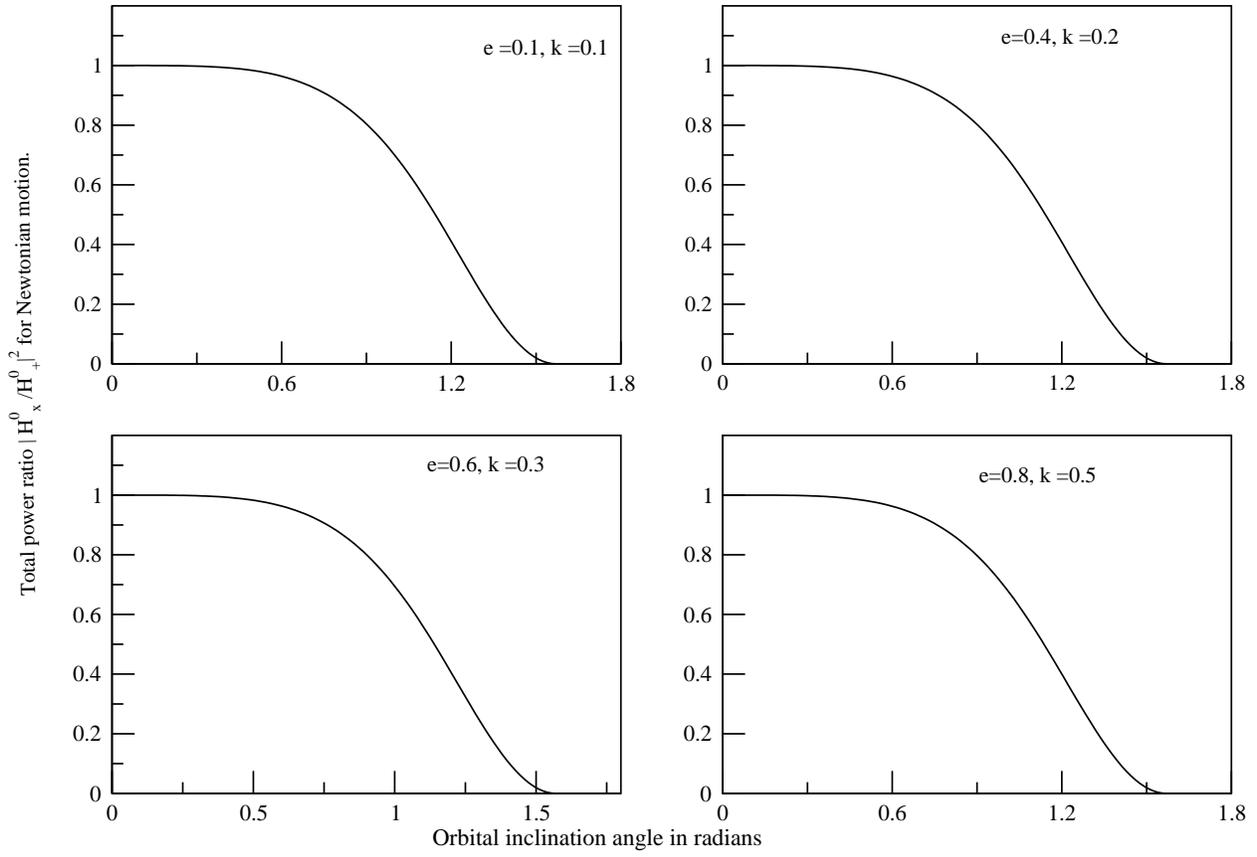}
\bigskip
\caption{
The ratio of the total power   
measured in $\times$ and $+$ polarization for Newtonian motion
as a function of orbital inclination angle $i$ for various
values of  periastron precession constant $k$  and   
eccentricity $e $.
From the  plots, it is clear that the ratio is independent of the orbital elements like $e$ and $k$.
}
\label{fig:i_N}
\end{figure}
\newpage       

\newpage
\begin{figure}
\epsfxsize=6.5in \epsfbox{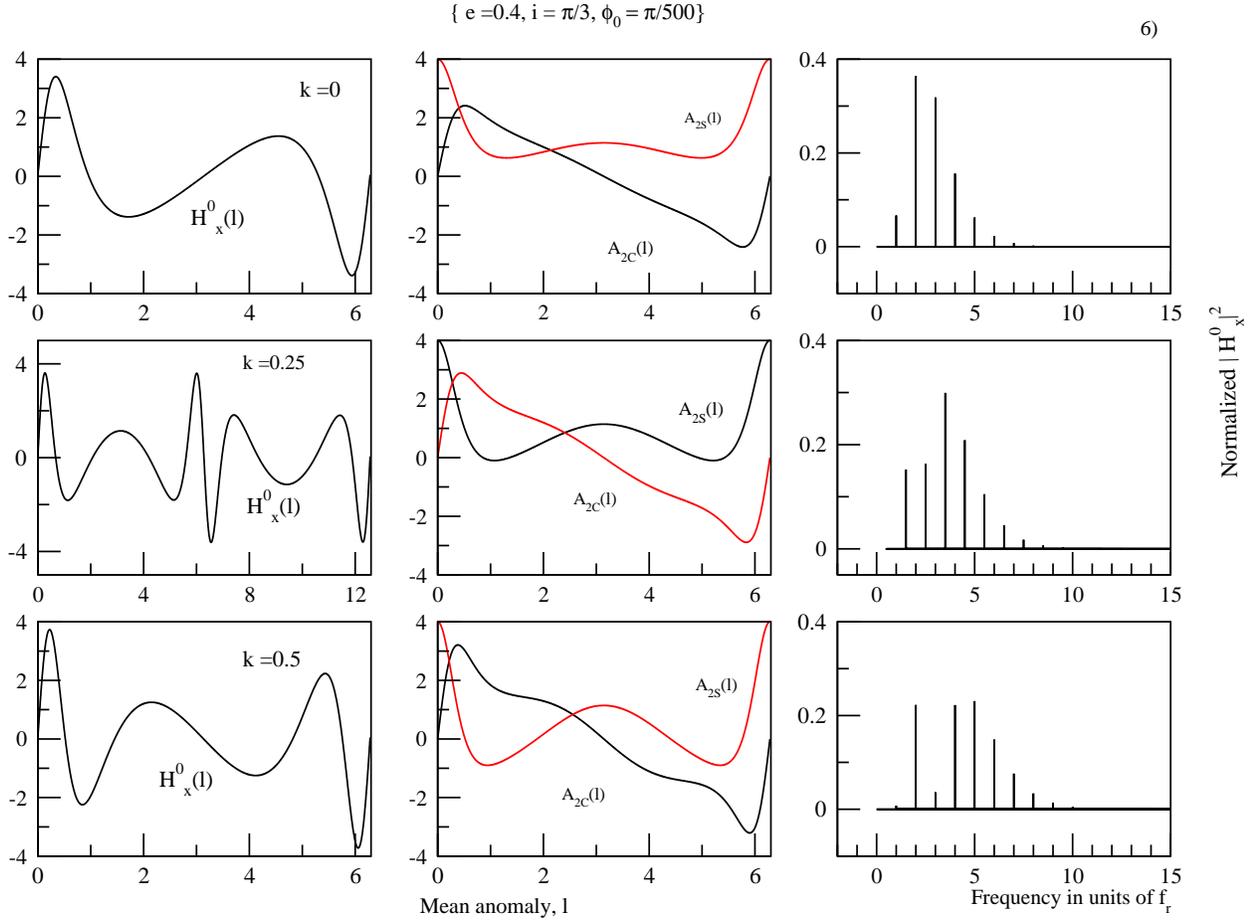}
\bigskip
\caption{
Plots of $H^{0}_{\times}{l}$, $ {\rm A_{2C}}(l), {\rm A_{2S}}(l)$ 
as a function of mean anomaly, $l$ 
and their relative power spectra constructed 
using Eq. (\ref{sp9}) and directly using $H^{0}_{\times}(l)$.
The orbital motion is Newtonian accurate and we employ certain
specific values of $k =0,0.5 $ and $0.25$.
The relative power spectra, plotted in third column, are 
numerically identical, hence indistinguishable.   
Note that plots in second column are always $2\,\pi$ periodic,
while those in the  first column  are $2\,\pi, 2\,\pi $ and $4\,\pi$ periodic
for $k=0$, $k=0.5$ and $k=0.25$ respectively
}
\label{fig:c_check}
\end{figure}
\newpage

\begin{figure}
\epsfxsize=6.5in \epsfbox{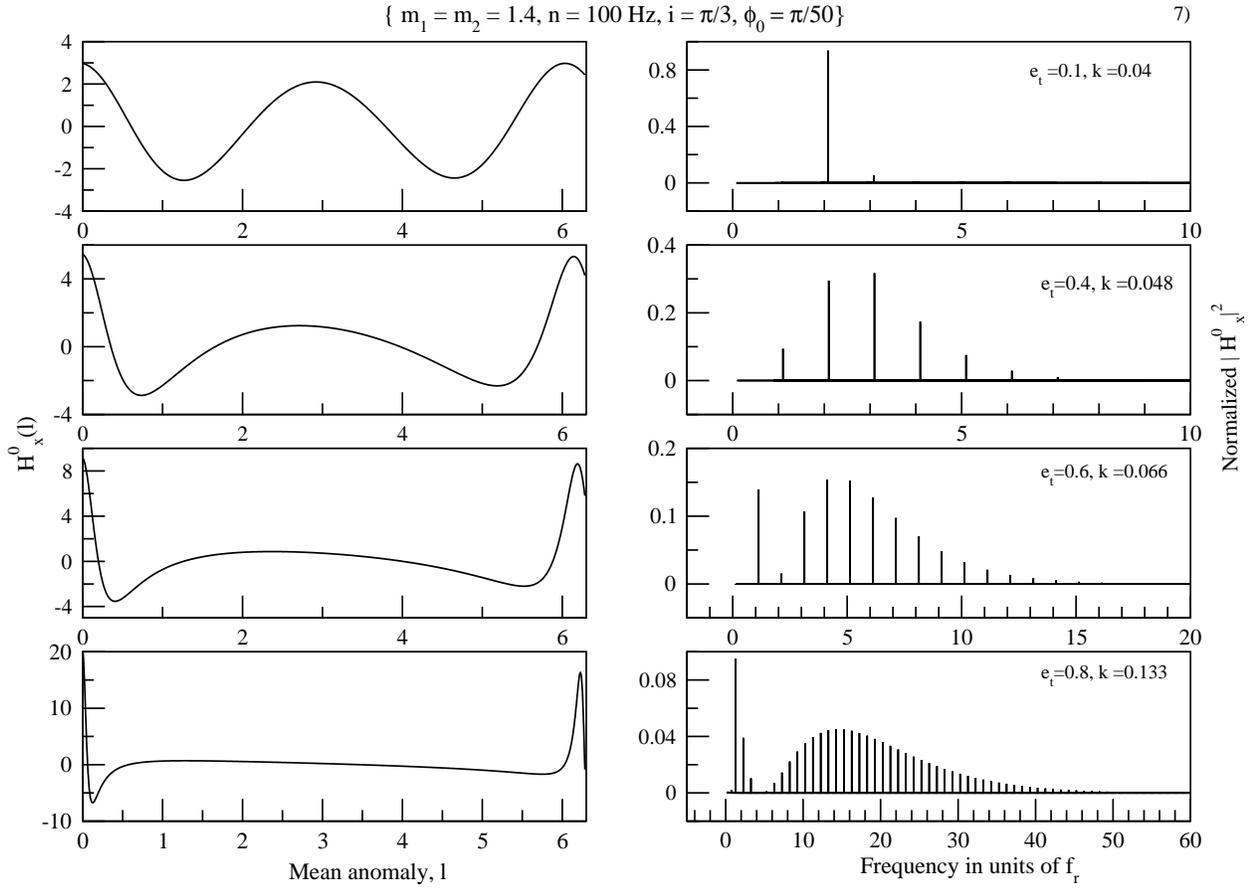}
\bigskip
\caption{
Plots for scaled GW polarization waveform, $H^{0}_{\times} (l)$
and  corresponding
normalized relative power spectrum using
2PN accurate orbital motion, for various values of $e_t$. 
Unlike in Fig. \ref{fig:c_N_k_0.1},
the value of $k$ cannot be independently
chosen since it is uniquely determined by the values of
$m_1, m_2, n $ and $e_t$.
The observations from plots are similar to Fig. \ref{fig:c_N_k_0.1}, as 
variation in $k$ is small compared to that in $e_t$.
In all frames, $m_1$ = $m_2 $ = 1.4 $M_{\odot}$,    
orbital inclination angle $i = {\pi \over 3} $ 
and mean motion $n = 100\,$ radians per second. 
}
\label{fig:c_2_1.4_1.4_100}
\end{figure}

\newpage
\begin{figure}
\epsfxsize=6.5in \epsfbox{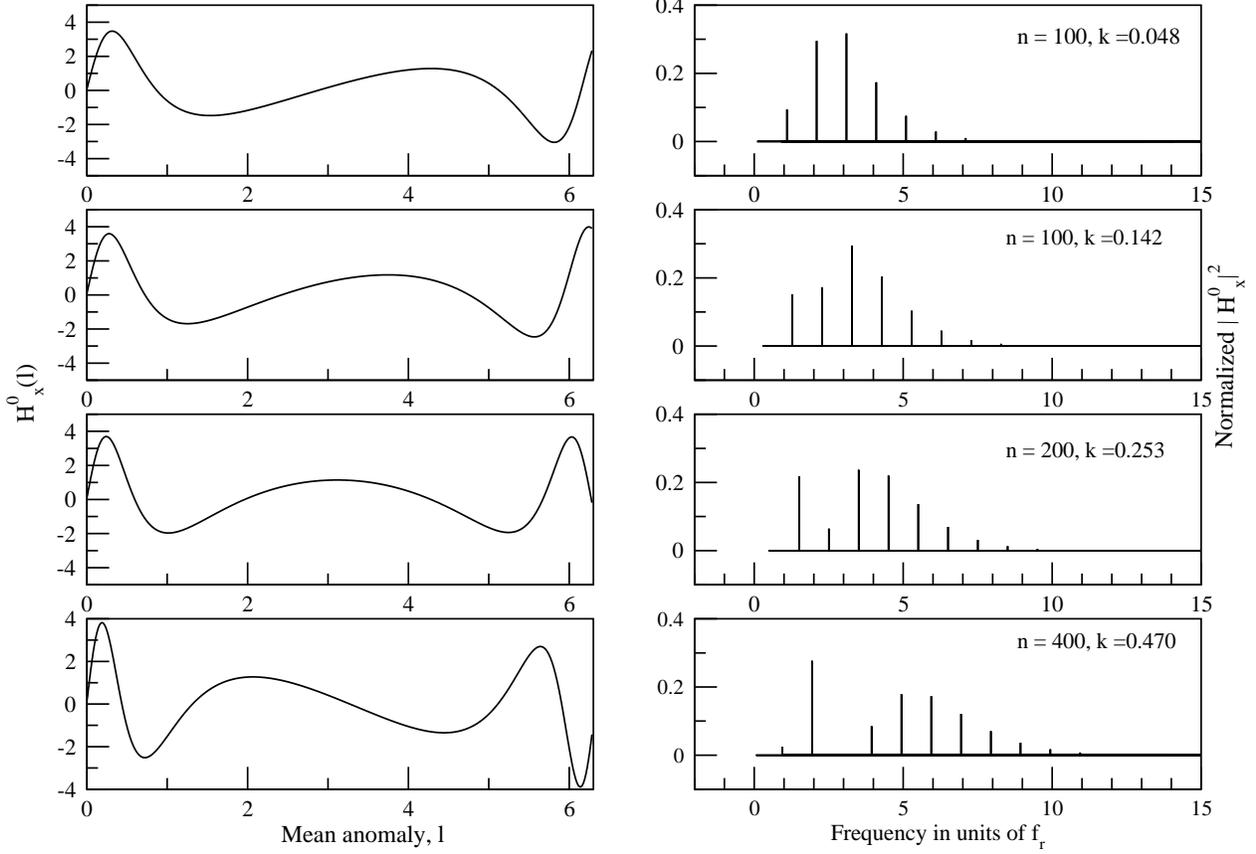}
\bigskip
\caption{
The configuration is similar to Fig. \ref{fig:c_2_1.4_1.4_100}, but
in the panels we vary masses and  mean motion     
to get  the  different values of $k$, rather than $e$. 
For all plots $e_t= 0.4 $ and $i ={\pi \over 3}$.
Qualitatively, conclusions are similar to Fig. \ref{fig:c_N_e_0.5}.
Unit for $n$ will be radians per second.
}
\label{fig:c_2_e_0.4}
\end{figure}

\newpage
\begin{figure}
\epsfxsize=6.5in \epsfbox{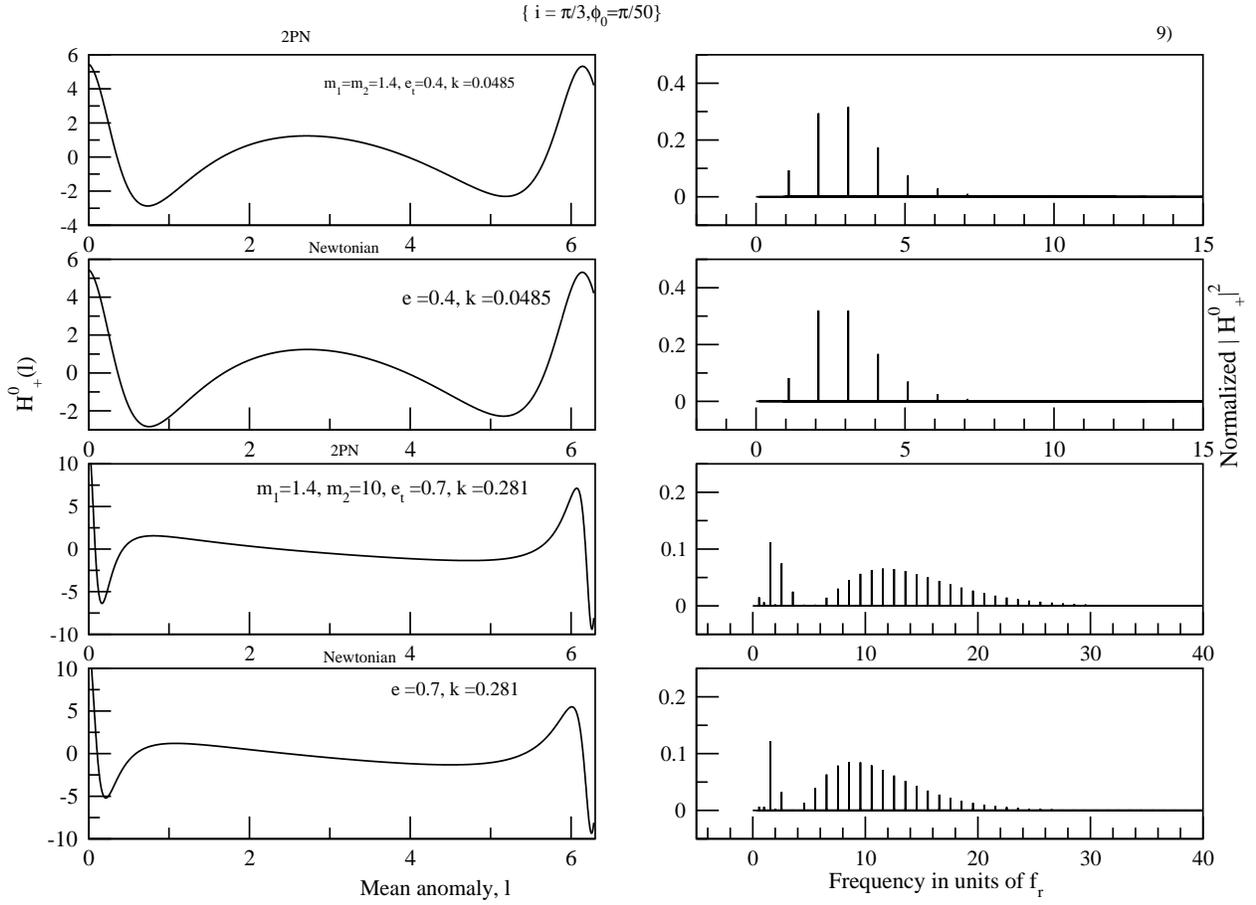}
\bigskip
\caption{
Plots of scaled GW polarization waveform, $H^{0}_{+} (l)$
and corresponding
normalized relative power spectrum for 
Newtonian and  2PN accurate orbital  motion. 
Given $e_t$, we vary 
values for $m_1$, $m_2 $ and $n$, to make 
$k$ values the same in the  2PN and the  Newtonian 
accurate orbital motion.
In all frames,  orbital inclination angle 
is $ { \pi \over 3} $.  
Panels in  the  2nd and 4th rows  are for Newtonian orbital motion, whereas
panels in the  1st and 3rd rows are for 2PN accurate orbital motion. 
We see  minor  quantitative differences in the position and 
strength of spectral lines at these two orders. 
}
\label{fig:p_2_N_comp}
\end{figure}

\newpage
\begin{figure}
\epsfxsize=6.5in \epsfbox{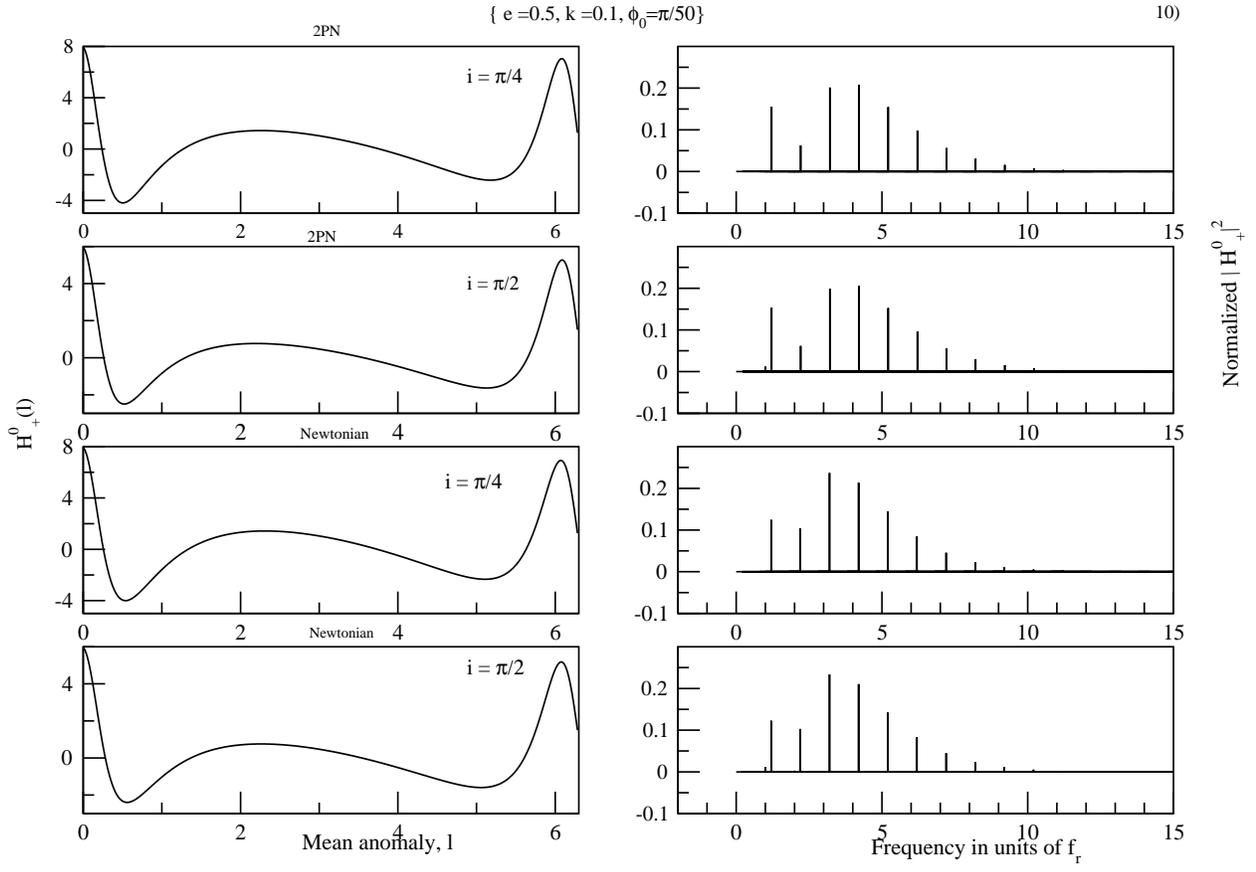}
\bigskip
\caption{
Plots for the  scaled GW polarization waveform, $H^{0}_{+}$
as a function of mean anomaly, $l$  and the 
corresponding
normalized relative power spectrum using 2PN and Newtonian
accurate orbital motion.                          
We vary the orbital inclination angle, $i$  keeping
other orbital elements constant at Newtonian and 2PN level.
Conclusions are similar to Fig. \ref{fig:p_N_0.5_0.1}.
}  
\label{fig:I_p_2_N_comp}
\end{figure}


\begin{references}
\bibitem{fn3}
Following conventions used
in celestial mechanics,
the orbital phase was measured from
the line of nodes in \cite{BIWW96}, which coincide with 
 $+$ve y-axis in Fig. \ref{fig:config}.
 However, we measure the orbital 
phase from $+$ve x-axis.
This is because the    
2PN accurate
generalized quasi-Keplerian parametrization is given in the usual
polar coordinates, $r $ and $\phi$.
The angular variable $\phi$, appearing in polar 
coordinates, is usually measured from $+$ve x-axis 
as shown in Fig. \ref{fig:config}.  
This is why we use angular transformation relation
$ \phi \equiv \lambda  + {\rm W}(l) 
 = \left ( \phi_{\rm BIWW} -{\pi \over 2} \right ) $  to compare
circular limit of our expressions for $ h_{+} $ and $ h_{\times} $ 
with those presented in \cite{BIWW96}, modulo the tail terms.
\bibitem{fn1}
At Newtonian order, units for $f_r$ is  arbitrary.
This is because the only relevant parameters
 $e$ and $k$ don't
define $n$ uniquely when we vary $l = n\,(t -t_0)$ from $0$ to $2\,\pi$
to get $H^{0}_{\times,+}(l)$.
However, at 2PN order, units for $f_r$ is Hertz,
since to get realistic values of
$k$, defined in terms of $m_1, m_2, n$ and $e$, we choose
$f_r$ in Hertz along with $m_1 $ and $m_2 $ in solar mass units.
\bibitem{fn2}
We consider $k =0, 0.25, 0.5$ values as they make $H^{(0)}_{\times} (l)$
periodic within a couple of orbits.
Note that $k =0 $ may be thought to represent a binary with vanishing
periastron advance.
\end{references}
\end{document}